\documentclass[11pt,a4paper]{article}
\pdfoutput=1
\usepackage{graphicx}
\usepackage{enumitem}
\usepackage{booktabs}
\usepackage{jcappub}
\usepackage{aas_macros_JCAP}
\newcommand{\Om}{\Omega_\mathrm{m}}
\newcommand{\Ob}{\Omega_\mathrm{b}}
\newcommand{\Ol}{\Omega_\Lambda}
\newcommand{\Ok}{\Omega_\mathrm{k}}

\newcommand{\hMpc}{h^{-1}{\rm Mpc}}

\newcommand{\kmsMpc}{\frac{\rm km}{\rm s\,Mpc}}

\newcommand{\void}{\mathrm{v}}

\newcommand{\rs}{r_s}
\newcommand{\dc}{\delta_c}
\newcommand{\xibar}{\overline{\xi}}

\newcommand{\C}{\mathbf{C}}
\newcommand{\DA}{D_\mathrm{A}}

\title{Precision cosmology with voids in the final BOSS data}

\author[a]{Nico Hamaus,}
\author[b]{Alice Pisani,}
\author[a]{Jin-Ah Choi,}
\author[c,d]{Guilhem Lavaux,}
\author[c,d,e]{Benjamin D. Wandelt,}
\author[a,f]{and Jochen Weller}

\affiliation[a]{Universit\"ats-Sternwarte M\"unchen, Fakult\"at f\"ur Physik, Ludwig-Maximilians Universit\"at, Scheinerstr. 1, 81679 M\"unchen, Germany}
\affiliation[b]{Princeton University, Department of Astrophysical Sciences, 4 Ivy Lane, Princeton, NJ 08544, USA}
\affiliation[c]{CNRS, Sorbonne Universit\'e, UMR 7095, Institut d'Astrophysique de Paris, 98bis bd Arago, 75014 Paris, France}
\affiliation[d]{Sorbonne Universit\'e, Institut Lagrange de Paris (ILP), 98bis bd Arago, 75014 Paris, France}
\affiliation[e]{Center for Computational Astrophysics, Flatiron Institute, 162 5th Avenue, New York, NY 10010, USA}
\affiliation[f]{Max Planck Institute for Extraterrestrial Physics, Giessenbachstr.~1, 85748 Garching, Germany}

\emailAdd{n.hamaus@physik.lmu.de}

\abstract{
We report novel cosmological constraints obtained from cosmic voids in the final BOSS DR12 dataset. They arise from the joint analysis of geometric and dynamic distortions of average void shapes (i.e., the stacked void-galaxy cross-correlation function) in redshift space. Our model uses tomographic deprojection to infer real-space void profiles and self-consistently accounts for the Alcock-Paczynski (AP) effect and redshift-space distortions (RSD) without any prior assumptions on cosmology or structure formation. It is derived from first physical principles and provides an extremely good description of the data at linear perturbation order. We validate this model with the help of mock catalogs and apply it to the final BOSS data to constrain the RSD and AP parameters $f/b$ and $\DA H/c$, where $f$ is the linear growth rate, $b$ the linear galaxy bias, $\DA$ the comoving angular diameter distance, $H$ the Hubble rate, and $c$ the speed of light. In addition, we include two nuisance parameters in our analysis to marginalize over potential systematics. We obtain $f/b=0.540\pm0.091$ and $\DA H/c=0.588\pm0.004$ from the full void sample at a mean redshift of $z=0.51$. In a flat $\Lambda$CDM cosmology, this implies $\Om=0.312\pm0.020$ for the present-day matter density parameter. When we use additional information from the survey mocks to calibrate our model, these constraints improve to $f/b=0.347\pm0.023$, $\DA H/c=0.588\pm0.003$, and $\Om=0.310\pm0.017$. However, we emphasize that the calibration depends on the specific model of cosmology and structure formation assumed in the mocks, so the calibrated results should be considered less robust. Nevertheless, our calibration-independent constraints are among the tightest of their kind to date, demonstrating the immense potential of using cosmic voids for cosmology in current and future data.
}

\date{\today}


\arxivnumber{2007.07895}

\begin{document}
\maketitle

\section{Introduction\label{sec:intro}}
With the advent of modern sky surveys that map out significant contiguous fractions of the observable Universe in ever greater detail~(e.g.,~\cite{6dFGS,BOSS,DES,DESI,EBOSS,EUCLID,LSST,SDSS,SPHEREX,VIPERS,WFIRST}), it has become possible to investigate its least luminous and most extended constituents: cosmic voids, vast regions of relatively empty space. Voids are not only fascinating objects in their own right, they may also hold the keys to resolving some of today's open problems in cosmology, a fact that has come into focus only recently (see references~\cite{vdWeygaert2009,vdWeygaert2016,Pisani2019} for an overview). Cosmic voids can be thought of pocket universes in which dark energy became important much earlier than elsewhere in the cosmos~\cite{Bos2012,Pisani2015a,Verza2019}. Making up the bulk of large-scale structure, they play a major role in the formation of its web-like pattern~\cite{Gregory1978,Zeldovich1982,deLapparent1986,Sheth2004,vdWeygaert2009}. This pattern contains a wealth of information on the fundamental properties of the Universe and voids have been shown to be sensitive probes thereof, such as its initial conditions~\cite{vdWeygaert1993,Chan2019}, its matter~\cite{Peebles2001,Nusser2005,Park2007,Lavaux2010,Sutter2012b,Sutter2014b,Hamaus2016,Mao2017} and energy components~\cite{Granett2008,Biswas2010,Ilic2013,Cai2014,Planck2014,Kovacs2019}. Moreover not only cosmology, but the very nature of gravity can be investigated with voids~\cite{Clampitt2013,Zivick2015,Cai2015,Achitouv2016,Falck2018,Paillas2019,Perico2019}, because it is gravity that gives rise to their formation and evolution in the first place. This happens via gravitational collapse of initially over-dense regions in the mass distribution into sheets, filaments, and clusters where galaxies form. The remaining space is occupied by voids that are characterized by the coherent flow of (predominantly dark) matter~\cite{Shandarin2011,Abel2012,Hahn2015}. Baryonic matter is even more scarce inside voids~\cite{Paillas2017,Pollina2017}, implying a significant advantage in the attempt to model their evolution when compared to the other structure types. This opens up the opportunity to use voids as laboratories for the physics of dark matter~\cite{Yang2015,Reed2015,Baldi2018} and other elusive particles, such as neutrinos, that freely permeate their interiors~\cite{Massara2015,Banerjee2016,Kreisch2019,Schuster2019}.

On the whole, cosmic voids offer radically novel avenues towards probing the fundamental laws of physics that govern our Universe. General Relativity (GR) relates the distribution of matter and energy to the geometry of spacetime via Einstein's field equations. Consequently, observations of the cosmic expansion history allow constraining the material components of the Universe. In this manner supernova distance measurements have inferred the existence of dark energy (in the form of a cosmological constant $\Lambda$) that dominates the cosmic energy budget today and is responsible for the observed accelerated expansion~\cite{Riess1998,Perlmutter1999}. Yet, the fundamental nature of dark energy remains mysterious and further efforts are necessary towards explaining its origin. This has been attempted in studying the expansion history by employing standard rulers, such as the Baryon Acoustic Oscillation~(BAO) feature imprinted in the spatial distribution of galaxies on scales of $\sim105\hMpc$~\cite{Eisenstein2005}. Because the physics of recombination is well understood, the BAO feature can be modeled from first principles and therefore provides a scale of known extent: a standard ruler. Observations of the BAO in the pairwise distribution of galaxies have been successful in constraining the expansion history and so far consistently confirmed the $\Lambda$CDM paradigm (e.g.,~\cite{Alam2017,SanchezA2017,Beutler2017a}).

A similar approach can be adopted for objects of known shape: standard spheres. Both methods are based on the cosmological principle, stating the Universe obeys statistical isotropy and homogeneity. A relatively novel technique is the use of cosmic voids in this context. After averaging over all orientations, their shape obeys spherical symmetry, even though individual voids may not~\cite{Park2007,Platen2008}. Therefore, stacked voids can be considered as standard spheres~\cite{Ryden1995,Lavaux2012,Sutter2012b,Sutter2014b,Hamaus2015,Hamaus2016,Mao2017} with sizes typically ranging from $10\hMpc$ to $100\hMpc$. This means that in a finite survey volume one can find a substantially larger number of such spheres than rulers in the form of BAO, allowing a significant reduction of statistical uncertainties and to probe a wider range of scales. Standard spheres can be used to constrain the expansion history: only if the fiducial cosmological model in converting redshifts to distances is correct, stacked voids appear spherically symmetric, a technique known as the Alcock-Paczynski (AP) test~\cite{Alcock1979}. In principle this test merely involves a trivial rescaling of coordinates. However, in observational data the spherical symmetry is broken by redshift-space distortions (RSD), which are caused by the peculiar motions of galaxies along the line of sight. Therefore, a successful application of the AP test to constrain cosmological parameters from voids crucially relies on the ability to robustly model their associated RSD~\cite{Ryden1996}. The latter are notoriously complex and difficult to model in the clustering statistics of galaxies, especially on intermediate and small scales, where non-linear clustering and shell crossing occurs. It has been shown that these limitations can be mitigated in voids, which are dominated by a laminar, single-stream flow of matter that is well described even by linear theory~\cite{Paz2013,Hamaus2014b,Hamaus2014c,Pisani2015b,Hamaus2015,Hamaus2016}. This, and the additional virtue of enabling constraints on the growth rate of structure, has sparked the recent interest for void RSD in the literature~\cite{Cai2016,Chuang2017,Achitouv2017a,Hawken2017,Hamaus2017,Correa2019,Achitouv2019,Nadathur2019a,Nadathur2019b,Hawken2020}.

In this paper we present a first cosmological analysis of voids from the combined galaxy sample of the final BOSS~\cite{BOSS} data. Our model self-consistently accounts for RSD and the AP effect, without the need for any external inputs from simulations or mock catalogs. The detailed derivation of the underlying theory is outlined in section~\ref{sec:theory}, along with a definition of all relevant observables. Section~\ref{sec:analysis} presents the observed and simulated data sets considered and our method for the identification and characterization of voids therein. Our analysis pipeline is then validated based on mock data in the first part of section~\ref{sec:analysis}, the second part is devoted to process the real data. We demonstrate that the AP test with voids offers cosmological constraints that are competitive with other large-scale structure probes. Section~\ref{sec:discussion} is used to summarize our constraints and to discuss them in the light of previous works on voids (see figure~\ref{fig:comparison}), representing the strongest such constraints in the literature. Finally, we draw our conclusions in section~\ref{sec:conclusion}.

\section{Theory \label{sec:theory}}

\subsection{Dynamic distortion \label{subsec:dynamic}}
In cosmology, our observables are the redshifts $z$ and angular sky coordinates $\boldsymbol{\theta}=(\vartheta,\varphi)$ of an astronomical object. The comoving distance of this object is defined as
\begin{equation}
\chi_\parallel(z)=\int_0^z\frac{c}{H(z')}\mathrm{d}z'\;,
\label{chi_par}
\end{equation}
where $H(z)$ is the Hubble rate and $c$ the speed of light. The observed redshift $z$ can contain contributions from many different physical effects, but the most important ones are the cosmological Hubble expansion $z_h$ and the Doppler effect $z_d$. The total observed redshift $z$ is then given by~\cite{Davis2014}
\begin{equation}
1+z = (1+z_h)(1+z_d)\;. \label{z_tot}
\end{equation}
The Doppler effect is caused by peculiar motions along the line of sight, $z_d=v_\parallel/c$. Because $z_d$ is typically small compared to $z_h$, we can write
\begin{equation}
\chi_\parallel(z)\simeq\chi_\parallel(z_h) + \frac{c(1+z_h)}{H(z_h)}z_d\;.
\label{chi_rsd}
\end{equation}
The transverse comoving distance for an observed angle $\theta\equiv|\boldsymbol{\theta}|$ on the sky is defined as
\begin{equation}
\chi_\perp(z) = \DA(z)\,\theta\;,
\label{chi_per}
\end{equation}
where the comoving angular diameter distance is given by
\begin{equation} \DA(z) = \frac{c}{H_0\sqrt{-\Ok}}\sin\left(\frac{H_0\sqrt{-\Ok}}{c}\chi_\parallel(z)\right)\;,
\label{D_A}
\end{equation}
with the Hubble constant $H_0\equiv H(z=0)$ and present-day curvature parameter $\Ok$. In a flat universe with $\Ok=0$, equation~(\ref{D_A}) reduces to $\DA(z)=\chi_\parallel(z)$. Now, given the observed coordinates $(z,\vartheta,\varphi)$, we can transform to the comoving space vector $\mathbf{x}$ via
\begin{equation}
\mathbf{x}(z,\vartheta,\varphi) = \DA(z)\begin{pmatrix}\cos\vartheta\cos\varphi\\\sin\vartheta\cos\varphi\\\sin\varphi\end{pmatrix}\;,
\label{x_comoving}
\end{equation}
where $\DA(z)\simeq \DA(z_h)+cz_d(1+z_h)/H(z_h)$, analogously to equation~(\ref{chi_rsd}). Hence, using $z_d=v_\parallel/c$, we can write
\begin{equation}
\mathbf{x}(z) \simeq \mathbf{x}(z_h) + \frac{1+z_h}{H(z_h)}\mathbf{v}_\parallel\;,
\label{x_rsd}
\end{equation}
where $\mathbf{v}_\parallel$ is the component of the velocity vector $\mathbf{v}$ along the line-of-sight direction. We describe the location and motion of tracers by vectors in comoving space, upper-case letters are used for void centers, lower-case letters for galaxies. The observer's location is chosen to be at the origin of our coordinate system, the void center position is denoted by $\mathbf{X}$ with redshift $Z$ and the galaxy position by $\mathbf{x}$ with redshift $z$. The redshift $Z$ of the void center is not a direct observable, but it is constructed via the redshifts of all the surrounding galaxies that define it (see section~\ref{subsec:voids}). Moreover, we pick the direction of the void center as our line of sight, i.e. $\mathbf{X}/|\mathbf{X}|$, and adopt the distant-observer approximation, assuming that $\mathbf{x}$ and $\mathbf{X}$ are parallel.

\begin{figure}[t]
	\centering
	\resizebox{\hsize}{!}{
		\includegraphics[trim= 0 0 0 40]{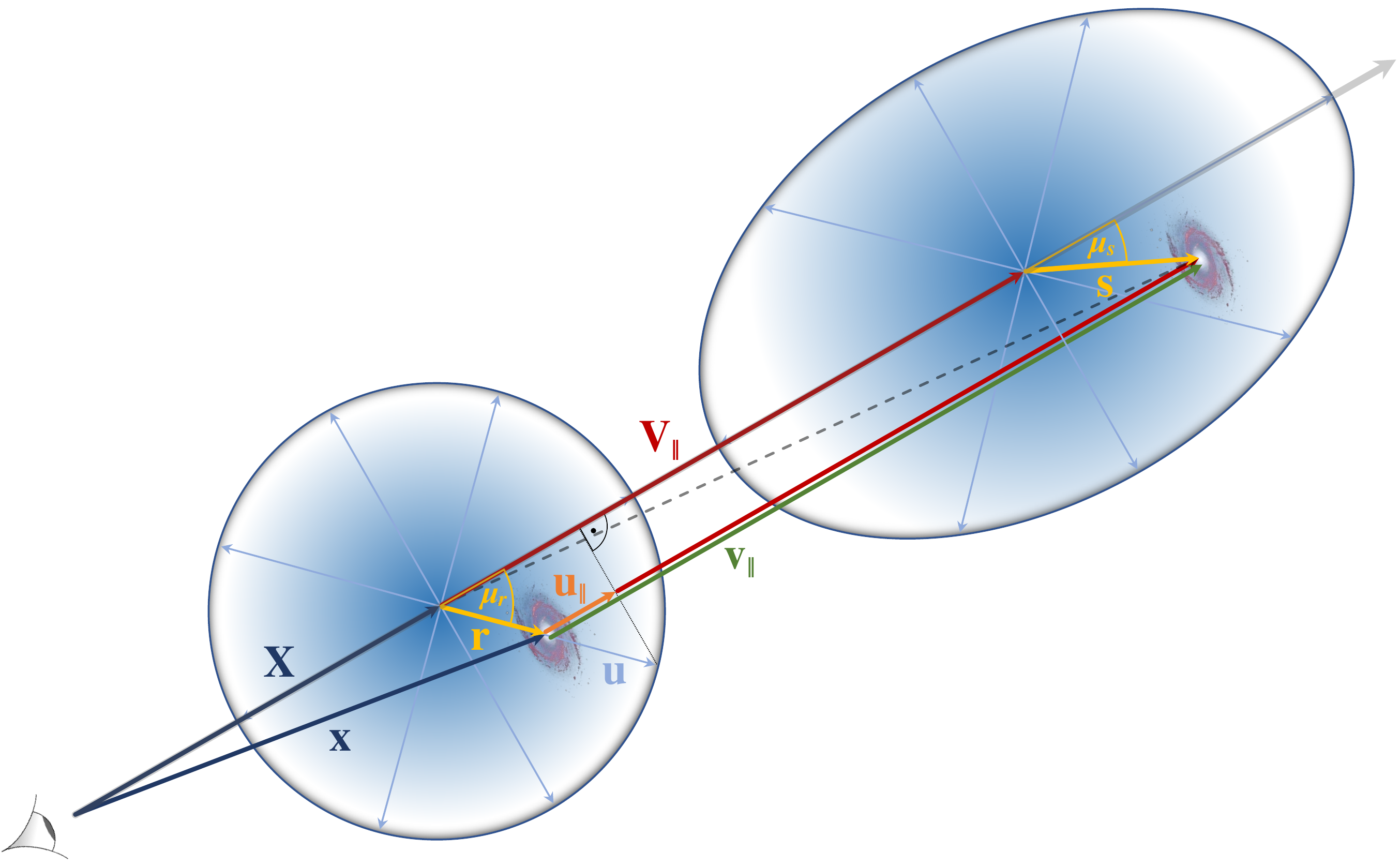}}
	\caption{Separation vector between the comoving void center location $\mathbf{X}$ and the galaxy location $\mathbf{x}$ in real space ($\mathbf{r}$, left) and in redshift space ($\mathbf{s}$, right). The peculiar line-of-sight velocity $\mathbf{v}_\parallel$ of every galaxy that defines the void can be decomposed into the peculiar velocity of the void center $\mathbf{V}_\parallel$ and the galaxy's relative velocity $\mathbf{u}_\parallel$ with respect to this center. For simplicity, the illustration displays $\mu$ instead of $\cos^{-1}(\mu)$ to indicate line-of-sight angles and shows velocity displacements in units of $(1+z_h)/H(z_h)$. This yields the relation $\mathbf{s}=\mathbf{r}+\mathbf{u}_\parallel$ between real- and redshift-space separations.}
	\label{fig:voidstretch}
\end{figure}

Let us first consider real space, where the Doppler effect is neglected ($z_d=0$) and hence $\mathbf{x}(z)=\mathbf{x}(z_h)$. The vector $\mathbf{r}\equiv\mathbf{x}-\mathbf{X}$ connects the two positions at a comoving distance of $r=|\mathbf{r}|$. Similarly, we define the relative velocity $\mathbf{u}$ between a void center of velocity $\mathbf{V}$ and a galaxy of velocity $\mathbf{v}$ as $\mathbf{u}\equiv\mathbf{v}-\mathbf{V}$. Now, if we consider redshift space with $z_d\ne0$ and use equation~(\ref{x_rsd}), the separation vector between void center and galaxy becomes
\begin{equation}
\mathbf{x}(z)-\mathbf{X}(Z) \simeq \mathbf{x}(z_h)-\mathbf{X}(Z_h) + \frac{1+z_h}{H(z_h)}\left(\mathbf{v}_\parallel-\mathbf{V}_\parallel\right) = \mathbf{r} + \frac{1+z_h}{H(z_h)}\mathbf{u}_\parallel \equiv \mathbf{s}\;.
\label{s(r)}
\end{equation}
Here we have used the approximation $z_h\simeq Z_h$ for the Doppler term, which is accurate to $\mathcal{O}(10^{-3})$ on scales of $r\sim\mathcal{O}(10^{1})\hMpc$ and velocities of $u_\parallel\sim\mathcal{O}(10^2)\kmsMpc$. This approximation becomes exact when we consider the average of this vector over many different voids -- a so-called void stack. In that case the cosmological principle ensures statistical homogeneity and isotropy such that $\langle\mathbf{V}\rangle=\langle\mathbf{V}_\parallel\rangle=0$, since there is no preferred observer (or void, for that matter) in the Universe. In other words: the average void moves with the background Hubble flow. Thus, the vector $\mathbf{s}$, which connects void centers and galaxies in redshift space, does not depend on the individual motions of galaxies or void centers, but only on their relative velocity $\mathbf{u}_\parallel$ along the line of sight. Of course this only applies to those galaxies that are part of the same void, not to the relative velocities of galaxies that are associated with distinct voids at larger separation (see~\cite{Hamaus2014a,Chan2014,Hamaus2014c,Liang2016,Chuang2017,Voivodic2020} for an account on large-scale void-galaxy cross-correlations and~\cite{Sutter2014c,Ruiz2015,Lambas2016,Ceccarelli2016,Wojtak2016,Lares2017b} on the motions and pairwise velocity statistics of voids). An illustration of this is shown in figure~\ref{fig:voidstretch}: voids experience a translation and a deformation between real and redshift space, but the translational component does not enter in the separation vector $\mathbf{s}$ for galaxies that belong to the void. As long as voids can be considered as coherent extended objects, their centers move along with the galaxies that define them from real to redshift space. This property distinguishes voids from galaxies, which are typically treated as point-like particles in the context of large-scale structure. The same reasoning applies to galaxy clusters, the overdense counterparts of voids: the center of mass is defined by the cluster member galaxies, which are observed in redshift space. Their relative (virial) motion with respect to the center of mass results in an elongated stacked cluster shape, irrespective of the movement of the entire cluster~\cite{Croft1999,Zu2013,Marulli2017,Farahi2016,Cai2017b}.

\subsection{Geometric distortion \label{subsec:geometric}}
Given the observed angular sky-coordinates and redshifts of galaxies and void centers, the comoving distance $s = (s_\parallel^2 + s_\perp^2)^{1/2}$ in redshift space between any void-galaxy pair is determined by equations~(\ref{chi_par}) and~(\ref{chi_per}),
\begin{equation}
s_\parallel = \frac{c}{H(z)}\delta z\quad\mathrm{and}\quad s_\perp = \DA(z)\delta\theta\;, \label{comoving}
\end{equation}
where $\delta z$ and $\delta\theta$ are the redshift and angular separation of the pair, respectively. This transformation requires the Hubble rate $H(z)$ and the comoving angular diameter distance $\DA(z)$ as functions of redshift, so a particular cosmological model has to be adopted. We assume a flat $\Lambda$CDM cosmology with
\begin{equation}
H(z) = H_0\sqrt{\Om(1+z)^3+\Ol}\;, \label{H(z)}
\end{equation}
where $\Om$ and $\Ol=1-\Om$ are today's density parameters for matter and a cosmological constant, respectively. For the low redshift range considered in this paper, we can neglect the radiation density for all practical purposes. Given a cosmology with parameters $\boldsymbol{\Omega}$ we can now convert angles and redshifts into comoving distances according to equation~(\ref{comoving}). However, the precise value of these parameters is unknown, as is the underlying cosmology, so we can only estimate the true $H(z)$ and $\DA(z)$ with a fiducial model and parameter set $\boldsymbol{\Omega}'$, resulting in the estimated distances
\begin{equation}
s_\parallel' = \frac{H(z)}{H'(z)}s_\parallel\equiv q_\parallel^{-1} s_\parallel\;\mathrm{,}\qquad s_\perp' = \frac{\DA'(z)}{\DA(z)}s_\perp\equiv q_\perp^{-1} s_\perp\;,
\end{equation}
where the primed quantities are evaluated in the fiducial cosmology and $(q_\parallel, q_\perp)$ are defined as the ratios between true and fiducial values of $H^{-1}(z)$ and $\DA(z)$, respectively~\cite{SanchezA2017}. Therefore, both magnitude and direction of the separation vector $\mathbf{s}$ may differ from the truth when a fiducial cosmology is assumed. Defining the cosine of the angle between $\mathbf{s}$ and the line of sight $\mathbf{X}/|\mathbf{X}|$ as
\begin{equation}
\mu_s\equiv\frac{\mathbf{s}\cdot\mathbf{X}}{|\mathbf{s}||\mathbf{X}|}=\frac{s_\parallel}{s}\;,
\label{mu_s}
\end{equation}
one can obtain the true $s$ and $\mu_s$ from the fiducial $s'$ and $\mu_s'$ via
\begin{gather}
s = \sqrt{q_\parallel^2 s_\parallel'^2+q_\perp^2s_\perp'^2} = s'\mu_s'q_\parallel\sqrt{1+\varepsilon^2(\mu_s'^{-2}-1)}\;,
\label{s_fid}
\\
\mu_s = \frac{\mathrm{sgn}(\mu_s')}{\sqrt{1+\varepsilon^2(\mu_s'^{-2}-1)}}\;,
\label{mu_s_fid}
\end{gather}
where
\begin{equation}
\varepsilon \equiv \frac{q_\perp}{q_\parallel} =  \frac{\DA(z)H(z)}{\DA'(z)H'(z)}\;.
\label{epsilon}
\end{equation}
If the fiducial cosmology agrees with the truth, $\varepsilon=q_\parallel=q_\perp=1$ and $s=s'$, $\mu_s=\mu_s'$. Conversely, if one of these parameters is measured to be different from unity, one may iteratively vary the fiducial cosmology until the true parameter values $\boldsymbol{\Omega}$ are found. As apparent from equations~(\ref{s_fid}) and~(\ref{mu_s_fid}), absolute distances $s$ in redshift space depend on both $q_\parallel$ and $q_\perp$, whereas angles $\mu_s$ merely depend on their ratio~$\varepsilon$. Exploiting the spherical symmetry of stacked voids via the AP effect therefore constrains $\varepsilon$, but $q_\parallel$ and $q_\perp$ remain degenerate without calibration of $s$ with a known scale (such as the BAO scale, for example). However, void-centric distances are typically expressed in units of the effective void radius $R$, which is defined via the cubic root of the void volume in redshift space (see section~\ref{subsec:voids}). The observed volume is proportional to $s'_\parallel s'^2_\perp$, implying $R=q_\parallel^{1/3}q_\perp^{2/3}R'$ to relate true with fiducial void radii. Then, the ratio $s/R$ only depends on $\varepsilon$, as it is the case for $\mu_s$,
\begin{equation}
\frac{s}{R}=\frac{s'}{R'}\mu_s'\varepsilon^{-2/3}\sqrt{1+\varepsilon^2(\mu_s'^{-2}-1)}\;.
\label{s_R_fid}
\end{equation}

\subsection{Void-galaxy cross-correlation function \label{subsec:correlation}}
The probability of finding a galaxy at comoving distance $r$ from a void center in real space is given by $1+\xi(r)$, where $\xi(r)$ is the void-galaxy cross-correlation function. Due to statistical isotropy, it only depends on the magnitude of the separation vector $\mathbf{r}$, not its orientation. This is no longer the case in redshift space, where peculiar motions break isotropy via the Doppler effect. However, since this causes RSD exclusively along the line-of-sight direction, we can eliminate their impact by projecting the correlation function onto the plane of the sky. This yields the projected correlation function $\xi_p$,
\begin{equation}
1+\xi_p(r_\perp) = \frac{\int\left[1+\xi(r)\right]\mathrm{d}r_\parallel}{\int\mathrm{d}r_\parallel} = \frac{\int\left[1+\xi^s(\mathbf{s})\right]\mathrm{d}s_\parallel}{\int\mathrm{d}s_\parallel} = 1+\xi^s_p(s_\perp)\;,
\label{xi_p}
\end{equation}
where $r=(r_\parallel^2+r_\perp^2)^{1/2}$ and $\xi^s(\mathbf{s})$ is the redshift-space correlation function, which can now be expressed as
\begin{equation}
1+\xi^s(\mathbf{s}) = \left[1+\xi(r)\right]\frac{\mathrm{d}r_\parallel}{\mathrm{d}s_\parallel}\;.
\label{xi^s}
\end{equation}
Equation~(\ref{s(r)}) provides the relation between $\mathbf{s}$ and $\mathbf{r}$. In particular, its line-of-sight component can be obtained via taking the dot product with $\mathbf{X}/|\mathbf{X}|$,
\begin{equation}
s_\parallel = r_\parallel + \frac{1+z_h}{H(z_h)}u_\parallel\;,
\label{s_par(r_par)}
\end{equation}
and hence
\begin{equation}
\frac{\mathrm{d}r_\parallel}{\mathrm{d}s_\parallel} = \left(1 + \frac{1+z_h}{H(z_h)}\,\frac{\mathrm{d}u_\parallel}{\mathrm{d}r_\parallel}\right)^{-1}\;.
\label{dr_par/ds_par}
\end{equation}
The relative peculiar velocity $\mathbf{u}$ between void centers and their surrounding galaxies can be derived by imposing local mass conservation. At linear order in the matter-density contrast~$\delta$ and assuming spherical symmetry in real space, the velocity field is given by~\cite{Peebles1980}
\begin{equation}
\mathbf{u}(\mathbf{r}) = -\frac{f(z_h)}{3}\frac{H(z_h)}{1+z_h}\Delta(r)\,\mathbf{r}\;,
\label{u(r)}
\end{equation}
where $f(z)\equiv-\frac{\mathrm{d}\!\ln D(z)}{\mathrm{d}\!\ln(1+z)}$ is the linear growth rate, defined as the logarithmic derivative of the linear growth factor $D(z)$ with respect to the scale factor. In $\Lambda$CDM, the linear growth rate is well approximated by
\begin{equation}
f(z)=\left[\frac{\Omega_\mathrm{m}(1+z)^3}{H^2(z)/H_0^2}\right]^\gamma,
\label{growth_rate}
\end{equation}
with a growth index of $\gamma\simeq0.55$~\cite{Lahav1991,Linder2005}. Furthermore, $\Delta(r)$ is the average matter-density contrast inside a spherical region of comoving radius~$r$,
\begin{equation}
\Delta(r) = \frac{3}{r^3}\int_0^r\delta(r')r'^2\,\mathrm{d}r'\;. \label{Delta(r)}
\end{equation}
Although the matter-density contrast in the vicinity of void centers is not necessarily in the linear regime (i.e., $|\delta|\ll1$), contrary to over-dense structures (such as galaxy clusters and their dark matter halos) it is bounded from below by the value of $-1$. In simulations it has been shown that equation~(\ref{u(r)}) provides an extremely accurate description of the local velocity field in and around most voids~\cite{vdWeygaert1993,Hamaus2014b}. While peculiar velocities at the void boundaries can be due to very non-linear structures, spherical averaging over large sample sizes helps to restore the validity of linear theory to a high degree. Only the smallest and most underdense voids exhibit a non-linear behavior close to their centers and may in fact collapse anisotropically under their external mass distribution~\cite{vdWeygaert1993,Sheth2004}. We can now evaluate the derivative term in equation~(\ref{dr_par/ds_par}) as
\begin{equation}
\frac{1+z_h}{H(z_h)}\frac{\mathrm{d}u_\parallel}{\mathrm{d}r_\parallel} = -\frac{f(z_h)}{3}\Delta(r)-f(z_h)\mu_r^2\left[\delta(r)-\Delta(r)\right]\;,
\label{du_par/dr_par}
\end{equation}
where $\mu_r=r_\parallel/r$ and the identity $\frac{\mathrm{d}\Delta(r)}{\mathrm{d}r} = \frac{3}{r}\left[\delta(r)-\Delta(r)\right]$ was used. Plugging this back into equation~(\ref{xi^s}) we obtain
\begin{equation}
1+\xi^s(\mathbf{s}) = \frac{1+\xi(r)}{1-\frac{f}{3}\Delta(r)-f\mu_r^2\left[\delta(r)-\Delta(r)\right]}\;.
\label{xi^s_nonlin}
\end{equation}
In order to evaluate this equation at a given observed separation $\mathbf{s}$, we make use of equations~(\ref{s_par(r_par)}) and (\ref{u(r)}),
\begin{equation}
r_\parallel = \frac{s_\parallel}{1 - \frac{f}{3}\Delta(r)}\;,
\label{r_par(s_par)}
\end{equation}
and calculate $r=(r_\parallel^2+r_\perp^2)^{1/2}$ with $r_\perp=s_\perp$. However, equation~(\ref{r_par(s_par)}) already requires knowledge of $r$ in the argument of $\Delta(r)$, so it can only be evaluated by iteration. We therefore start with using $\Delta(s)$ as initial step, and iteratively calculate $r_\parallel$ and $\Delta(r)$ until convergence is reached. In practice we find $5$ iterations to be fully sufficient for that purpose.

Furthermore, in equation~(\ref{xi^s_nonlin}) both the void-galaxy cross-correlation function $\xi(r)$, as well as the void-matter cross-correlation function $\delta(r)$ are required in real space. The former can be obtained via deprojection of equation~(\ref{xi_p}),
\begin{equation}
\xi(r) = -\frac{1}{\pi}\int_r^\infty\frac{\mathrm{d}\xi^s_p(s_\perp)}{\mathrm{d}s_\perp}\frac{\mathrm{d}s_\perp}{\sqrt{s_\perp^2-r^2}}\;,
\label{xi_d}
\end{equation}
making use of the inverse Abel transform~\cite{Pisani2014,Hawken2017}. The latter function $\delta(r)$, also referred to as the void density profile, is not directly observable. However, it has been investigated in $N$-body simulations and can be inferred via the gravitational lensing effect in imaging surveys~\cite{Krause2013,Higuchi2013,Melchior2014}. The parametric form suggested in reference~\cite{Hamaus2014b} (HSW profile) has been shown to accurately describe both simulated~\cite{Sutter2014a,Hamaus2015,Barreira2015,Falck2018,Pollina2017,Perico2019}, as well as observational data~\cite{Hamaus2016,SanchezC2017,Pollina2019,Fang2019},
\begin{equation}
\delta_{\scriptscriptstyle\mathrm{HSW}}(r) = \dc\frac{1-(r/\rs)^\alpha}{1+(r/R)^\beta}\;. \label{HSW}
\end{equation}
Here $R$ is the effective void radius, the scale radius $\rs$ determines where $\delta_{\scriptscriptstyle\mathrm{HSW}}(\rs)=0$, the central underdensity is defined as $\dc\equiv\delta_{\scriptscriptstyle\mathrm{HSW}}(r=0)$, and the power-law indices $\alpha$ and $\beta$ control the inner and outer slopes of the profile. In equation~(\ref{xi^s_nonlin}) these quantities can then be included as free parameters to be constrained by the observed $\xi^s(\mathbf{s})$. This approach has been pursued in the framework of the Gaussian streaming model (GSM)~\cite{Hamaus2015,Hamaus2016}, which incorporates an additional parameter for the velocity dispersion $\sigma_v$ of galaxies. In the limit of $\sigma_v\rightarrow0$, the GSM recovers the result of equation~(\ref{xi^s_nonlin}) at linear order in $\delta$~\cite{Cai2016}. We note that equation~(\ref{HSW}) describes the spherically averaged density profile with respect to the void center, but a similar parametrization exists for profiles centered on the void boundary~\cite{Cautun2016}.

Another option to constrain the void density profile $\delta(r)$ is through its relation to $\xi(r)$, which is equivalent to the void density profile in galaxies. Both simulations~\cite{Pollina2017,Ronconi2019,Contarini2019} and observational approaches~\cite{Pollina2019,Fang2019} have established robust evidence for the relationship between $\delta(r)$ and $\xi(r)$ to be a linear one, such that
\begin{equation}
\xi(r) = b\delta(r)\;,
\label{xi(delta)}
\end{equation}
with a single proportionality constant $b$. This is similar to the relation between the overdensity of tracers and the underlying matter distribution on large scales, where $|\delta|\ll1$~\cite{Desjacques2018}. That condition is not necessarily satisfied in the interiors and immediate surroundings of voids, and the large-scale linear bias does not coincide with the value of $b$ in general, even for the same tracer population. However, the two bias values approach each other for voids of increasing size, and converge in the limit of large effective void radius $R$~\cite{Pollina2017,Pollina2019,Contarini2019}. Using equation~(\ref{xi(delta)}) for $\delta(r)$, we can simply exchange it with $\xi(r)$ by making the replacements $f\rightarrow f/b$ and $\Delta(r)\rightarrow\xibar(r)$, with
\begin{equation}
\xibar(r) = \frac{3}{r^3}\int_0^r\xi(r')r'^2\mathrm{d}r'\;.
\label{xibar}
\end{equation}
Now equation~(\ref{xi^s_nonlin}) can be written as
\begin{equation}
1+\xi^s(\mathbf{s}) = \frac{1+\xi(r)}{1-\frac{1}{3}\frac{f}{b}\xibar(r)-\frac{f}{b}\mu_r^2\left[\xi(r)-\xibar(r)\right]}\;.
\label{xi^s_nonlin2}
\end{equation}
Moreover, we can expand this to linear order in $\delta$ (or equivalently, $\xi$) for consistency with the perturbative level of the mass conservation equation~(\ref{u(r)})~\cite{Cai2016,Hamaus2017},
\begin{equation}
\xi^s(\mathbf{s}) \simeq \xi(r) + \frac{1}{3}\frac{f}{b}\xibar(r) + \frac{f}{b}\mu_r^2\left[\xi(r)-\xibar(r)\right]\;.
\label{xi^s_lin}
\end{equation}
The function $\xi^s(\mathbf{s})$ can be decomposed into independent multipoles via
\begin{equation}
\xi^s_\ell(s) = \frac{2\ell+1}{2}\int\limits_{-1}^1\xi^s(s,\mu_s)\mathcal{L}_\ell(\mu_s)\mathrm{d}\mu_s\;,
\label{multipoles}
\end{equation}
with the Legendre polynomials $\mathcal{L}_\ell(\mu_s)$ of order $\ell$. For equation~(\ref{xi^s_lin}) the integral can be performed analytically and the only non-vanishing multipoles at linear order in $\xi$ and $\xibar$ are the monopole ($\ell=0$) and quadrupole ($\ell=2$) with
\begin{eqnarray}
\xi^s_0(s) &=& \left(1+\frac{f/b}{3}\right)\xi(r)\;,\label{xi_0} \\
\xi^s_2(s) &=& \frac{2f/b}{3}\left[\xi(r)-\xibar(r)\right]\;.\label{xi_2}
\end{eqnarray}
This can be recast into the following form~\cite{Cai2016,Hamaus2017},
\begin{equation}
\xi^s_0(s) - \xibar^s_0(s) = \xi^s_2(s)\frac{3+f/b}{2f/b}\;,
\label{xi_0_2}
\end{equation}
providing a direct link between monopole and quadrupole in redshift space without reference to any real-space quantity. However, note that equations~(\ref{xi_0}),~(\ref{xi_2}) and (\ref{xi_0_2}) only hold for the case of $\varepsilon=1$, and multipoles of higher order can be generated via geometric distortions when assuming a fiducial cosmology that is different from the truth, as discussed in section~\ref{subsec:geometric}.

\section{Analysis \label{sec:analysis}}

\subsection{BOSS galaxies and mocks \label{subsec:galaxies}}
We consider galaxy catalogs from the final data release 12 (DR12) of the SDSS-III~\cite{Eisenstein2011} Baryon Oscillation Spectroscopic Survey (BOSS)~\cite{Dawson2013}. In particular, we make use of the combined sample of the individual target selections denoted as LOWZ and CMASS~\cite{Reid2016}. With a total sky area of about $10\,000$ square degrees from both the northern and southern Galactic hemispheres the sample contains $1\,198\,006$ galaxies in a redshift range of $0.20<z<0.75$ with a linear bias of $b=1.85$~\cite{Alam2017}, making it the largest spectroscopic galaxy catalog available to date. In order to test and validate our analysis pipeline, we additionally consider the MultiDark PATCHY mock galaxy catalogs~\cite{Kitaura2016a} that are specifically designed to match the properties of the BOSS DR12 galaxy samples and have been calibrated to $N$-body simulations~\cite{Klypin2016}. The fiducial cosmology of the mocks assumes a flat $\Lambda$CDM model with parameter values $\boldsymbol{\Omega}'=(\Om,\Ob,\sigma_8,n_s,h)=(0.307115,0.048206,0.8288,0.9611,0.6777)$ adopted from Planck 2013~\cite{Planck2013} and a linear bias of $b=2.20$. For both real and mock samples we also make use of the random catalogs provided by the BOSS collaboration, which provide an unclustered realization of the galaxy distribution with the same survey geometry, but $50$ times its density at any given redshift. This facilitates an accurate estimation of correlation functions (see section~\ref{subsec:estimators}).

\subsection{VIDE voids \label{subsec:voids}}
We identify voids in the galaxy distribution using \textsc{vide}\footnote{\url{https://bitbucket.org/cosmicvoids/vide_public/}}, a publicly available software repository that can handle simulation snapshots, as well as masked light cones from observations~\cite{Sutter2015}. \textsc{vide} is based on \textsc{zobov}~\cite{Neyrinck2008}, which implements a watershed transform~\cite{Platen2007} to delineate local basins in the three-dimensional density field of tracer particles. Assuming a fiducial cosmology $\boldsymbol{\Omega}'$, equation~(\ref{x_comoving}) is used to transform angles and redshifts to comoving space. Then the density field is estimated via Voronoi tessellation, assigning each tracer particle $i$ a unique cell of volume $\mathcal{V}_i$. A Voronoi cell is defined as the sub-set of space that is closer to its associated tracer particle than to any other particle in the entire catalog. The tessellation of space into Voronoi cells is volume filling and hence allows the assignment of a tracer density to any location inside that volume. The density can be estimated as $1/\mathcal{V}_i$ anywhere inside the cell of tracer particle $i$, providing a piecewise constant density field over the entire domain. Local density minima serve as starting points to search for extended watershed basins whose density monotonically increases among their neighboring cells. The extent of a basin is determined as soon as a saddle point is encountered, indicating the occurrence of an adjacent basin whose cell densities start decreasing again. All the Voronoi cells that together make up such a basin already define a void region. The construction of a nested hierarchy of voids and sub-voids can then be defined by merging adjacent basins~\cite{Neyrinck2008}. For simplicity, in this paper we only consider the non-overlapping leaf nodes of this hierarchy.
\begin{figure}[t]
	\centering
	\includegraphics[scale=0.95, trim=32 0 50 60]{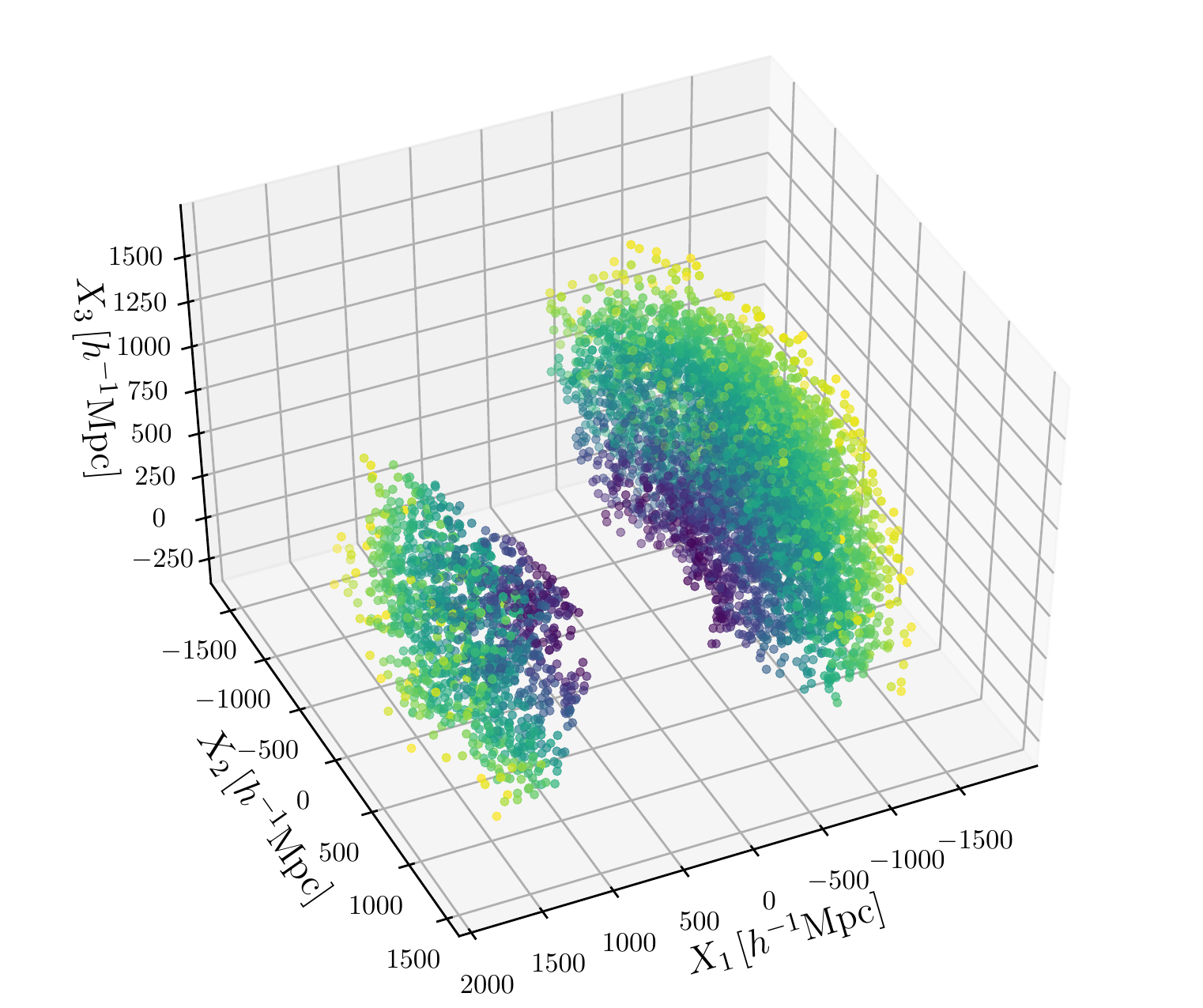}
	\includegraphics[scale=1.15, trim=90 90 90 50, clip]{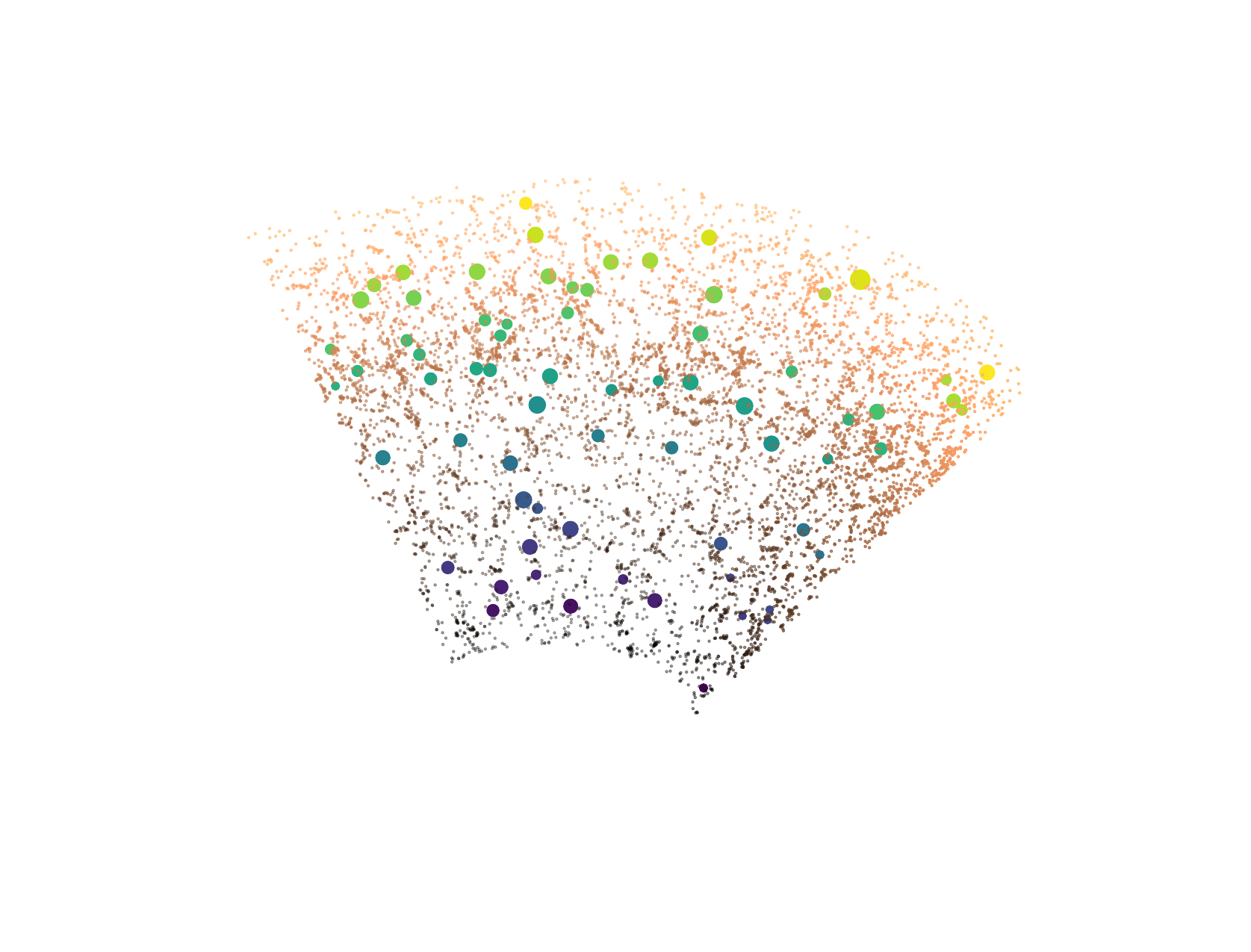}
	\caption{TOP: \textsc{vide} void centers ($N_\mathrm{v}=5952$) in comoving space from the BOSS DR12 combined sample. The observer is located at the origin, redshift is color-coded between $0.20$ (blue) and $0.75$ (yellow). BOTTOM: \textsc{vide} void centers (circles scaled by the effective void radius) and BOSS galaxies (dots) inside a slice of about one degree width in declination within the northern Galactic hemisphere.}
	\label{fig:box}
\end{figure}

We define the center of each void as the volume-weighted barycenter among all of its constituent Voronoi cells at tracer locations $\mathbf{x}_i$,
\begin{equation}
\mathbf{X} = \frac{\sum_i\mathbf{x}_i\mathcal{V}_i}{\sum_i\mathcal{V}_i}\;.
\end{equation}
The barycenter can be thought of as the geometric center of the void, as it is mostly constrained by its boundaries where the majority of tracers reside. This implies that it does not generally coincide with the minimum density inside the void, due to its non-spherical geometry. Conversely, the location of the minimum density is largely independent of the void boundaries, making it a poor indicator of the void geometry. Thus, the optimal center definition typically depends on the desired void observables. In this paper we are interested in measurements of the shape of stacked voids, so we want to maximize the sensitivity to their boundaries~\cite{Cautun2016}. To retain knowledge of this defining property of watershed voids, we therefore choose the volume-weighted barycenter. Moreover, in contrast to center definitions that are based on merely a single or a few tracers, the barycenter is more robust against discreteness noise and peculiar motions of individual tracers~\cite{Hamaus2014c}. In order to calculate redshifts and sky coordinates from our void centers $\mathbf{X}$, we simply invert the coordinate transformation from equation~(\ref{x_comoving}). Although watershed basins may exhibit arbitrary geometries, we assign an effective radius
\begin{equation}
R = \left(\frac{3}{4\pi}\sum\nolimits_i\mathcal{V}_i\right)^{1/3}\;,
\end{equation}
to each void, which corresponds to the radius of a sphere of equal volume. Running \textsc{vide} on the entire light cone of the BOSS DR12 combined sample we find a total of $N_\void=5952$ voids. This number already accounts for voids that have been discarded due to intersection with the survey mask, redshift boundaries, and other quality cuts (see reference~\cite{Sutter2015} for more details). Moreover, we impose a purity selection cut on voids based on their effective radius,
\begin{equation}
R > N_\mathrm{s}\left[\frac{4\pi}{3}n(z)\right]^{-1/3}\;,
\label{ats}
\end{equation}
where $n(z)$ is the number density of tracers at redshift $z$ and $N_\mathrm{s}$ determines the minimum considered void size in units of the average tracer separation. The smaller $N_\mathrm{s}$, the larger the contamination by spurious voids that may arise from Poisson fluctuations~\cite{Neyrinck2008,Cousinou2019}. This cut also preferentially removes voids that may have been misidentified due to RSDs~\cite{Pisani2015b}. As a default we assume a conservative value of $N_\mathrm{s}=4$, which yields a minimum effective void radius of $R=34.9\hMpc$ in our catalog. We note that this criterion depends on the specific tracer sample considered for void identification. It is known that Poisson point processes exhibit highly non-Gaussian Voronoi cell volume distributions~\cite{vdWeygaert2009}, which can cause spurious void detections. Reference~\cite{Cousinou2019} finds a very low contamination fraction of spurious voids in the BOSS DR12 CMASS sample based on a multivariate analysis of void properties in training and test samples. This is attributed to the relatively high clustering bias of CMASS galaxies, which is very similar to the combined BOSS sample used here.

\begin{figure}[t]
	\centering
	\resizebox{\hsize}{!}{
		\includegraphics[trim=0 -3 0 0]{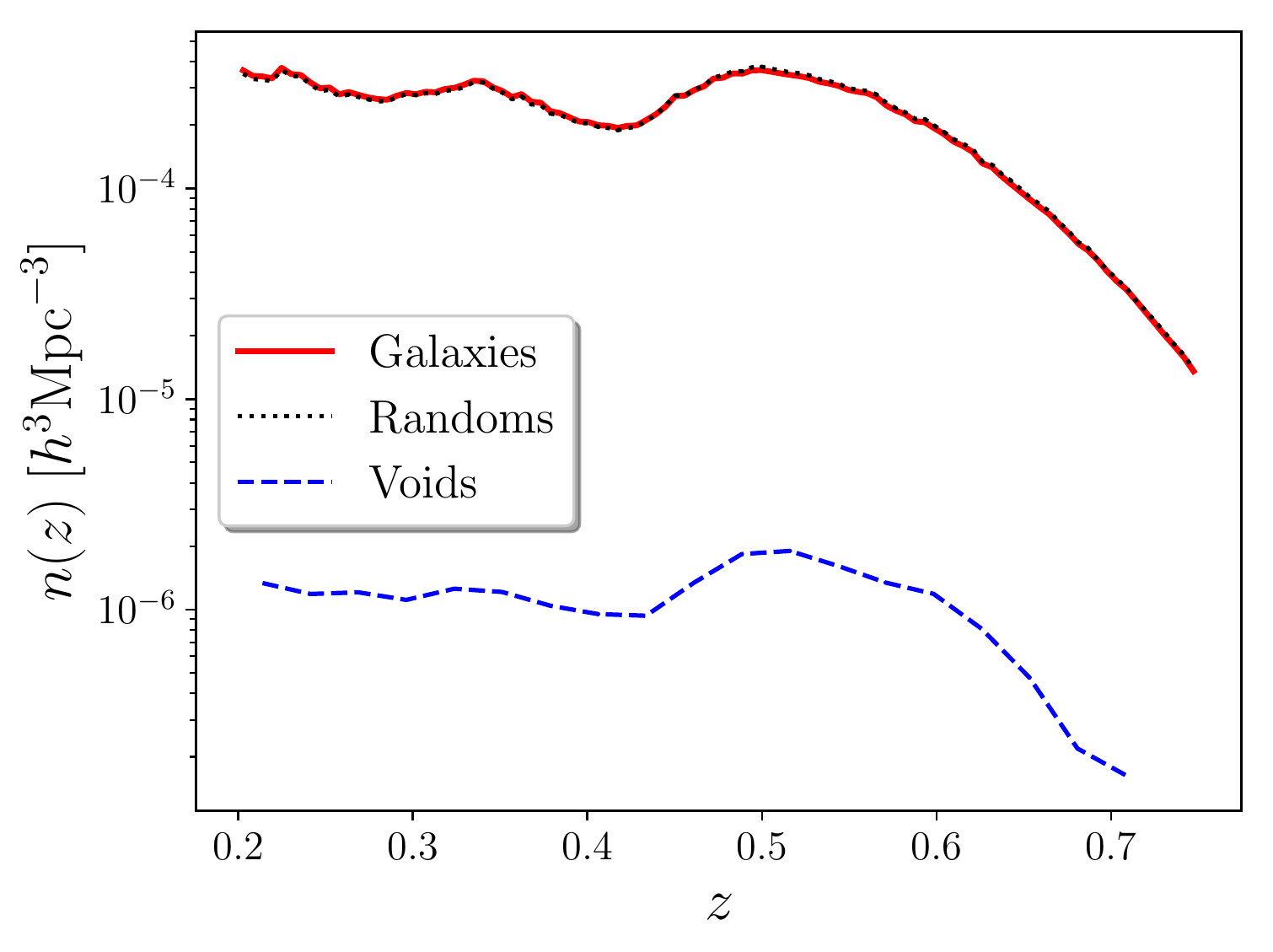}
		\includegraphics{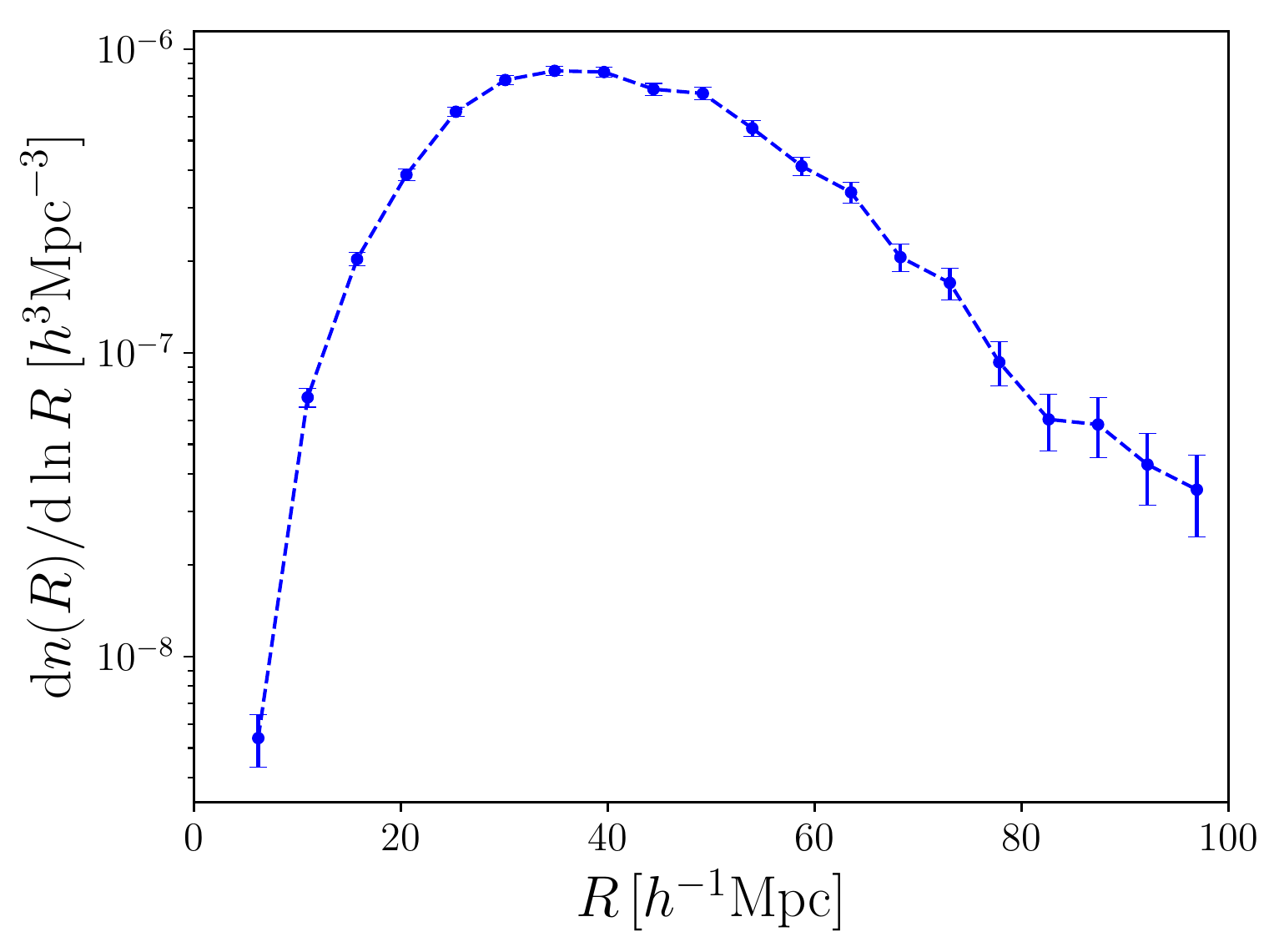}}
	\caption{LEFT: Number density of galaxies, randoms (scaled to the galaxy density), and \textsc{vide} voids as a function of redshift in the BOSS DR12 combined sample. RIGHT: Number density of all \textsc{vide} voids as a function their effective radius (void-size function, only shown for illustration).}
	\label{fig:nz}
\end{figure}

The top of figure~\ref{fig:box} presents a three-dimensional view of the selected void centers from the northern (right) and southern (left) Galactic hemispheres in comoving space, with the observer located at the origin. Below it, a narrow slice of about one degree in declination within the northern Galactic hemisphere visualizes the distribution of void centers together with their tracer galaxies. Despite the sparsity of the BOSS DR12 combined sample, intricate features of the cosmic web-like structure become apparent. Note that due to the extended three-dimensional geometry of voids, their centers do not necessarily intersect with the slice, leaving some seemingly empty regions without associated void center. The left panel of figure~\ref{fig:nz} shows the redshift distribution of galaxies, randoms, and voids from the data. For visibility, we have rescaled the total number of randoms to the number of galaxies (by a factor of $50$). Voids are roughly two orders of magnitude scarcer than galaxies, but their redshift distribution follows a similar trend. This is because higher tracer densities allow the identification of smaller voids, as expected from simulation studies~\cite{Jennings2013,Sutter2014a,Chan2014,Wojtak2016}. The right panel of figure~\ref{fig:nz} shows the distribution of effective void radii with Poisson error bars, also known as the void-size function. The latter is a quantity of interest for cosmology on its own~\cite{Pisani2015a,Ronconi2019,Contarini2019,Verza2019}, but in this paper we do not investigate it for that purpose any further, it is shown here only as supplementary information. We repeat the void finding procedure on each of the PATCHY mocks, allowing us to scale down all statistical uncertainties by roughly a factor of $\sqrt{N_\mathrm{m}}$, where $N_\mathrm{m}$ is the number of mock catalogs considered.

\subsection{Estimators \label{subsec:estimators}}
We need to define an estimator to measure the observed void-galaxy cross-correlation function $\xi^s(\mathbf{s})$ in redshift space. In order to take into account the survey geometry, we make use of a random catalog that samples the masked survey volume without intrinsic clustering.  We adopt the expression derived in reference~\cite{Hamaus2017},
\begin{equation}
\xi^s(\mathbf{s}) = \frac{\langle\mathbf{X},\mathbf{x}\rangle(\mathbf{s})}{\langle\mathbf{X}\rangle\langle\mathbf{x}\rangle} - \frac{\langle\mathbf{X},\mathbf{x}_r\rangle(\mathbf{s})}{\langle\mathbf{X}\rangle\langle\mathbf{x}_r\rangle}\;,
\label{estimator}
\end{equation}
with the void center, galaxy, and random positions $\mathbf{X}$, $\mathbf{x}$, and $\mathbf{x}_r$, respectively. Here, the angled brackets with two vectors represent the average pair count between objects at separation $\mathbf{s}$, whereas the brackets with only one vector indicate the mean number count of an object in the corresponding redshift slice. This estimator is a limiting case of Landy \& Szalay~\cite{Landy1993} and has been validated using mock data on the relevant scales for voids in BOSS data~\cite{Hamaus2017}. We will consider its two-dimensional variant $\xi^s(s_\parallel,s_\perp)$, with explicit dependence on distances along and perpendicular to the line of sight, as well as its multipoles $\xi^s_\ell(s)$ with $\ell=(0,2,4)$, i.e. monopole, quadrupole and hexadecapole. Calculating the multipoles is particularly simple with this estimator, because it allows the direct application of equation~(\ref{multipoles}) on the pair counts by using Legendre polynomials as weights in the integral without the need to define bins in $\mu_s$~\cite{Hamaus2017}. The mean number counts in the denominator of equation~(\ref{estimator}) can be pulled outside the integral, as they do not depend on $\mathbf{s}$. This way the estimation of multipoles becomes more accurate, especially at small separations $s$ and when $\mu_s$ approaches one, where the binning in both quantities inevitably results in a coarse spatial resolution~\cite{Cai2016}. We have explicitly compared this estimator with other common choices in the literature~\cite{Vargas2013} and refer the reader to section~\ref{subsec:systematics} for more details on this.

Watershed voids exhibit an angle-averaged density profile of universal character, irrespective of their absolute size~\cite{Hamaus2014b,Sutter2014a,Cautun2016}. In order to capture this unique characteristic in the two-point statistics of a void sample with a broad range of effective radii, it is beneficial to express all separations in units of $R$ (see section~\ref{subsec:comparison}). When this is done for every void individually in the estimation of $\xi^s$, it is commonly denoted as a void stack. Like the majority of void-related studies in the literature, we adopt this approach in our analysis and refer to $\xi^s(\mathbf{s}/R)$ as the stacked void-galaxy cross-correlation function. For simplicity, we will omit this explicit notation in the following and bear in mind that all separations $\mathbf{s}$ are expressed in units of~$R$.

The degree of uncertainty in a measurement of the void-galaxy cross-correlation function can be quantified by its covariance matrix. It is defined as the covariance of $\xi^s$ at separations $\mathbf{s}_i$ and $\mathbf{s}_j$ from $N$ independent observations,
\begin{equation}
\C_{ij} = \bigl<\bigl(\xi^s(\mathbf{s}_i)-\langle\xi^s(\mathbf{s}_i)\rangle\bigr)\bigl(\xi^s(\mathbf{s}_j)-\langle\xi^s(\mathbf{s}_j)\rangle\bigr)\bigr>\;,
\label{covariance}
\end{equation}
where the angled brackets indicate averages over the sample size. Although we can only observe a single universe, we have a large sample of voids at our disposal that enables an estimate of the covariance matrix as well. Note that we are considering mutually exclusive voids, each of which provide an independent patch of large-scale structure. As we are primarily interested in $\xi^s$ on scales up to the void extent, as opposed to inter-void scales, we can employ a jackknife resampling strategy to estimate $\C_{ij}$~\cite{Hamaus2017}. For this we simply remove one void at a time in the estimator of $\xi^s$ from equation~(\ref{estimator}), which provides $N_\void$ jackknife samples in total. These samples can then be used in equation~(\ref{covariance}) to calculate $\C_{ij}$, albeit with an additional factor of $(N_\void-1)$ to account for the statistical weight of the jackknife sample size. We use the square root of the diagonal elements of the covariance matrix to quote error bars on our measurements of $\xi^s$. The identical procedure can be applied to the multipoles $\xi^s_\ell$, except in that case one can use equation~(\ref{covariance}) to additionally calculate the covariance between multipoles of different order.

There are several advantages of this jackknife technique over other common methods for covariance estimation, which typically rely on simulated mock catalogs. Most importantly, it is based on the observed data itself and does not involve prior model assumptions about cosmology, structure formation, or galaxy evolution. In addition, a statistically reliable estimation of $\C_{ij}$ requires large sample sizes, which are expensive in terms of numerical resources when considering realistic mocks from $N$-body simulations. Our void catalog already provides $\mathcal{O}(10^3)$ spatially independent samples at no additional cost. It has been shown that the jackknife technique provides covariance estimates that are consistent with those obtained from independent mock catalogs in the limit of large jackknife sample sizes~\cite{Favole2020}.

\subsection{Likelihood \label{subsec:likelihood}}
Equipped with the theory from section~\ref{subsec:correlation} we can now define the likelihood $L(\hat{\xi}^s|\boldsymbol{\Omega})$ of the measurement given a model, which we approximate to be of Gaussian form,
\begin{equation}
\ln L(\hat{\xi}^s|\boldsymbol{\Omega}) = -\frac{1}{2N_\mathrm{m}}\sum\limits_{i,j}\Bigl(\hat{\xi}^s(\mathbf{s}_i)-\xi^s(\mathbf{s}_i,\boldsymbol{\Omega})\Bigr)\,\hat{\C}_{ij}^{-1}\Bigl(\hat{\xi}^s(\mathbf{s}_j)-\xi^s(\mathbf{s}_j,\boldsymbol{\Omega})\Bigr)\;.
\label{likelihood}
\end{equation}
The hat symbols indicate a measured quantity to be distinguished from the model, which explicitly depends on the parameters $\boldsymbol{\Omega}$. Here we have dropped the normalization term involving the determinant of $\hat{\C}_{ij}$, since it only adds a constant. The form of equation~(\ref{likelihood}) can be applied to either the two-dimensional void-galaxy cross-correlation function $\xi^s(s_\parallel,s_\perp)$, or its multipoles $\xi^s_\ell(s)$. We use $\xi^s(s_\parallel,s_\perp)$, which contains the information from the multipoles of all orders. However, we have verified that only including the multipoles of orders $\ell=(0,2,4)$ yields consistent results. When analyzing mock catalogs we scale their covariance in equation~(\ref{likelihood}) by the number of mock samples $N_\mathrm{m}$ used, allowing us to validate the statistical constraining power of the data. When analyzing the data itself, we set $N_\mathrm{m}=1$. We vary $\boldsymbol{\Omega}$ until a global maximum of the likelihood at the best-fit parameter set is found. The quality of the fit can be assessed by evaluation of the reduced chi-square statistic,
\begin{equation}
\chi^2_\mathrm{red} = -\frac{2N_\mathrm{m}}{N_\mathrm{dof}}\ln L(\hat{\xi}^s|\boldsymbol{\Omega}) \;,
\label{chi2}
\end{equation}
for $N_\mathrm{dof}=N_\mathrm{bin}-N_\mathrm{par}$ degrees of freedom, where $N_\mathrm{bin}$ is the number of bins for the data and $N_\mathrm{par}$ the number of parameters. Moreover, we explore the likelihood surface in the neighborhood of the global maximum using the Monte Carlo Markov Chain (MCMC) sampler \textsc{emcee}~\cite{Foreman-Mackey2019}, which enables us to access the posterior probability distribution of the model parameters.

\subsection{Parameters \label{subsec:parameters}}
Instead of using the fundamental cosmological parameters, we express $\boldsymbol{\Omega}$ in terms of derived parameters that directly affect the void-galaxy cross-correlation function, namely the linear growth rate to bias ratio $f/b$, and the AP parameter $\varepsilon$. To account for potential systematics in the data that can be caused by discreteness noise or selection effects, we further allow for two additional nuisance parameters $\mathcal{M}$ and $\mathcal{Q}$. The parameter $\mathcal{M}$ may adjust for possible inaccuracies arising in the deprojection technique and a contamination of the void sample by Poisson fluctuations, which can attenuate the amplitude of the monopole~\cite{Cousinou2019}. On the other hand, the parameter $\mathcal{Q}$ accounts for potential selection effects when voids are identified in anisotropic redshift space~\cite{Pisani2015b,Nadathur2019b}. A physical origin of this can be violent shell-crossing and virialization events that change the topology of void boundaries~\cite{Hahn2015}, causing a so-called Finger-of-God (FoG) effect~\cite{Jackson1972,Hamilton1998,Scoccimarro2004}. It appears around compact overdensities, such as galaxy clusters, generating elongated features along the line of sight that extend over several $\hMpc$~\cite{Peacock2001,Zehavi2011,Pezzotta2017}. A similar effect can be caused by cluster infall regions, leading to closed caustics in redshift space that may be misinterpreted as voids~\cite{Kaiser1987,Hamilton1998}. Therefore, this can have a non-trivial impact on the identification of voids with diameters of comparable size~\cite{Stanonik2010,Kreckel2011a,vdWeygaert2011b}, although the tracer sample we use consists of massive luminous red galaxies that typically reside in the centers of clusters and do not sample the entire cluster profile in its outer parts. $\mathcal{M}$ (monopole-like) is used as a free amplitude of the deprojected correlation function $\xi(r)$ in real space, and $\mathcal{Q}$ (quadrupole-like) is a free amplitude for the quadrupole term proportional to $\mu_r^2$. Hence, equations~(\ref{xi^s_nonlin2}) and (\ref{xi^s_lin}) can be extended by the replacements $\xi(r)\rightarrow\mathcal{M}\xi(r)$ and $\mu_r^2\rightarrow\mathcal{Q}\mu_r^2$, which results in the following form for the final parametrization of our model at linear perturbation order
\begin{equation}
\xi^s(\mathbf{s}) = \mathcal{M}\left\{\xi(r) + \frac{1}{3}\frac{f}{b}\xibar(r) + \frac{f}{b}\mathcal{Q}\mu_r^2\left[\xi(r)-\xibar(r)\right]\right\}\;,
\label{xi^s_lin2}
\end{equation}
together with the equivalent replacements in equation~(\ref{r_par(s_par)}) for the mapping from the observed separation $\mathbf{s}$ to $\mathbf{r}$: $r_\parallel = s_\parallel/[1-\frac{1}{3}\frac{f}{b}\mathcal{M}\xibar(r)]$. One can think of equation~(\ref{xi^s_lin2}) as an adaptive template that attempts to extract those anisotropic distortions that match the radial and angular shape as predicted by linear theory. For the nuisance parameters we assign default values of $\mathcal{M}=\mathcal{Q}=1$, as they are not known a priori. Note that this extension does not introduce any fundamental parameter degeneracies, due to the different functional forms of $\xi(r)$ and $\xibar(r)$. For both our model and nuisance parameters we assume uniform prior ranges of $\left[-10,+10\right]$, given that each of their expected values is of order unity. We checked that an extension of these boundaries has no impact on our results.

\subsection{Model validation \label{subsec:validation}}
The theory model derived in section~\ref{subsec:correlation} requires knowledge of the void density profile $\delta(r)$ and the void-galaxy cross-correlation function $\xi(r)$ in real space for equation~(\ref{xi^s_nonlin}). Its linear version, equation~(\ref{xi^s_lin}), only requires $\xi(r)$. $\delta(r)$ can be accurately modeled with the HSW profile (cf. equation~(\ref{HSW})), at the price of including additional parameters to be constrained by the data, an approach that has already been successfully applied to BOSS data~\cite{Hamaus2016}.

\begin{figure}[t]
	\centering
	\resizebox{\hsize}{!}{
		\includegraphics{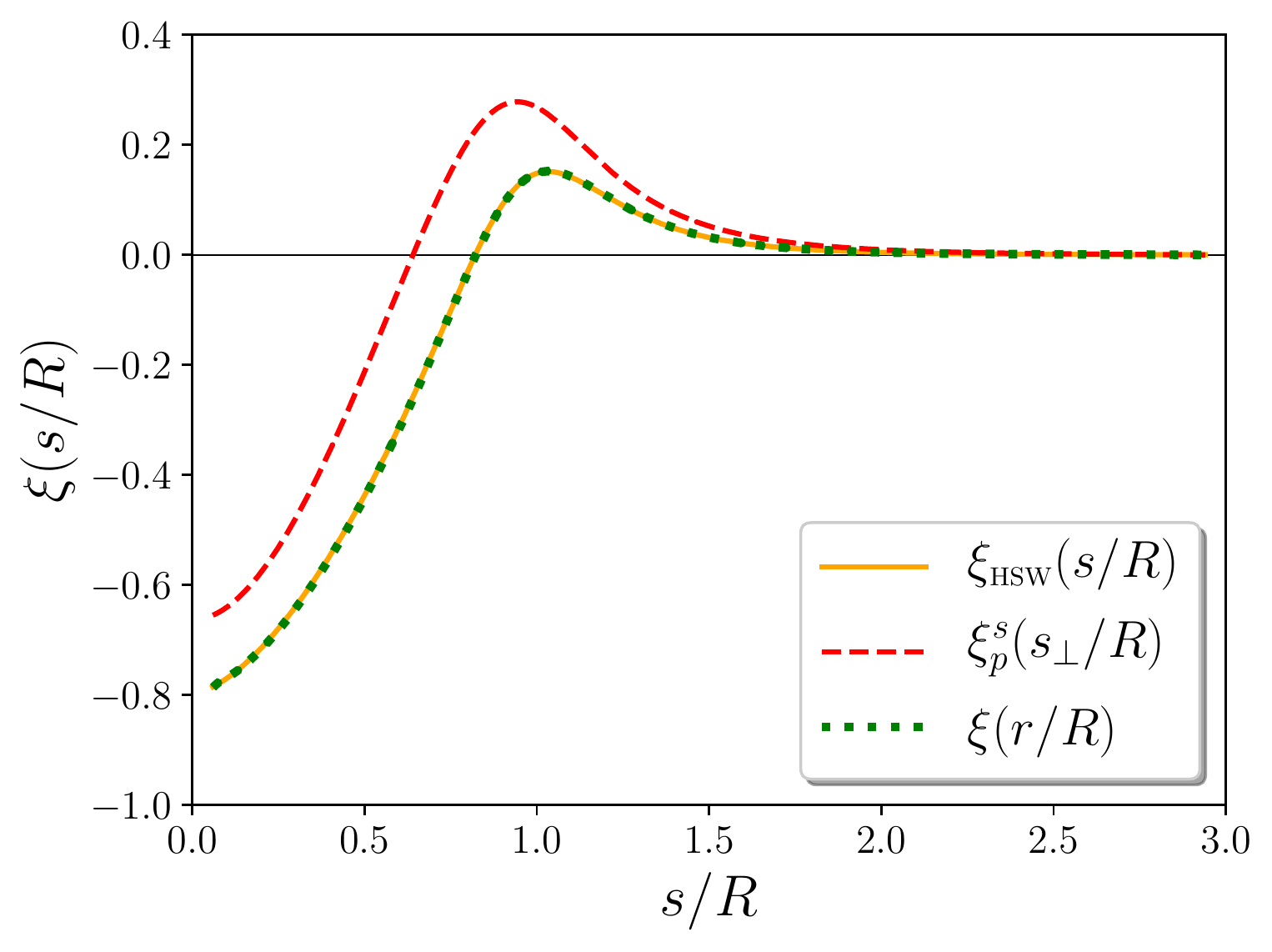}
		\includegraphics{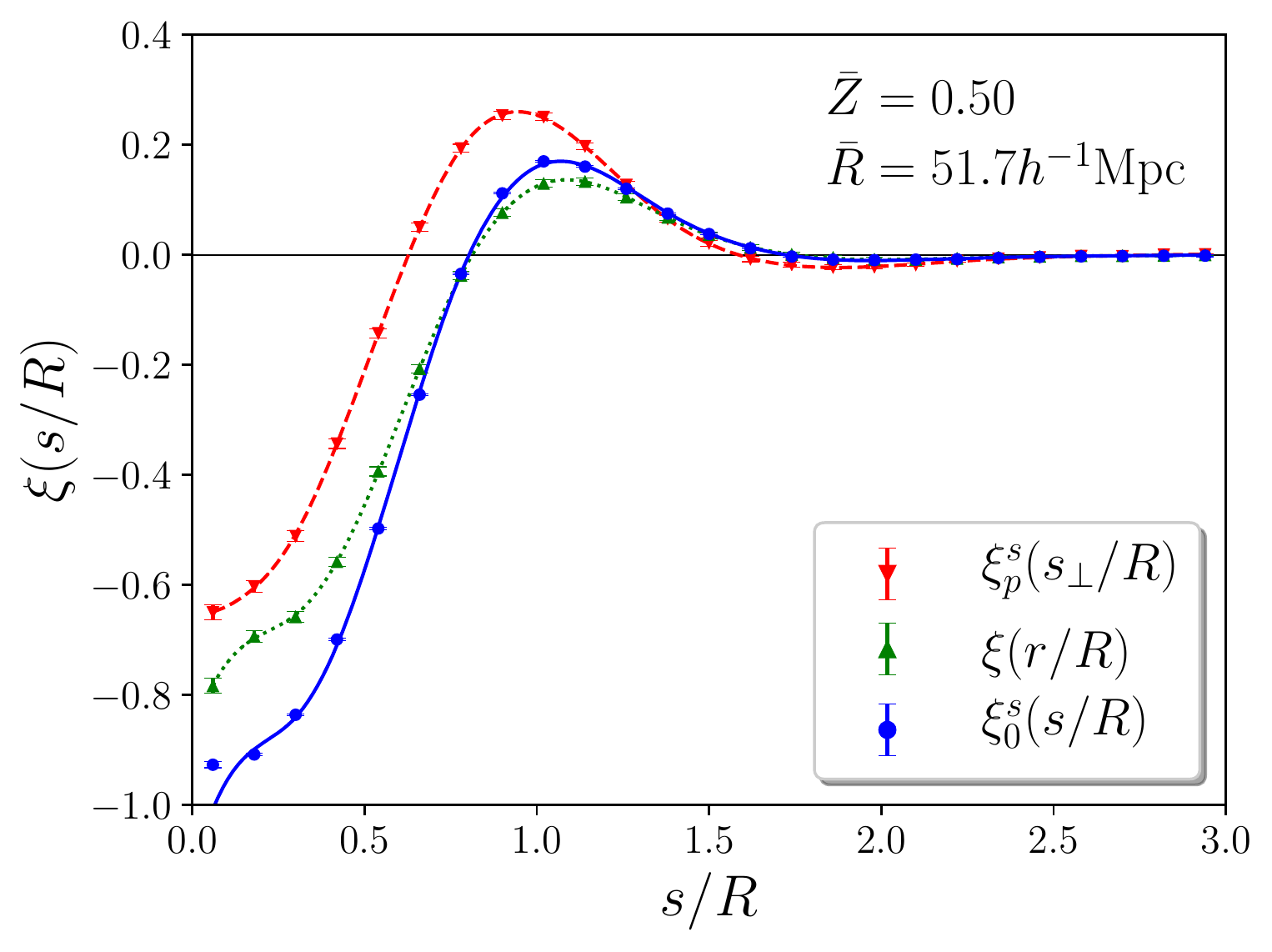}}
	\caption{LEFT: Projection (dashed red) and deprojection (dotted green) of a model void-galaxy cross-correlation function (solid orange) based on the HSW profile from equation~(\ref{HSW}), using the Abel transform. RIGHT: Projected void-galaxy cross-correlation function (red wedges, dashed line) of voids from $30$ PATCHY mock catalogs in redshift space, and its real-space counterpart after deprojection (green triangles, dotted line). The redshift-space monopole in the mocks (blue dots) follows the same functional form, in agreement with the linear model (blue solid line) from equation~(\ref{xi_0}).}
	\label{fig:xi_p_mock}
\end{figure}
In this paper we instead choose a more data-driven approach and use the real-space profile $\xi(r)$ obtained by deprojecting the measured projected void-galaxy cross-correlation function $\xi^s_p(s_\perp)$ using the inverse Abel transform, equation~(\ref{xi_d}). We first test this procedure based on the model template from equation~(\ref{HSW}) and use equation~(\ref{xi(delta)}) to define $\xi_{\scriptscriptstyle\mathrm{HSW}}(r)\equiv b\delta_{\scriptscriptstyle\mathrm{HSW}}(r)$. The left panel of figure~\ref{fig:xi_p_mock} shows $\xi_{\scriptscriptstyle\mathrm{HSW}}$ with the parameter values $(\rs/R,\dc,\alpha,\beta)\simeq(0.82,-0.36,1.6,9.1)$ and $b=2.2$. These values have been chosen to match the mock data reasonably well (see below). We then use the forward Abel transform to obtain the projected correlation function,
\begin{equation}
\xi^s_p(s_\perp) = 2\int_{s_\perp}^\infty\xi(r)\frac{r\mathrm{d}r}{\sqrt{r^2-s_\perp^2}}\;.
\label{xi_p2}
\end{equation}
Finally, we apply the inverse Abel transform of equation~(\ref{xi_d}) to infer the original void-galaxy cross-correlation function $\xi_{\scriptscriptstyle\mathrm{HSW}}$. As evident from the perfectly overlapping lines in the plot, this procedure works extremely well with noiseless data. In reality, however, we have to measure correlation functions with an estimator, which is unavoidably associated with a finite covariance. In order to estimate $\xi^s_p$ from the data, we adopt equation~(\ref{estimator}) for pairs on the plane of the sky, which can be achieved by exchanging the three-dimensional position $\mathbf{x}$ of an object by $x_\perp=(|\mathbf{x}|^2-x_\parallel^2)^{1/2}$ and counting pairs at a given projected separation $s_\perp$ over the redshift range. We restrict the line-of-sight projection range to $s_\parallel=3R$ at the near and far sides from the void center, where $\xi^s$ has well converged to zero. The right panel of figure~\ref{fig:xi_p_mock} shows the result when stacking $N_\void\simeq2\times10^5$ voids from $N_\mathrm{m}=30$ PATCHY mock catalogs. In this case the situation is very similar to the test case from the left panel. The deprojection via the inverse Abel transform results in a smooth curve, only close to the void center we observe mild fluctuations away from the expected shape, due to larger statistical uncertainties~\cite{Pisani2014}. We verified that reprojecting our result using equation~(\ref{xi_p2}) agrees well with the original $\xi^s_p$. We also plot the measured monopole $\xi^s_0$ from the same PATCHY mocks, including its best-fit model from equation~(\ref{xi^s_lin2}) using the deprojected $\xi^s_p$. It provides an excellent agreement with the mock data, and confirms the predicted proportionality between $\xi^s_0(s)$ and $\xi(r)$ of equation~(\ref{xi_0}).

Having validated the deprojection procedure to obtain $\xi(r)$ from the data, we are now ready to test our model for the void-galaxy cross-correlation function in redshift space.
\begin{figure}[h]
	\centering
	\resizebox{0.86\hsize}{!}{\includegraphics[trim=-40 0 0 10]{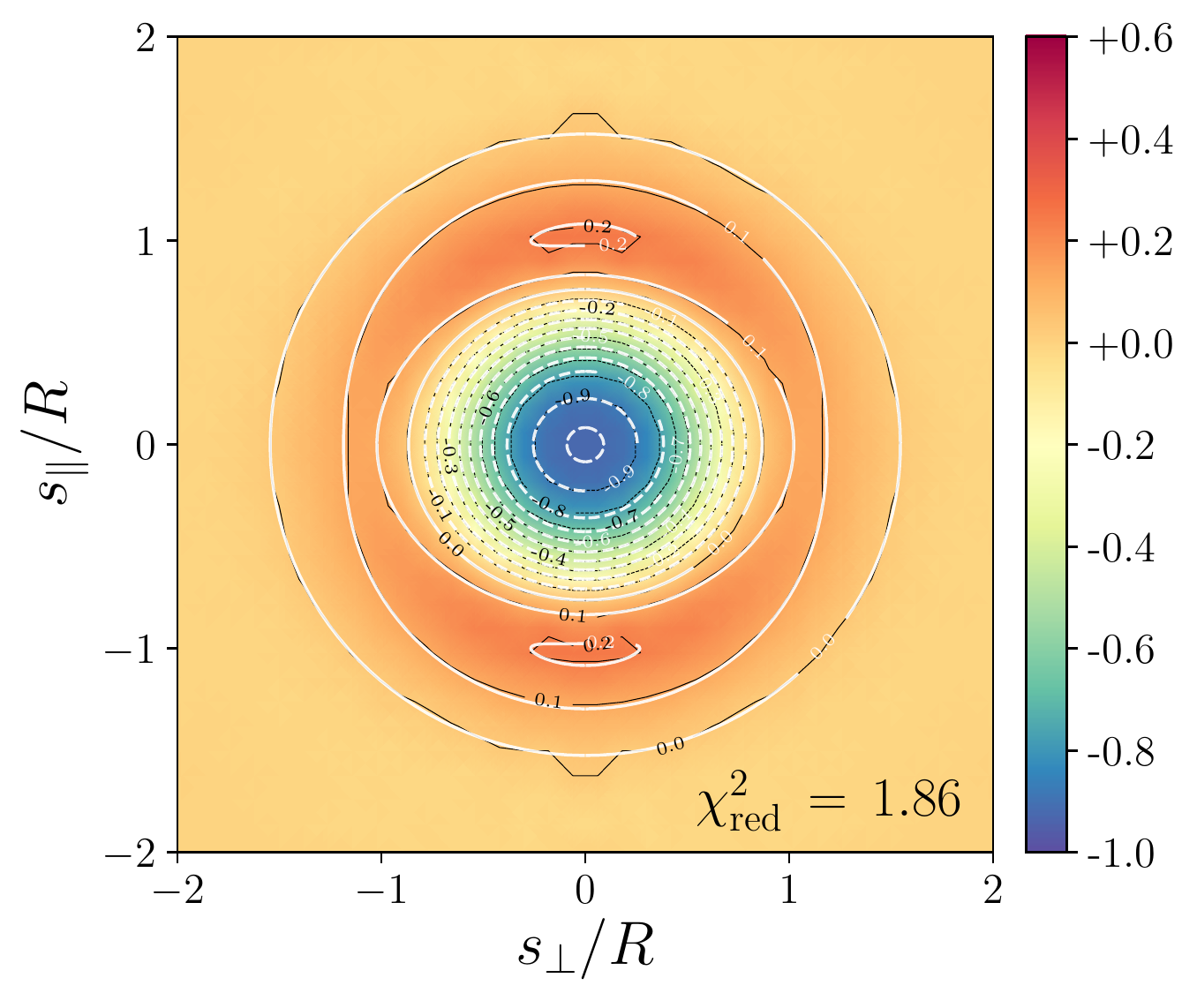}}
	\resizebox{0.86\hsize}{!}{\includegraphics[trim=0 10 0 10]{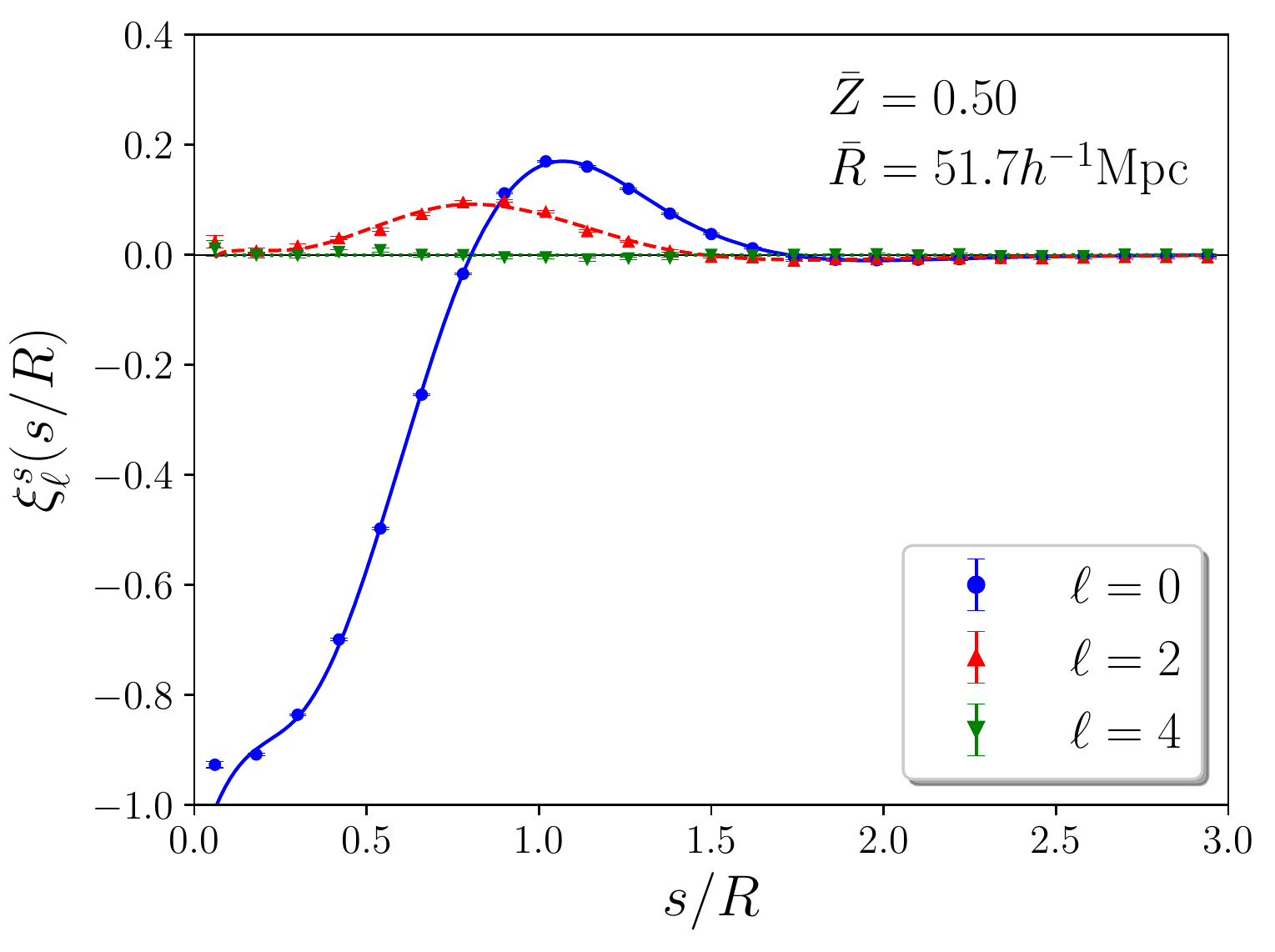}}
	\caption{TOP: Estimation of the stacked void-galaxy cross-correlation function $\xi^s(s_\parallel/R,s_\perp/R)$ from voids in 30 PATCHY mock catalogs (color scale with black contours) and the best-fit model (white contours) from equation~(\ref{xi^s_lin2}). BOTTOM: Monopole (blue dots), quadrupole (red triangles) and hexadecapole (green wedges) from the same mock data with corresponding model fits (solid, dashed, dotted lines). The mean redshift and effective radius of the void sample is shown at the top.}
	\label{fig:xi_mock}
\end{figure}
\begin{figure}[h]
	\centering
	\resizebox{0.7\hsize}{!}{\includegraphics{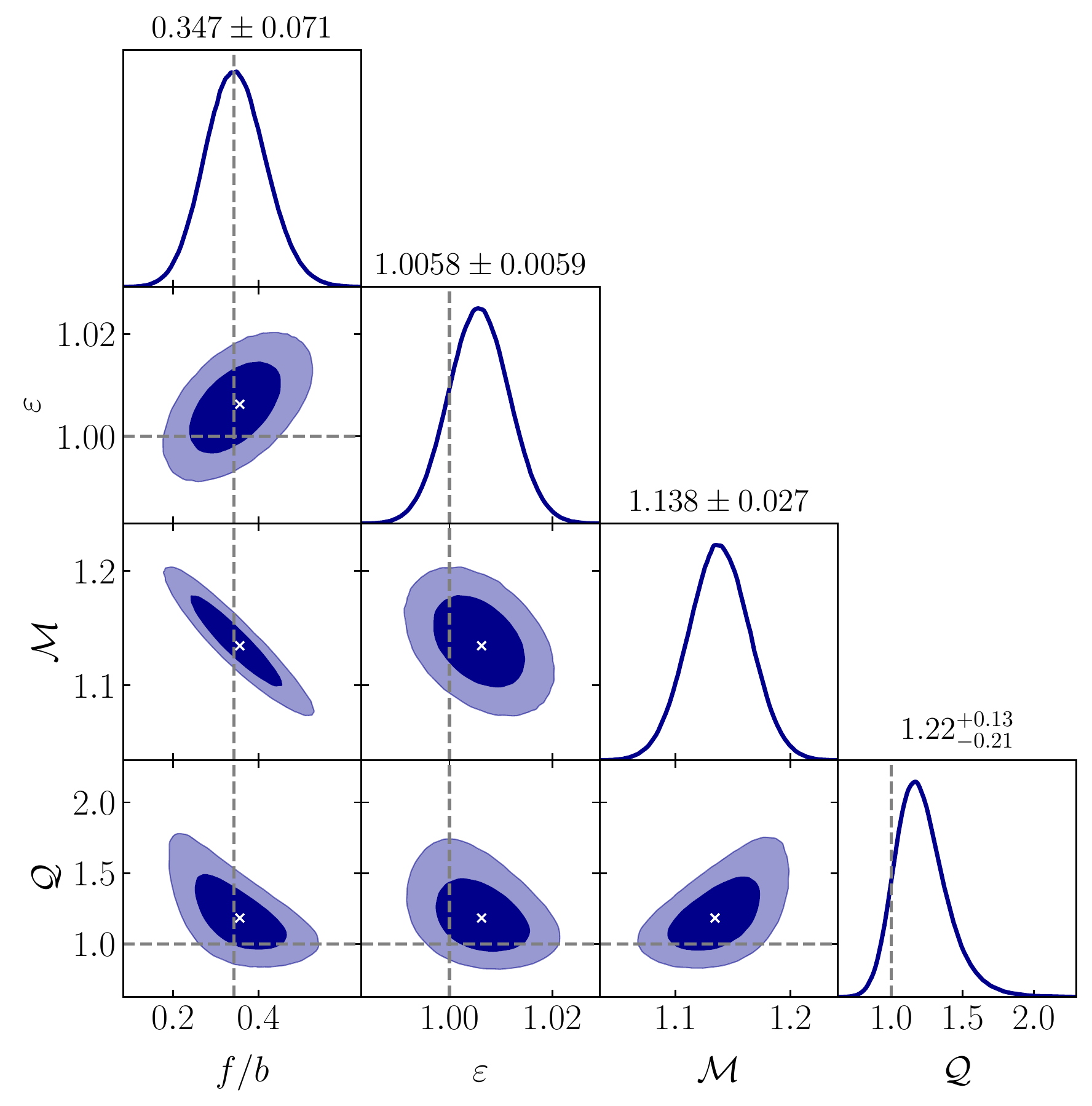}}
	\caption{Posterior probability distribution of the model parameters that enter in equation~(\ref{xi^s_lin2}), obtained via MCMC from the PATCHY mock data shown in figure~\ref{fig:xi_mock}. Dark and light shaded areas show $68\%$ and $95\%$ confidence regions with a cross marking the best fit, dashed lines indicate fiducial values of the RSD and AP parameters $(f/b=0.344,\,\varepsilon=1)$, and default values for the nuisance parameters $(\mathcal{M}=\mathcal{Q}=1)$. The top of each column states the mean and standard deviation of the 1D marginal distributions.}
	\label{fig:triangle_mock}
\end{figure}
Figure~\ref{fig:xi_mock} presents the corresponding measurement from voids in 30 PATCHY mock catalogs, both its 2D variant with separations along and perpendicular to the line of sight (top panel), as well as the multipoles of order $\ell=(0,2,4)$ (bottom panel). For the former we use $18$ bins per direction, resulting in $N_\mathrm{bin}=18^2=324$, whereas for the multipoles we use $25$ radial bins, which yields $N_\mathrm{bin}=3\times25=75$ in total. We apply the linear model from equation~(\ref{xi^s_lin2}) to fit this data (omitting the innermost radial bin) using the $N_\mathrm{par}=4$ parameters $\boldsymbol{\Omega}=(f/b,\varepsilon,\mathcal{M},\mathcal{Q})$. As apparent from figure~\ref{fig:xi_mock}, this yields a very good fit to the mock data, with a reduced chi-square value of $\chi^2_\mathrm{red}=1.86$. We note that this value corresponds to the statistical power of $30$ mock observations, so it is entirely satisfactory for the purpose of validating our model for a single BOSS catalog. An increase in the number of mock samples marginally affects our summary statistics, which exhibit a negligible amount of statistical noise compared to the real data. The anisotropy of the void-galaxy cross-correlation function is well captured close to the void center, as well as on the void boundaries at $s\simeq R$. The flattened contours around the void interior are a result of the quadrupole term in equation~(\ref{xi^s_nonlin}), respectively its linear version~(\ref{xi^s_lin})~\cite{Cai2016}. Outside the void boundaries, where the correlation function declines again, this results in elongated contours, as necessary in order to restore spherical symmetry in the limit of large separations. It is worth noting that the coordinate transformation from equation~(\ref{r_par(s_par)}) causes a line-of-sight elongation from $r_\parallel$ to $s_\parallel$ for negative $\Delta(r)$, acting in the opposite direction. However, this coordinate effect merely accounts for a small correction to the flattening caused by the quadrupole, which we explicitly checked in our analysis. Finally, we emphasize the strong evidence for a vanishing hexadecapole on all scales, in agreement with the theory prediction from section~\ref{subsec:correlation}.

We then run a MCMC to sample the posterior distribution of the model parameters. The result is presented in figure~\ref{fig:triangle_mock} using the \textsc{getdist} software package~\cite{Lewis2019}. We recover the fiducial values of the cosmologically relevant parameters $f/b$ and $\varepsilon$ to within the $68\%$ confidence regions, which validates the theory model we use. Moreover, the parameter contours reveal a nearly Gaussian shape of the posterior, only the nuisance parameter $\mathcal{Q}$ exhibits a slightly non-Gaussian behavior. While $\mathcal{Q}$ is marginally consistent with unity to within $1\sigma$, we find clear evidence for the parameter $\mathcal{M}$ to exceed unity by roughly $14\%$. This could be attributed to some degree of overdispersion beyond Poisson noise that has been used in the PATCHY algorithm to calibrate the clustering statistics of galaxies in BOSS~\cite{Kitaura2016a}, resulting in a higher amplitude of random fluctuations inside voids. We also note a strong anti-correlation between $\mathcal{M}$ and $f/b$, which can be understood from equation~(\ref{xi_2}) for the quadrupole, where both parameters enter via multiplication of $\xi(r)$ and $\xibar(r)$. However, the monopole in equation~(\ref{xi_0}) breaks their degeneracy.

\begin{figure}[t]
	\centering
	\resizebox{\hsize}{!}{
		\includegraphics[trim=0 30 0 5, clip]{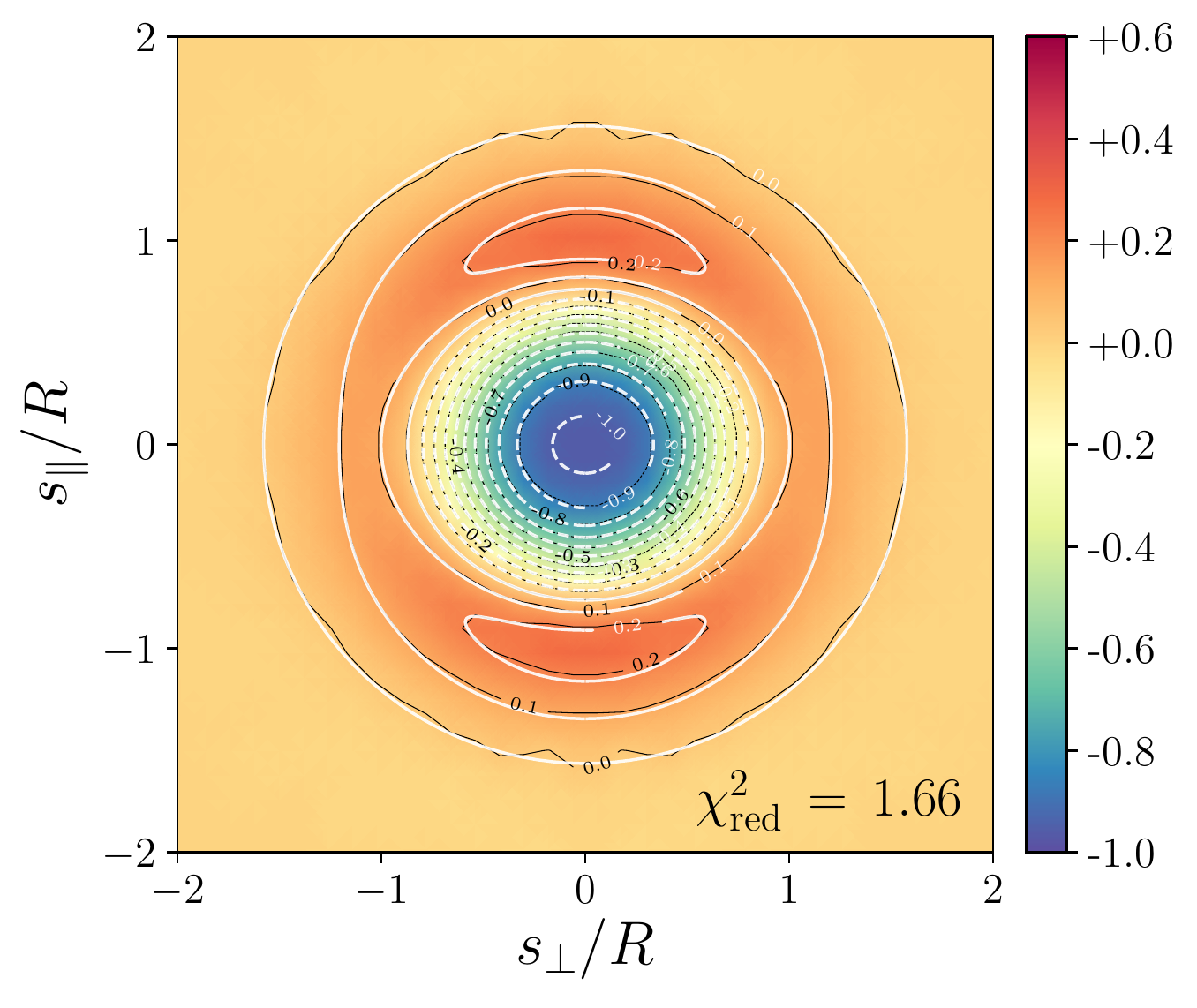}
		\includegraphics[trim=0 30 0 5, clip]{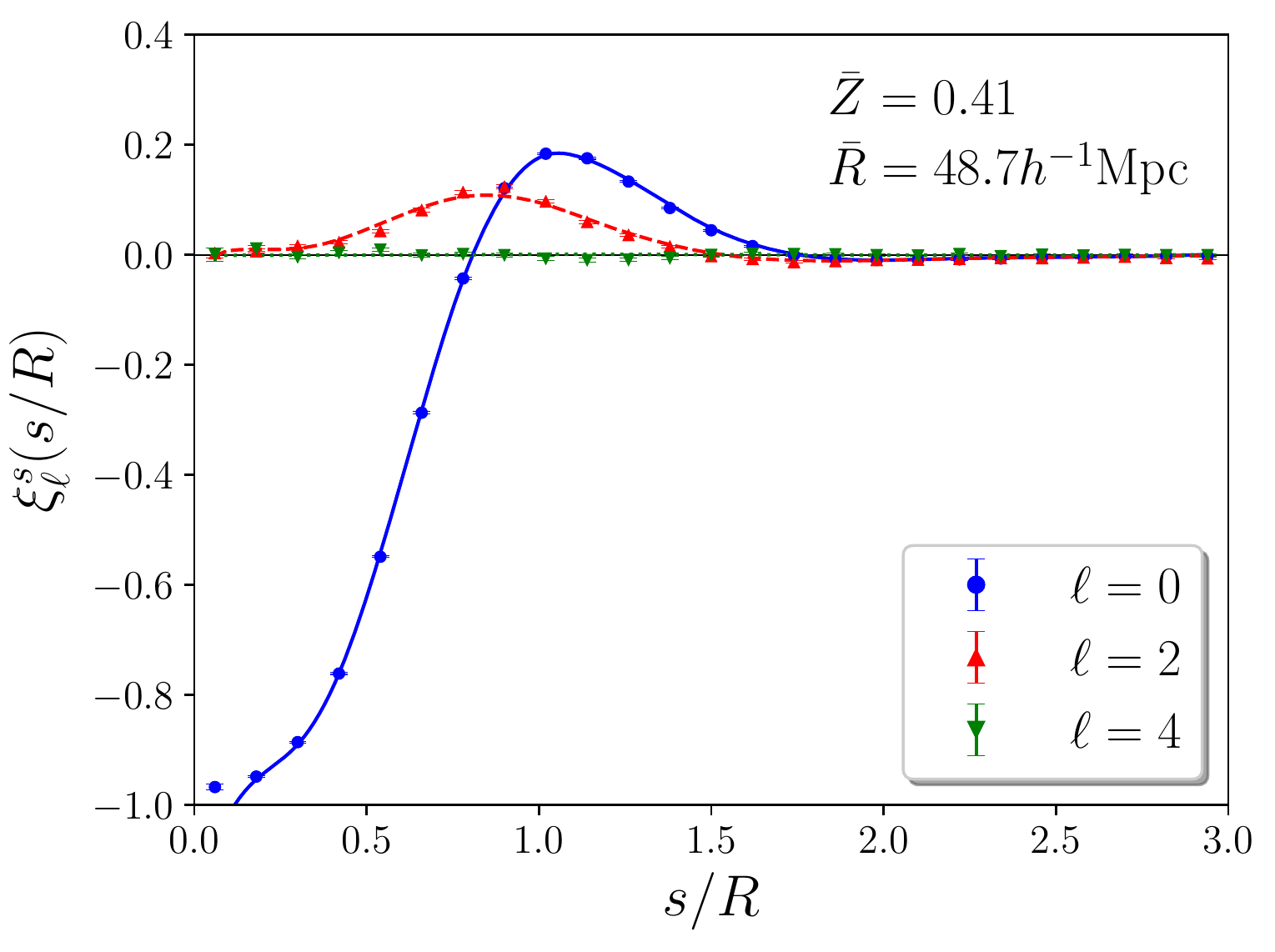}}
	\resizebox{\hsize}{!}{
		\includegraphics[trim=0 10 0 5, clip]{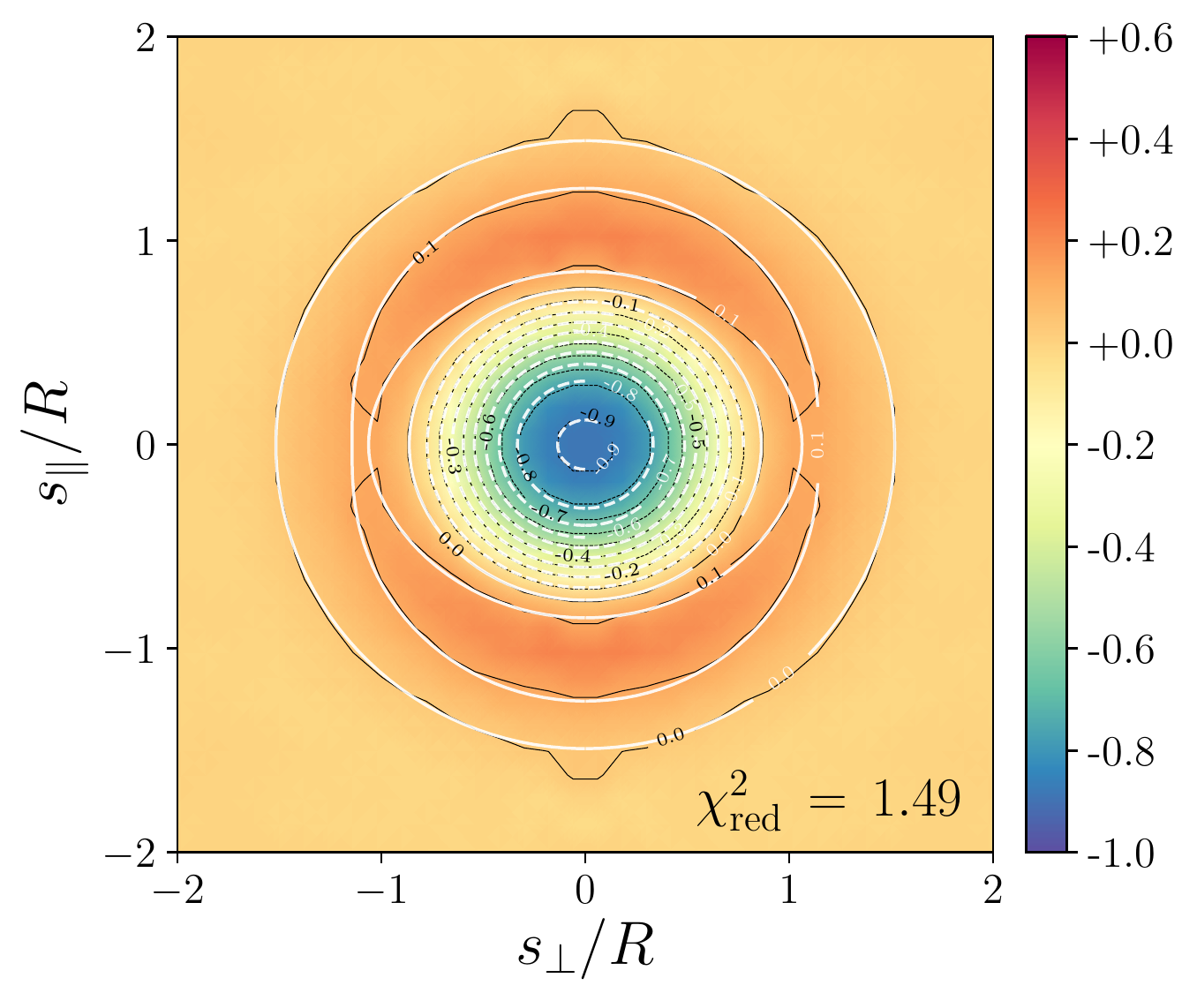}
		\includegraphics[trim=0 10 0 5, clip]{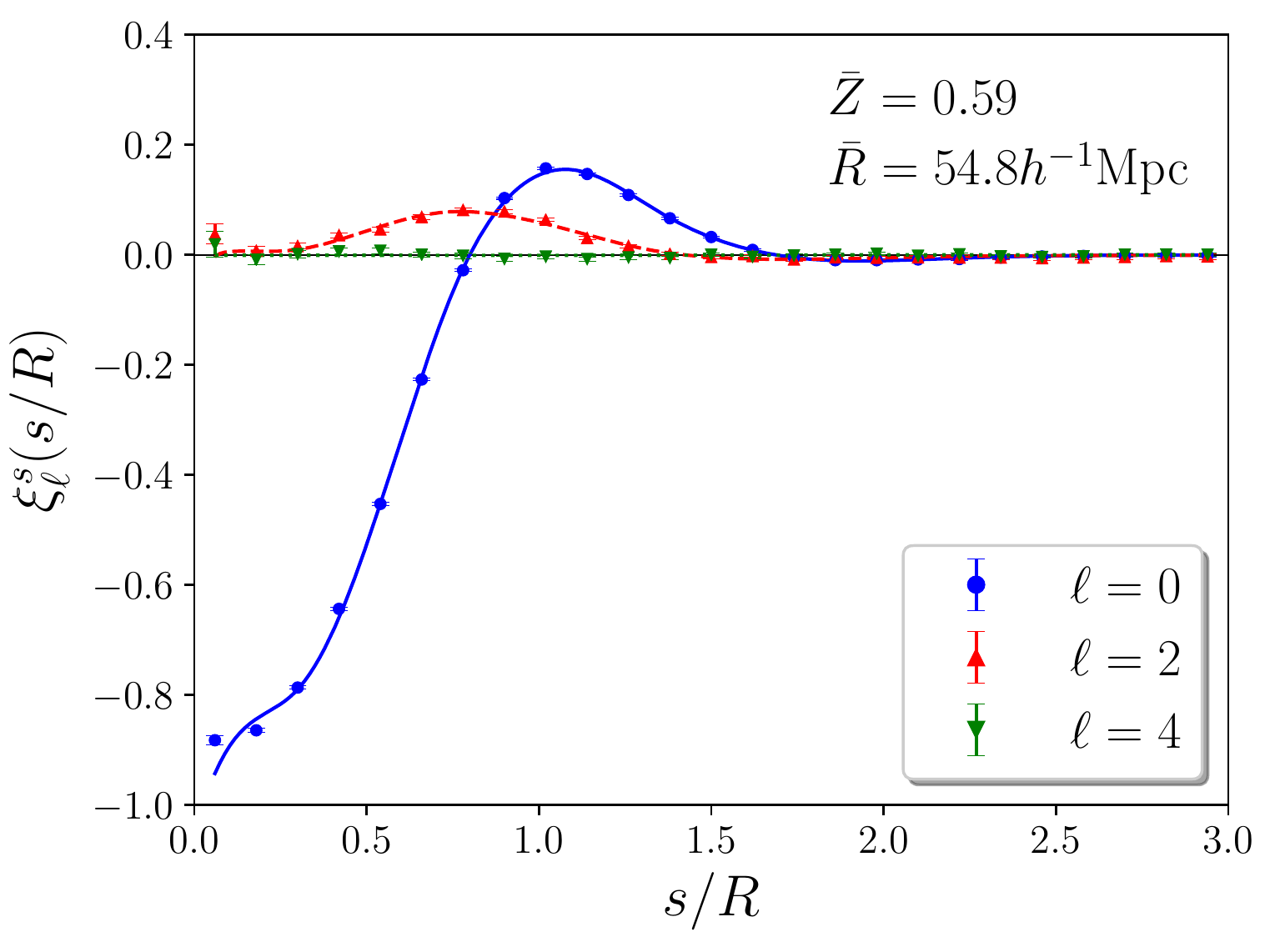}}
	\caption{As figure~\ref{fig:xi_mock} after splitting the PATCHY void sample at its median redshift of $Z=0.51$ into $50\%$ lowest-redshift (``low-z'', top row) and $50\%$ highest-redshift voids (``high-z'', bottom row).}
	\label{fig:xi_mock_Z}
\end{figure}

As a next step we investigate the dependence on void redshift $Z$. To this end we split our catalog into subsets that contain $50\%$ of all voids with redshifts below or above their median value. We will refer to these subsets as ``low-z'' and ``high-z'', respectively.  Figure~\ref{fig:xi_mock_Z} presents the corresponding correlation function statistics, revealing characteristic redshift trends. Note that the effective radii $R$ in our catalog are somewhat correlated with $Z$ due to the variation of tracer density $n(z)$ with redshift, as shown in figure~\ref{fig:nz}. When $n(z)$ decreases, fewer small voids can be identified, which means that at the high-redshift end our voids tend to be larger in size. On average, smaller voids are emptier and exhibit higher compensation walls at their edges~\cite{Sheth2004,vdWeygaert2011a,Hamaus2014b,vdWeygaert2016}, which in turn induces a higher amplitude of both monopole and quadrupole~\cite{Hamaus2017}. A similar trend is manifest in the evolution from high to low redshift, which reveals the growth of structure on the void boundaries over time while the void core continuously deepens~\cite{Sheth2004,vdWeygaert2011a,Hamaus2014b,vdWeygaert2016}. Both effects are supported by the mock data shown in figure~\ref{fig:xi_mock_Z}.
\begin{figure}[h]
	\centering
	\resizebox{\hsize}{!}{
		\includegraphics{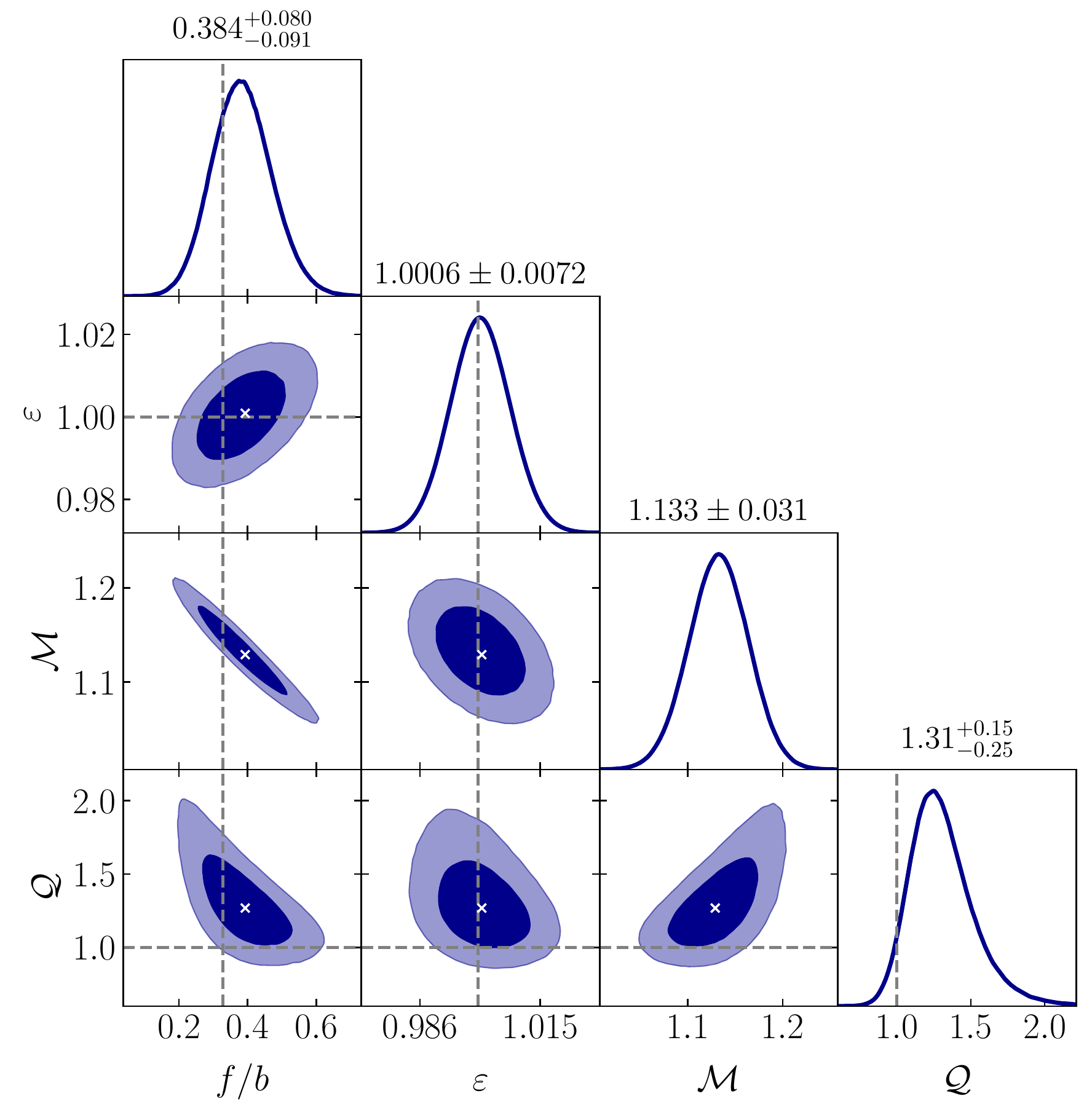}
		\includegraphics{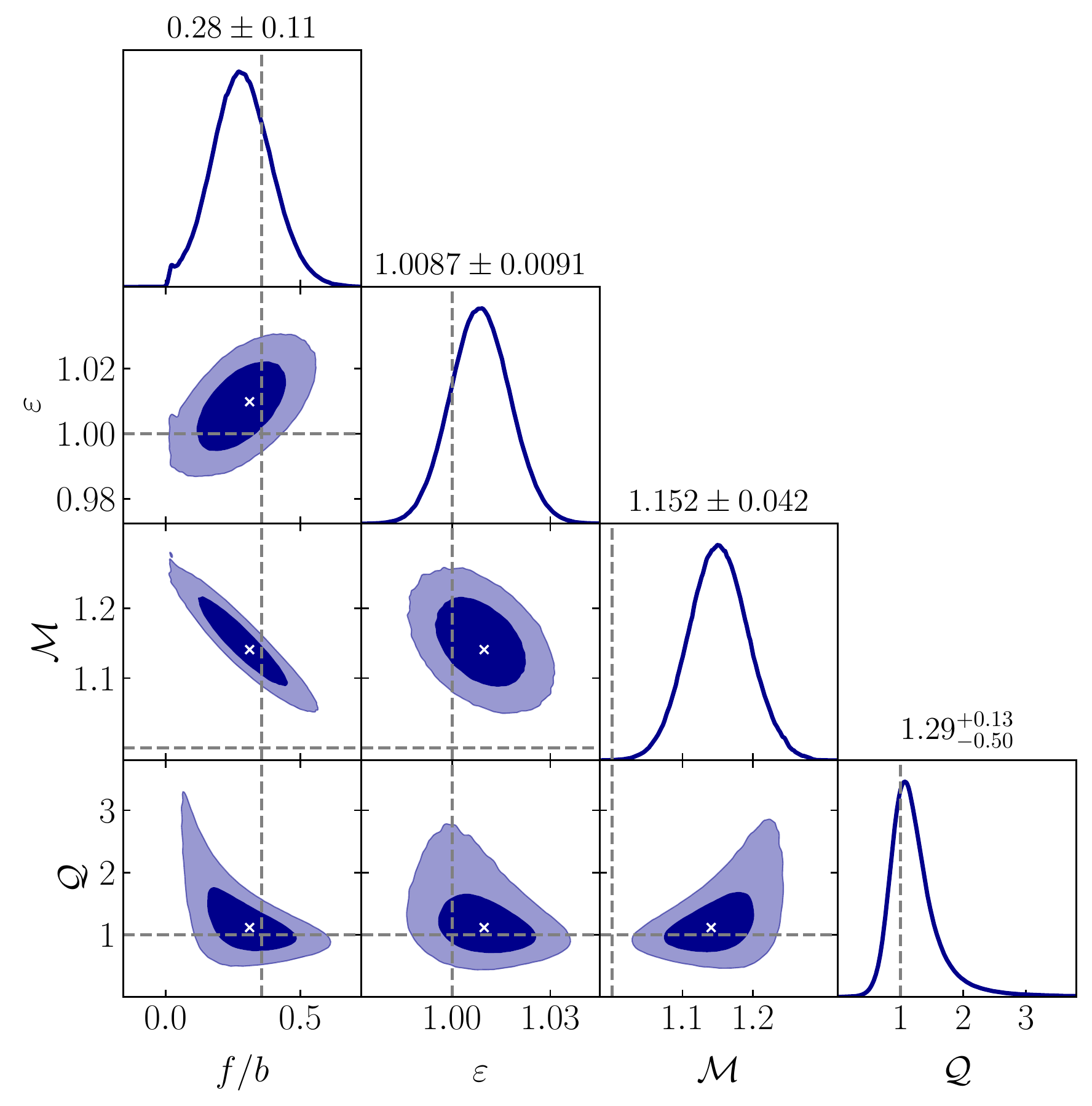}}
	\caption{As figure~\ref{fig:triangle_mock} for the ``low-z'' (left) and ``high-z'' (right) PATCHY void sample in figure~\ref{fig:xi_mock_Z}.}
	\label{fig:triangle_mock_Z}
\end{figure}

Finally, we repeat the model fits for the subsets in void redshift and run a full MCMC for each of them. The parameter posteriors are shown in figure~\ref{fig:triangle_mock_Z}. As for the full void sample from before we retrieve the input cosmology of the PATCHY mocks to within $68\%$ of the confidence levels for $f/b$ and $\varepsilon$. Also the posteriors of the nuisance parameters $\mathcal{M}$ and $\mathcal{Q}$ look similar as for the full void sample shown in figure~\ref{fig:triangle_mock}. However, we notice a very mild increase of $\mathcal{M}$ and a decrease of $\mathcal{Q}$ towards higher redshifts. Although the shifts remain well within the $1\sigma$ confidence intervals for these parameters, they may indicate a slightly lower contamination by Poisson noise, but a slightly stronger anisotropic selection effect for the smaller voids at lower redshift. This would indeed agree with previous simulation results that suggest the impact of RSDs on void identification to be more severe for voids with smaller effective radii~\cite{Pisani2015b}. Moreover, as large-scale structures develop more non-linear over time, we do expect the FoG effect to have a stronger impact on voids at lower redshift.

\subsection{Data analysis \label{subsec:fitting}}
\begin{figure}[t]
	\centering
	\resizebox{0.86\hsize}{!}{\includegraphics[trim=0 10 0 10]{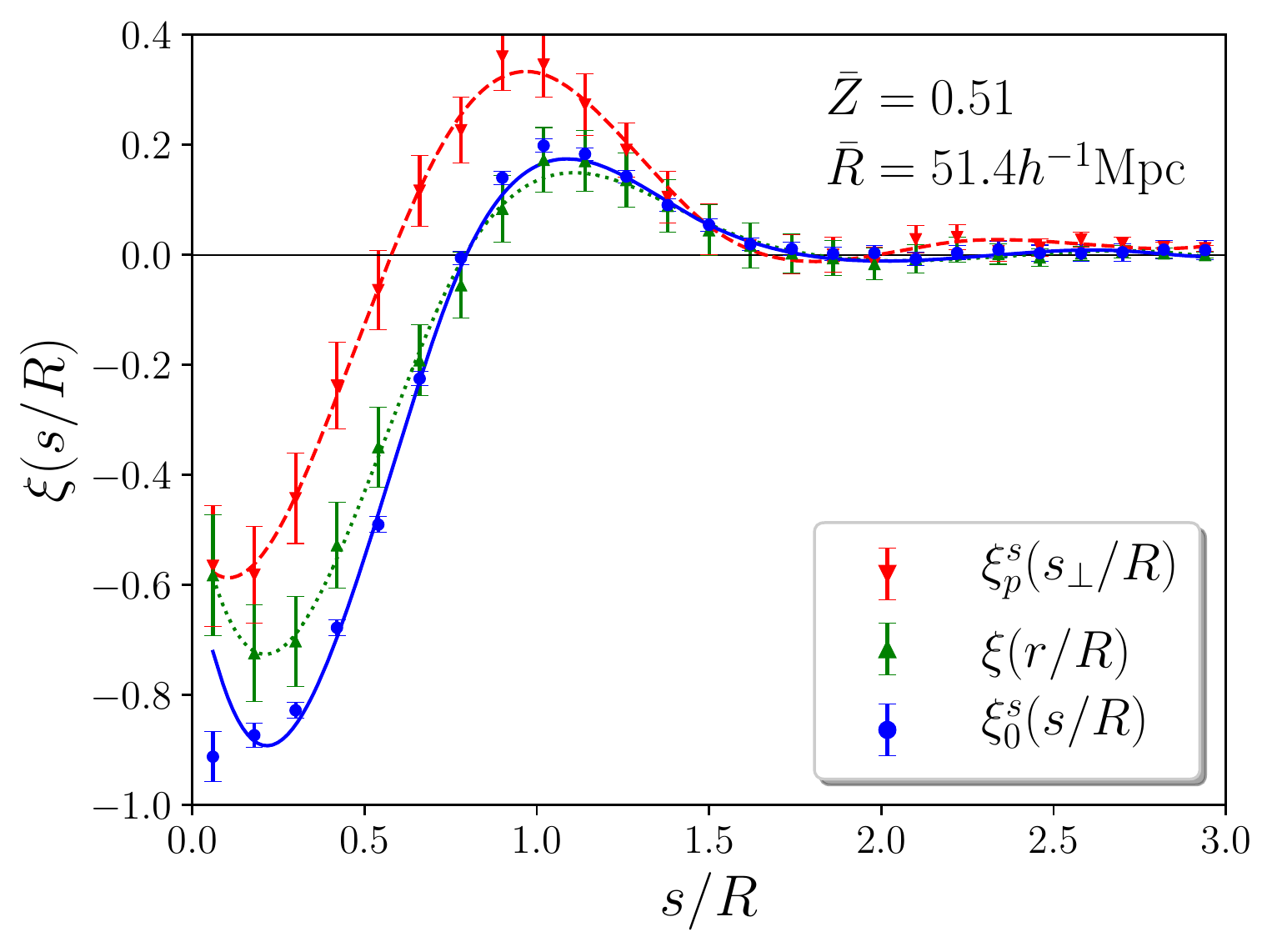}}
	\caption{Measurement of the projected void-galaxy cross-correlation function (red wedges, dashed line) from the BOSS DR12 combined sample in redshift space, and its real-space counterpart after deprojection (green triangles, dotted line). The measured redshift-space monopole (blue dots) follows the same functional form, in agreement with the linear model (blue solid line) from equation~(\ref{xi_0}).}
	\label{fig:xi_p_data}
\end{figure}

The successful model validation from the previous section now enables us to perform model fits on the real BOSS data. To this end we simply repeat the analysis steps that have already been performed on the PATCHY mocks above. We first measure the projected void-galaxy cross-correlation function $\xi^s_p(s_\perp)$ and apply the deprojection technique using the inverse Abel transform to obtain $\xi(r)$. The result is shown in figure~\ref{fig:xi_p_data}, along with the redshift-space monopole $\xi^s_0(s)$ and its best-fit model. We observe very similar trends as in the mocks, albeit with larger error bars as expected from the smaller sample size of voids available in the BOSS data.
\begin{figure}[h]
	\centering
	\resizebox{0.86\hsize}{!}{\includegraphics[trim=-40 0 0 10]{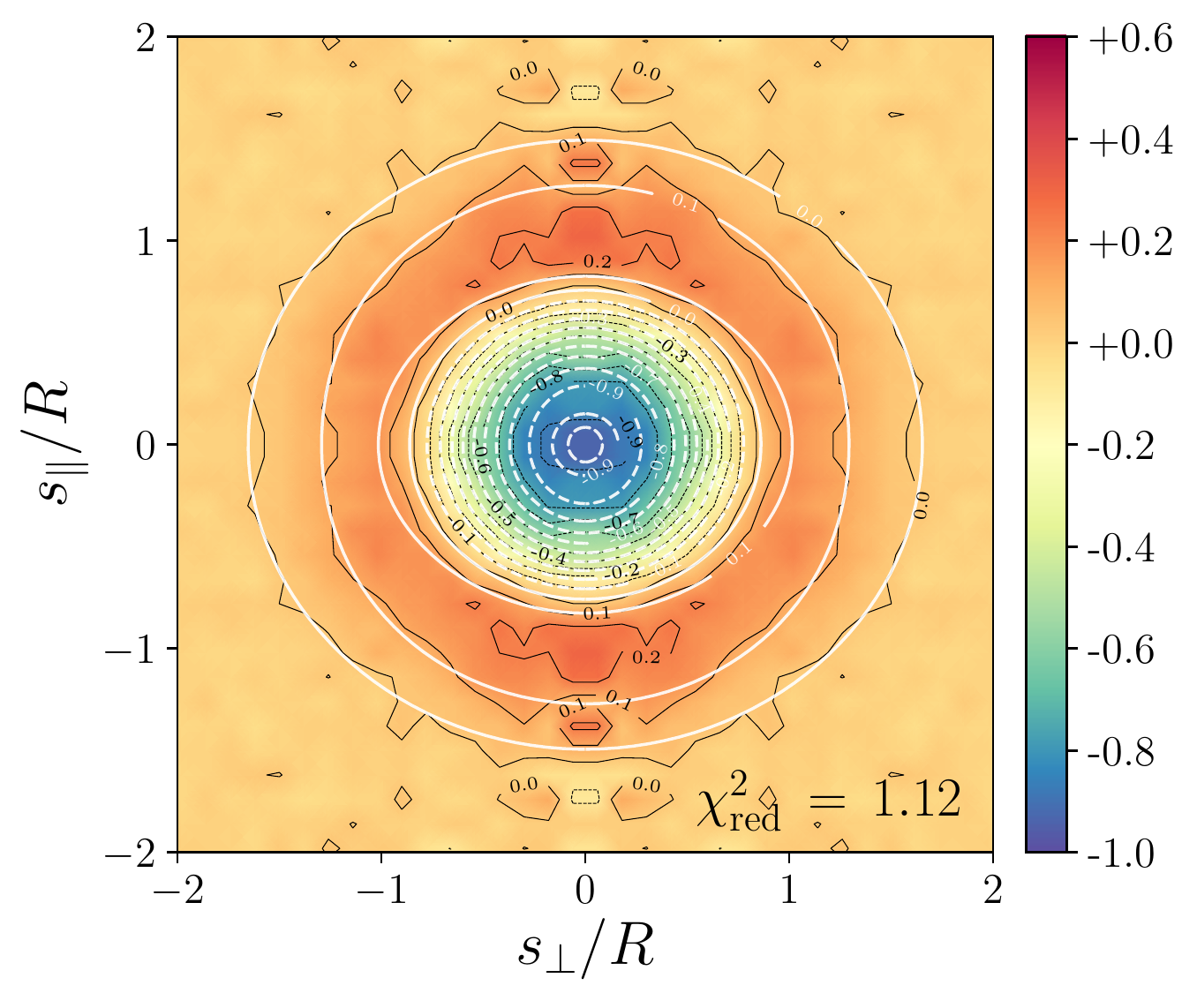}}
	\resizebox{0.86\hsize}{!}{\includegraphics[trim=0 10 0 10]{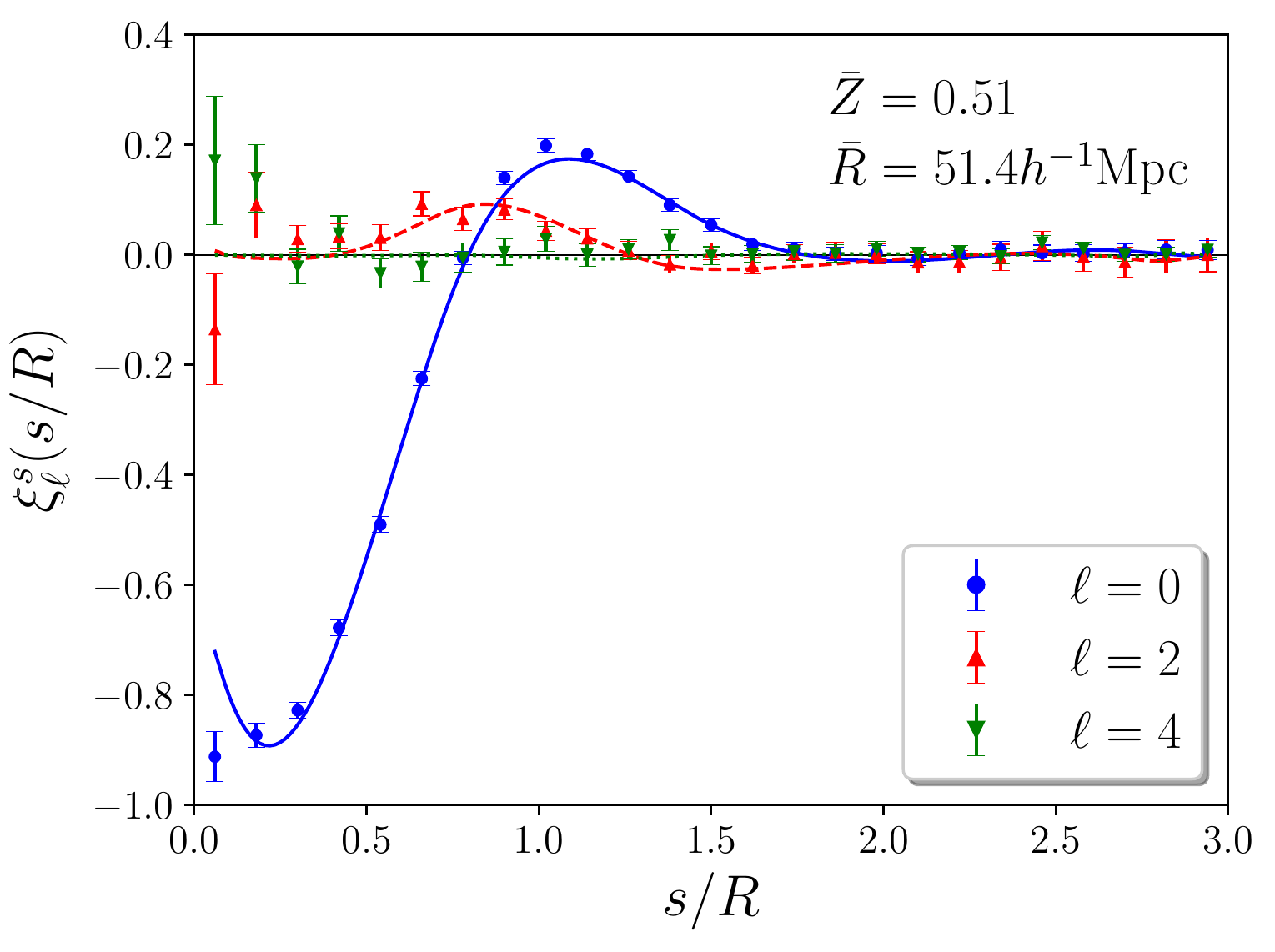}}
	\caption{TOP: Measurement of the stacked void-galaxy cross-correlation function $\xi^s(s_\parallel/R,s_\perp/R)$ from voids in the BOSS DR12 combined sample (color scale with black contours) and the best-fit model (white contours) from equation~(\ref{xi^s_lin2}). BOTTOM: Monopole (blue dots), quadrupole (red triangles) and hexadecapole (green wedges) from the same data with corresponding model fits (solid, dashed, dotted lines). The mean redshift and effective radius of the void sample is shown at the top.}
	\label{fig:xi_data}
\end{figure}
Figure~\ref{fig:xi_data} presents the two-point statistics for the void-galaxy cross-correlation function and its multipoles. Apart from the larger impact of statistical noise due to the substantially smaller sample size (by a factor of $N_\mathrm{m}=30$), the results are in excellent agreement with the mock data. Both amplitude and shape of $\xi^s(s_\parallel,s_\perp)$, as well as $\xi^s_\ell(s)$ are very consistent in comparison with figure~\ref{fig:xi_mock}. A mild but noticeable difference can be seen very close to the void center, which appears more flattened in the data. One can also perceive stronger fluctuations of the quadrupole and hexadecapole in this regime, but those are simply due to the sparser statistics of galaxies near the void center and thus fully consistent with the error bars. This fact is further supported by the accurate model fit to the data, resulting in a reduced chi-square value of $\chi^2_\mathrm{red}=1.12$.
\begin{figure}[t]
	\centering
	\resizebox{0.7\hsize}{!}{
		\includegraphics{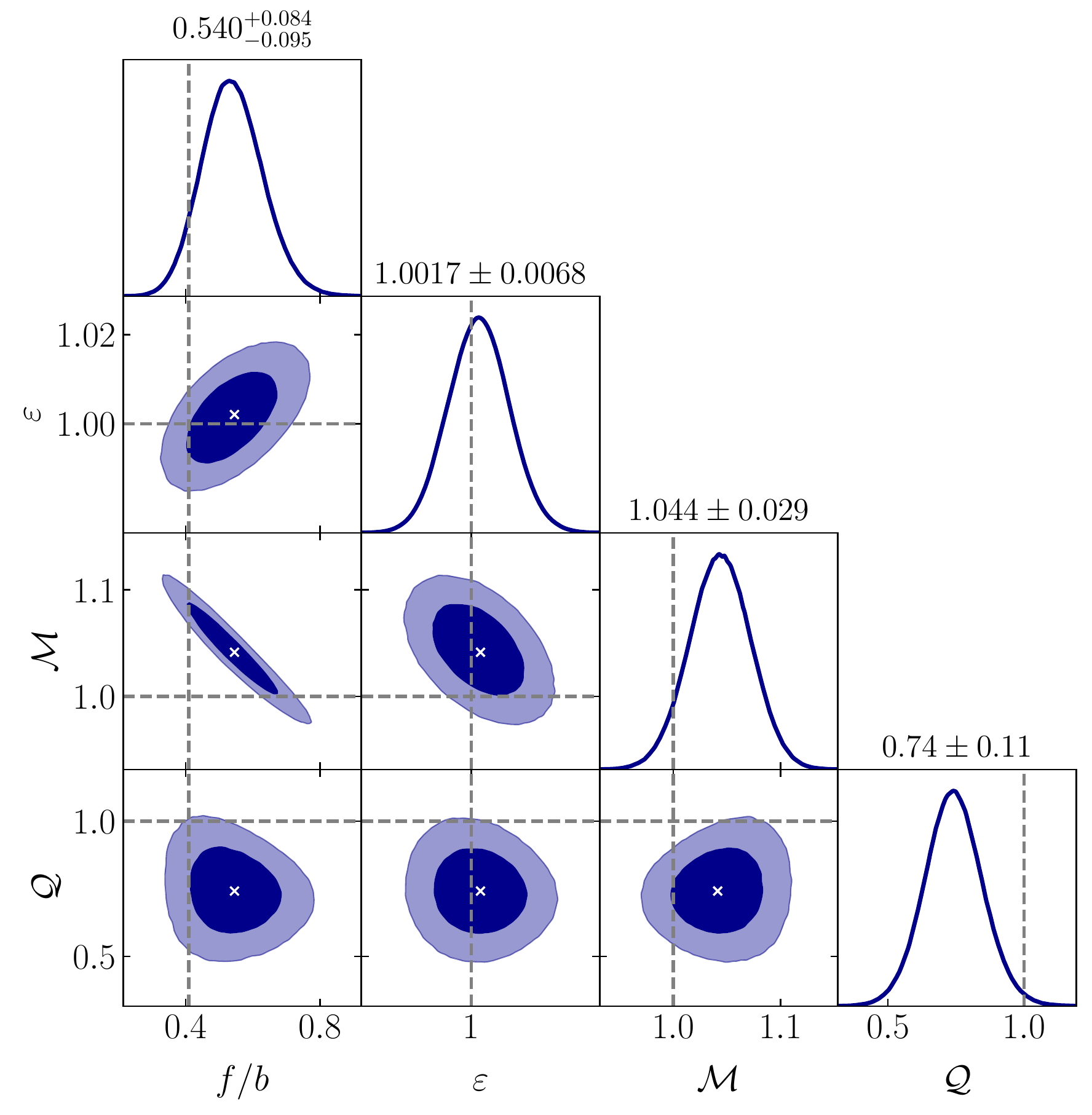}}
	\caption{Posterior probability distribution of the model parameters that enter in equation~(\ref{xi^s_lin2}), obtained via MCMC from the BOSS DR12 data shown in figure~\ref{fig:xi_data}. Dark and light shaded areas show $68\%$ and $95\%$ confidence regions with a cross marking the best fit, dashed lines indicate fiducial values of the RSD and AP parameters $(f/b=0.409,\,\varepsilon=1)$, and default values for the nuisance parameters $(\mathcal{M}=\mathcal{Q}=1)$. The top of each column states the mean and standard deviation of the 1D marginal distributions.}
	\label{fig:triangle_data}
\end{figure}

The full posterior parameter distribution obtained from the BOSS data is shown in figure~\ref{fig:triangle_data}, which qualitatively resembles the mock results from figure~\ref{fig:triangle_mock}. However, a few important differences are apparent. Firstly, the value of $f/b$ from the data is significantly higher than in the mocks, which is partly driven by the lower value of $b=1.85$ in the BOSS data, compared to $b=2.20$ of the mocks. Further, the nuisance parameters $\mathcal{M}$ and $\mathcal{Q}$ both take on lower values in the data than in the mocks. In particular, $\mathcal{M}$ is consistent with unity to within $68\%$ confidence, which could indicate that voids in the BOSS data are less affected by discreteness noise than what was expected from the PATCHY mocks. At the same time, $\mathcal{Q}$ is consistent with unity only at the $95\%$ confidence level from below, suggesting an attenuation of the quadrupole amplitude when compared to the mocks. This could be caused by systematics in the BOSS data that have not been taken into account at the same level of complexity in the mocks. One such example is the foreground contamination by stars~\cite{Reid2016}.

\begin{figure}[t]
	\centering
	\resizebox{\hsize}{!}{
		\includegraphics[trim=0 30 0 5, clip]{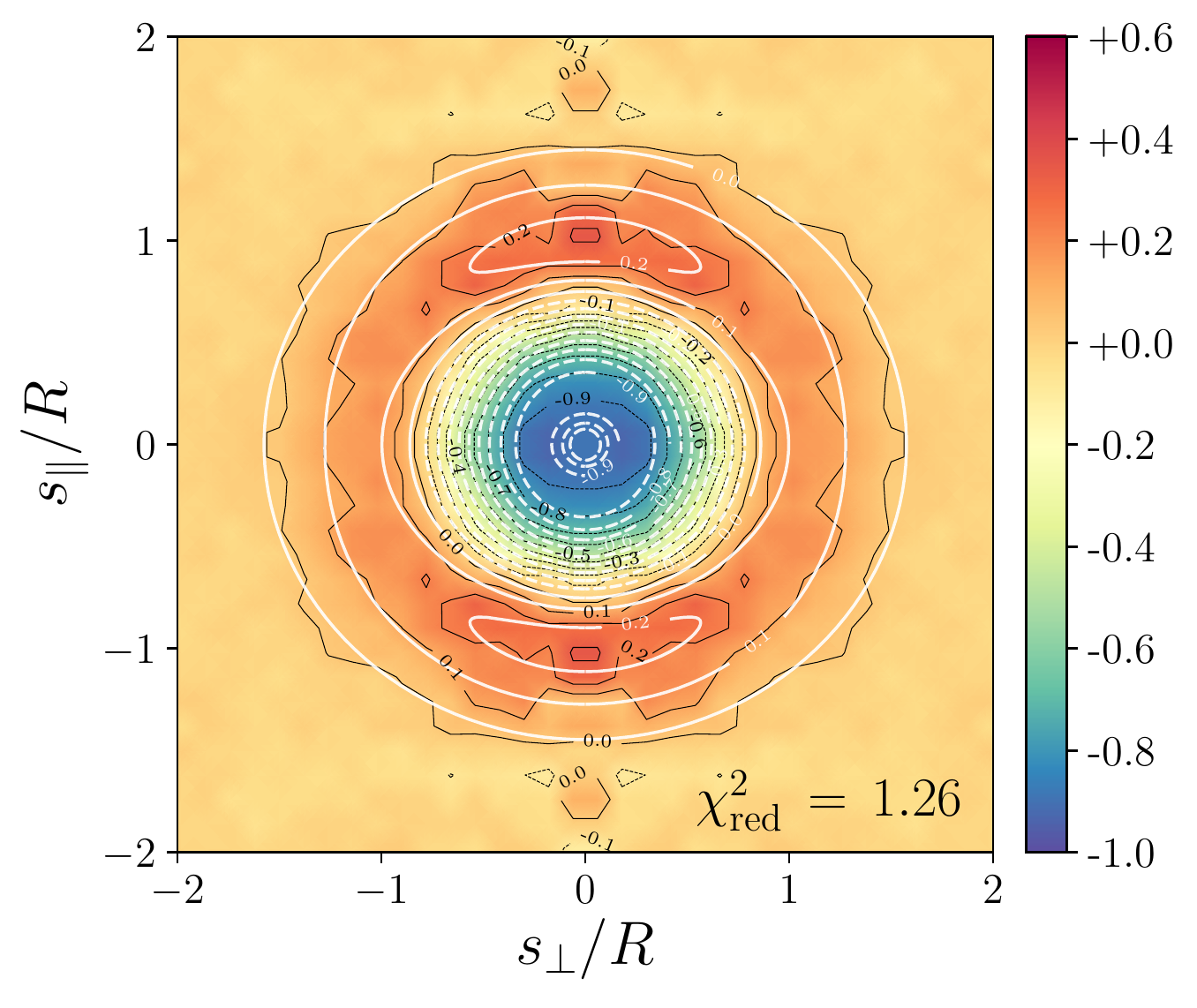}
		\includegraphics[trim=0 30 0 5, clip]{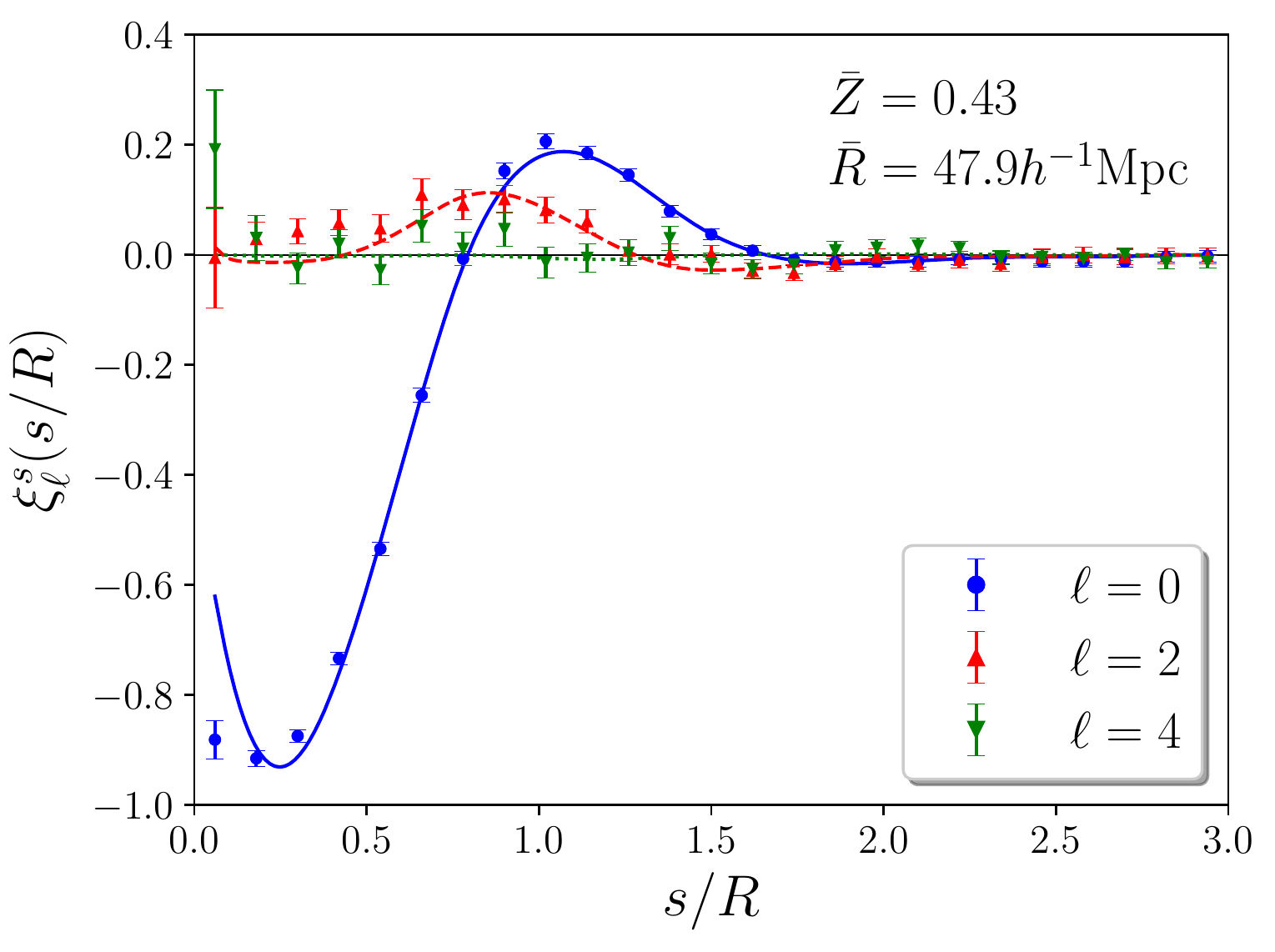}}
	\resizebox{\hsize}{!}{
		\includegraphics[trim=0 10 0 5, clip]{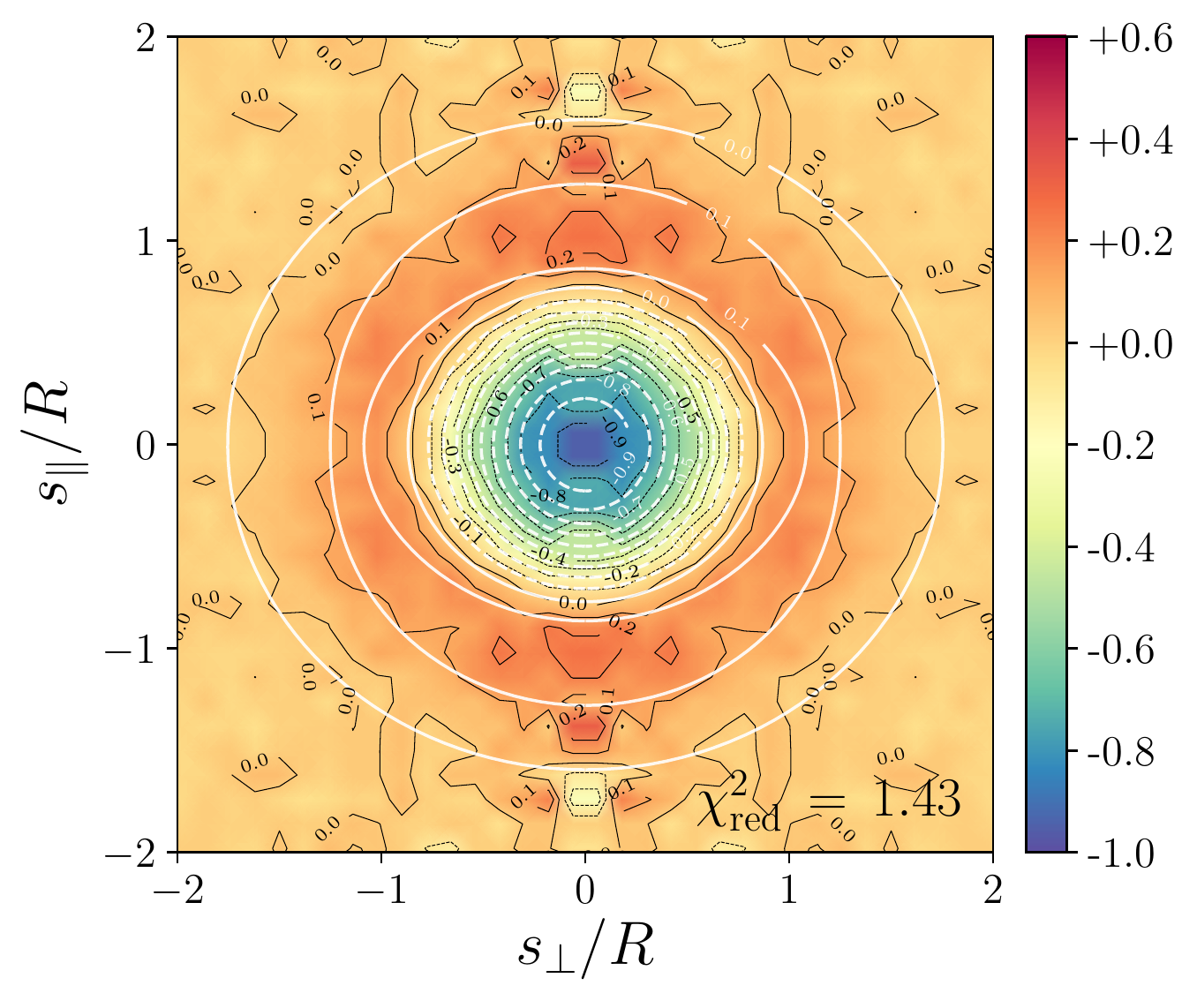}
		\includegraphics[trim=0 10 0 5, clip]{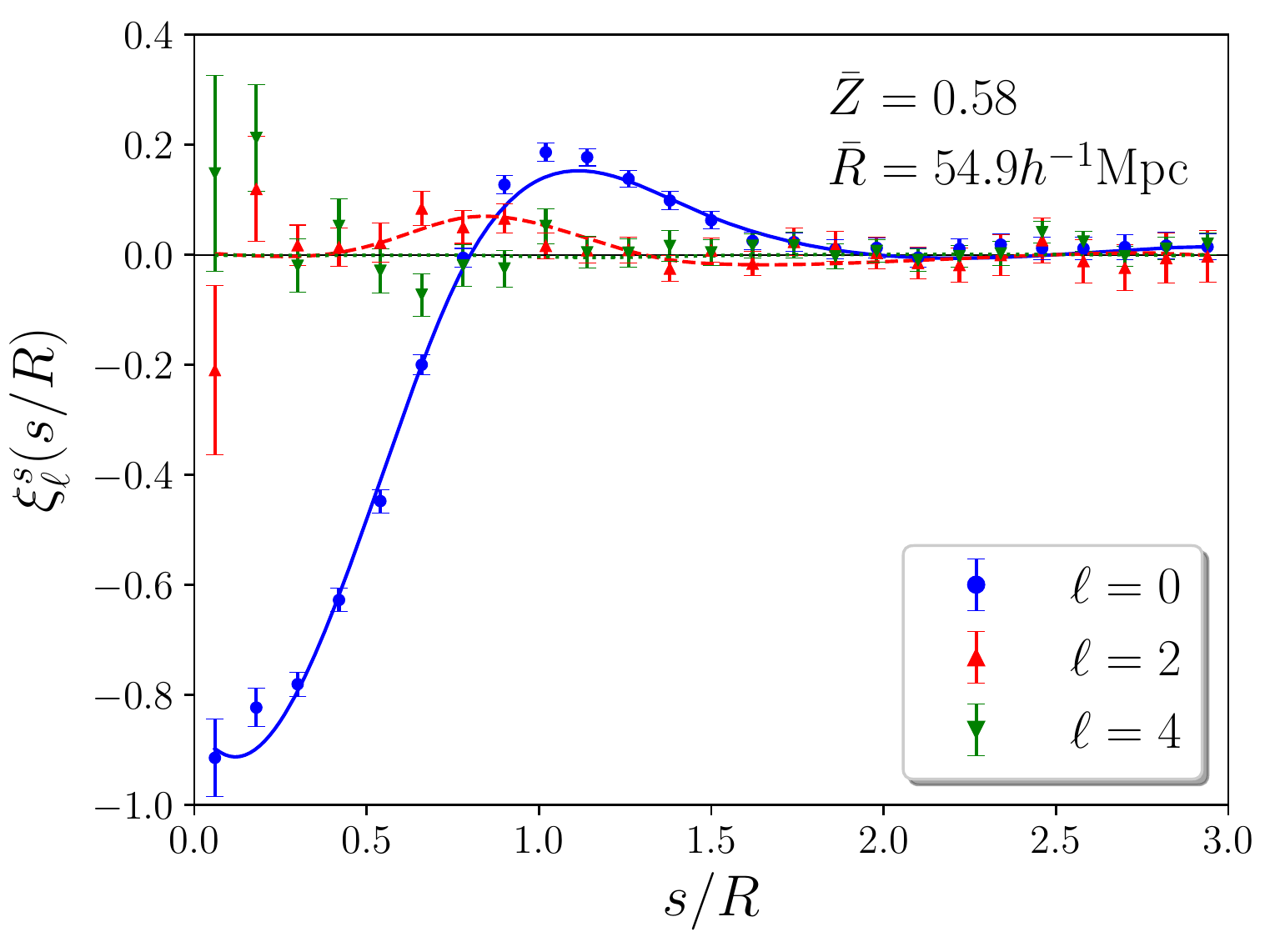}}
	\caption{As figure~\ref{fig:xi_data} after splitting the BOSS void sample at its median redshift of $Z=0.52$ into $50\%$ lowest-redshift (``low-z'', top row) and $50\%$ highest-redshift voids (``high-z'', bottom row).}
	\label{fig:xi_data_Z}
\end{figure}
\begin{figure}[h]
	\centering
	\resizebox{\hsize}{!}{
		\includegraphics{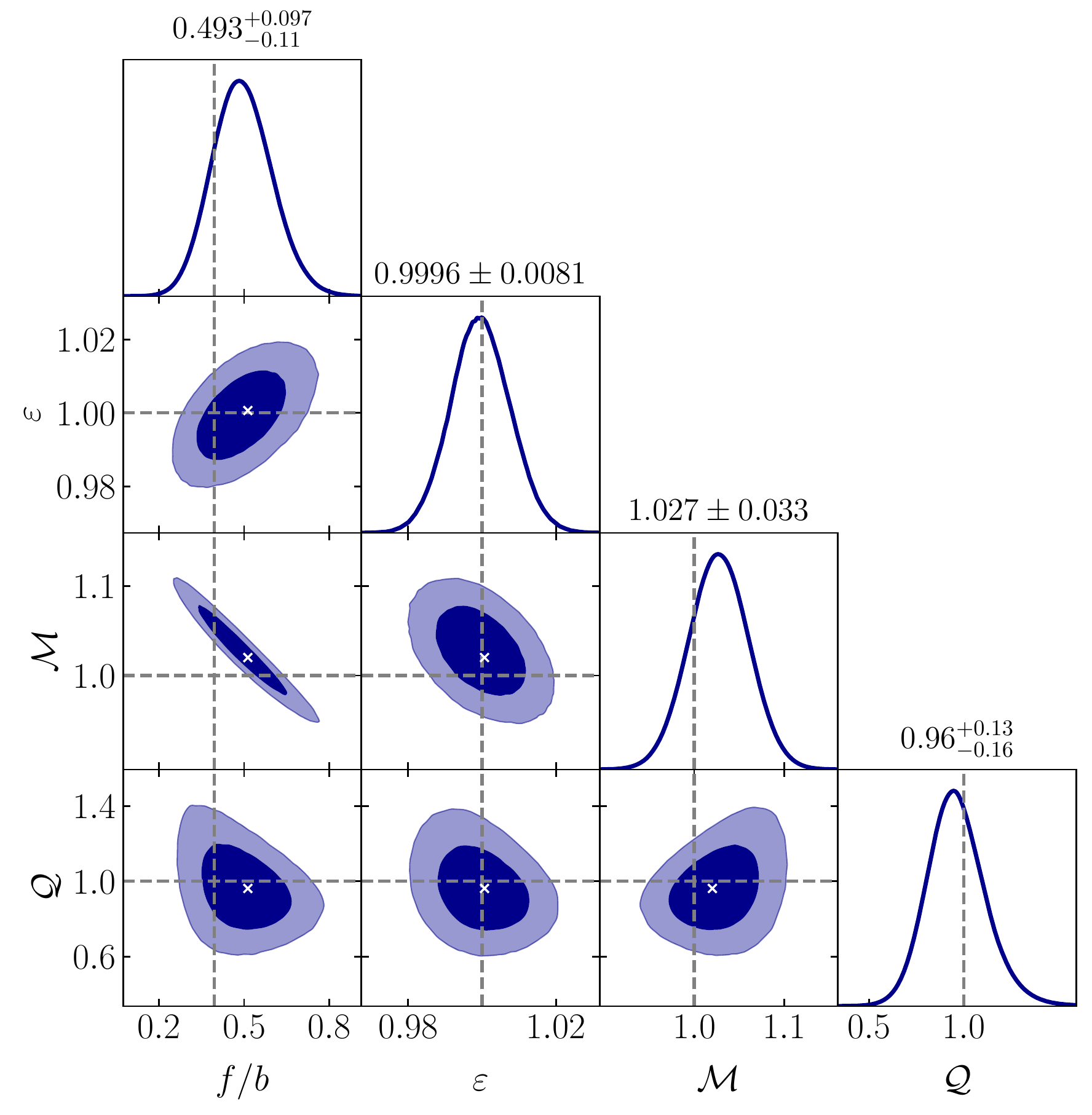}
		\includegraphics{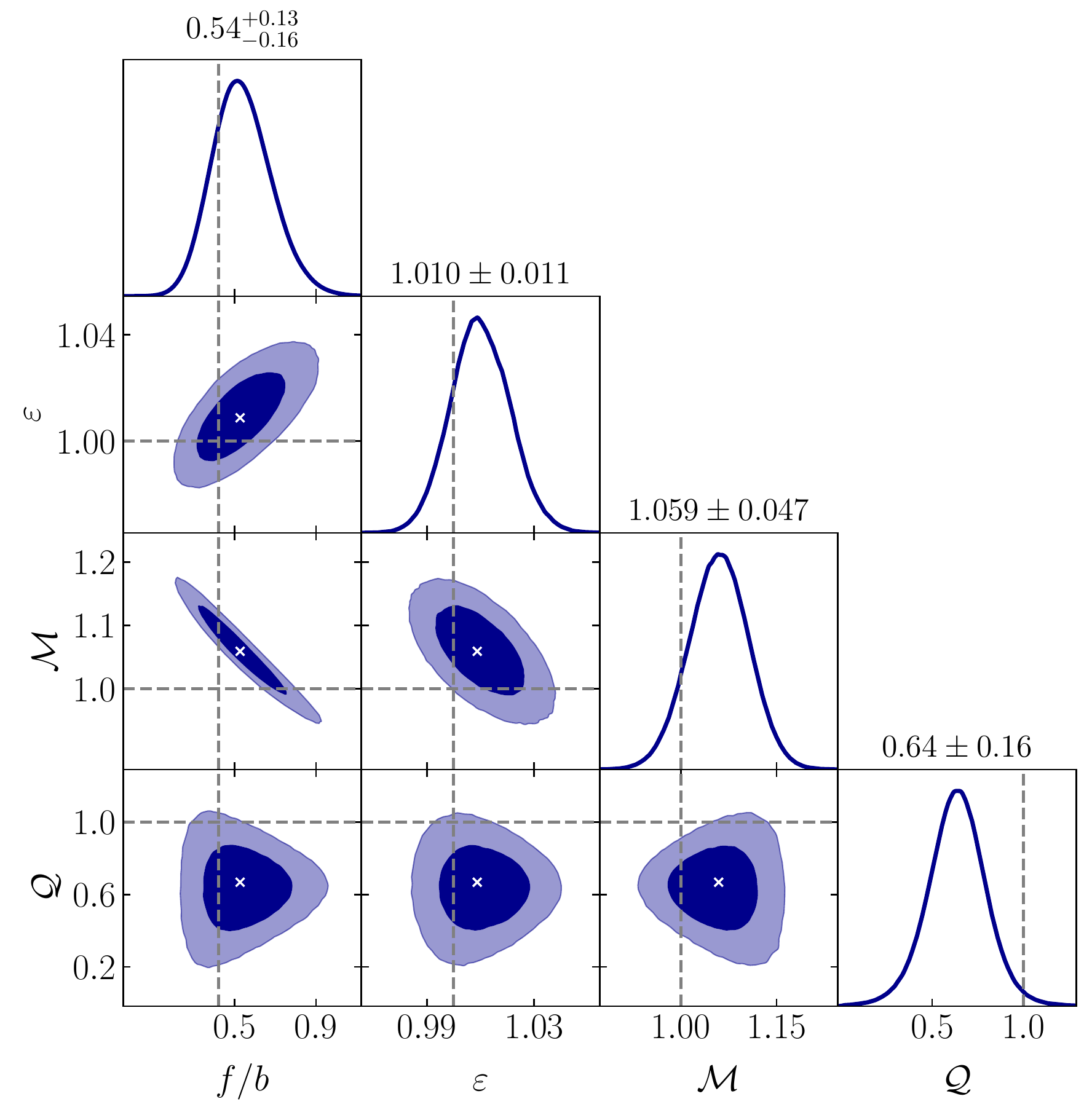}}
	\caption{As figure~\ref{fig:triangle_data} for the ``low-z'' (left) and ``high-z'' (right) BOSS void sample in figure~\ref{fig:xi_data_Z}.}
	\label{fig:triangle_data_Z}
\end{figure}

As done for the PATCHY mocks, we further investigate the redshift evolution of our constraints with the BOSS data. To this end we split our catalog into two equally sized low- and high-redshift bins again (``low-z'' and ``high-z'' sample), the resulting clustering statistics are shown in figure~\ref{fig:xi_data_Z}. We observe the same trends as before, namely a deepening of void interiors and an increase of the quadrupole towards lower redshift. Given the lower statistical power of these bins the data evidently look more noisy, but the linear model still provides a good fit overall. Note that we neglect any uncertainties in our theory model, which relies on a measurement of the projected correlation function $\xi^s_p(s_\perp)$. Thus, especially for noisy data, this may result in an underestimation of the full covariance and hence a higher reduced chi-square. Nevertheless, our $\chi^2_\mathrm{red}$ values are still reasonably close to unity. Figure~\ref{fig:triangle_data_Z} presents the parameter posteriors of the model fit. Evidently, even these subsets of voids can still provide interesting constraints with a good accuracy. We find our best-fit values for $f/b$ and $\varepsilon$ to be in agreement with the fiducial Planck cosmology to within $68\%$ of the confidence levels. Moreover, we notice that the low amplitude for the nuisance parameter $\mathcal{Q}$ is driven by the high-redshift bin, otherwise the best-fit values for both $\mathcal{M}$ and $\mathcal{Q}$ are consistent with unity to within the $68\%$ contours.

So far our analysis has exclusively been based on the observed data without using any prior information. The model ingredients $\xi(r)$, $\mathcal{M}$, and $\mathcal{Q}$ have been derived from this data self-consistently. One may argue, however, that these quantities are already available from the survey mocks to a much higher accuracy (see section~\ref{subsec:validation}). Hence, making use of the mocks to calibrate those model ingredients allows us to evade marginalization over nuisance parameters and to use the statistical power of the data solely to constrain cosmology. We implement this calibration approach by simply using the $30$ PATCHY mocks to estimate $\xi(r)$ via equation~(\ref{xi_d}) and fixing the nuisance parameters $\mathcal{M}$ and $\mathcal{Q}$ to their best-fit values from the corresponding void sample in the mock analysis. This leaves us with only two remaining free parameters $f/b$ and $\varepsilon$, for which we repeat the MCMC runs.

\begin{figure}[b]
	\centering
	\resizebox{\hsize}{!}{
		\includegraphics{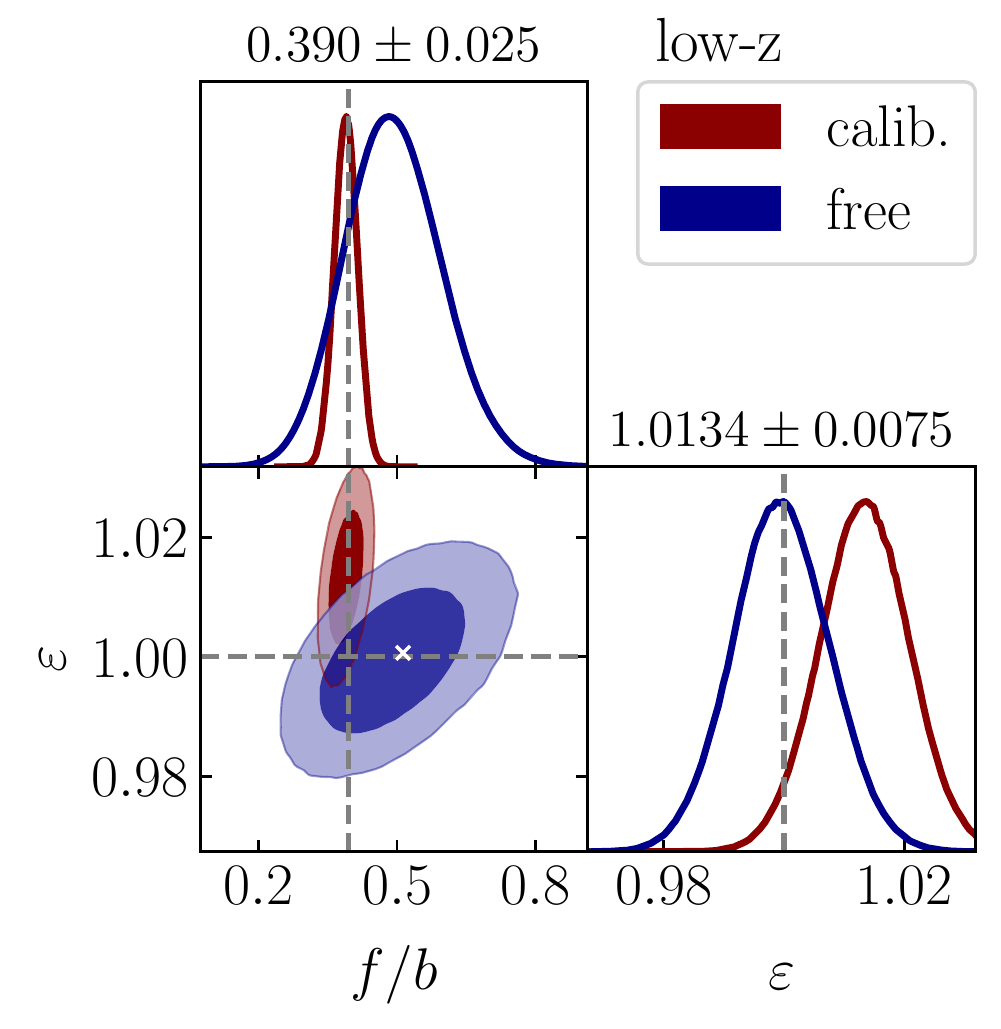}
		\includegraphics{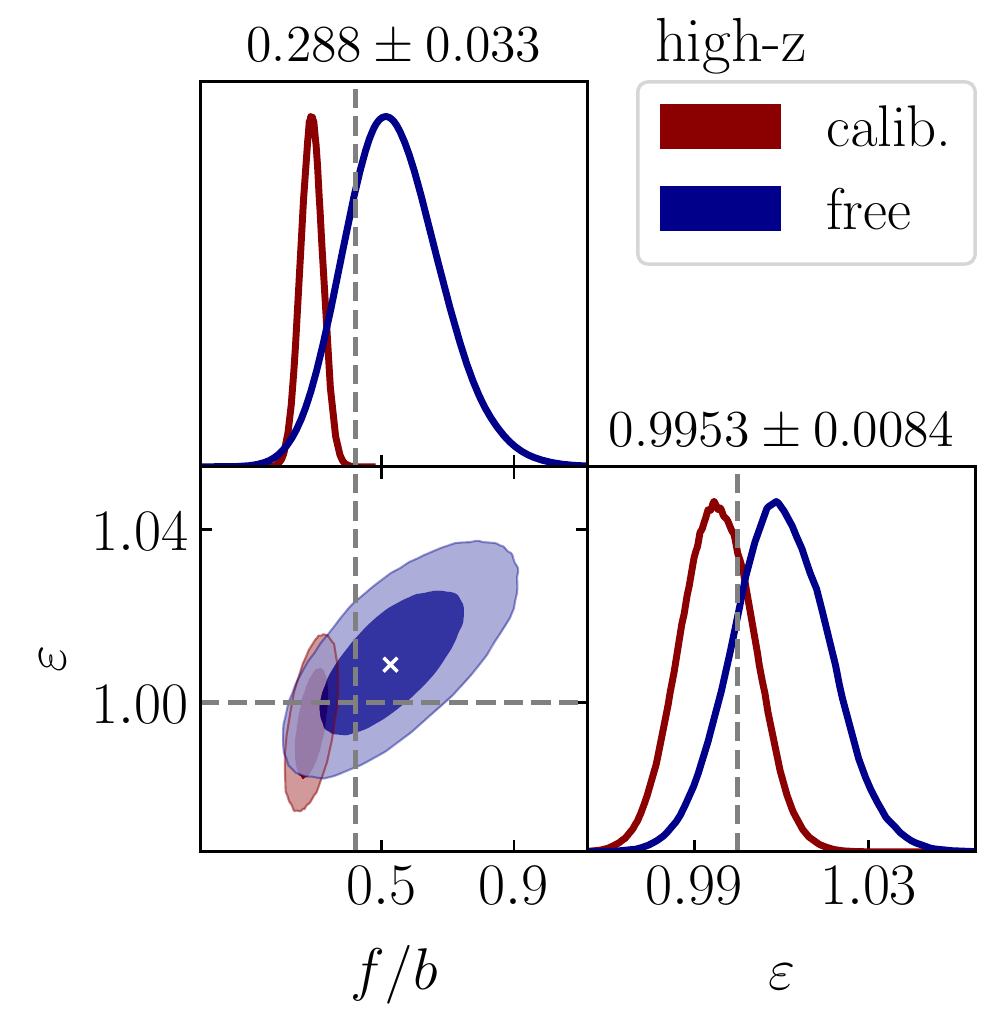}
		\includegraphics{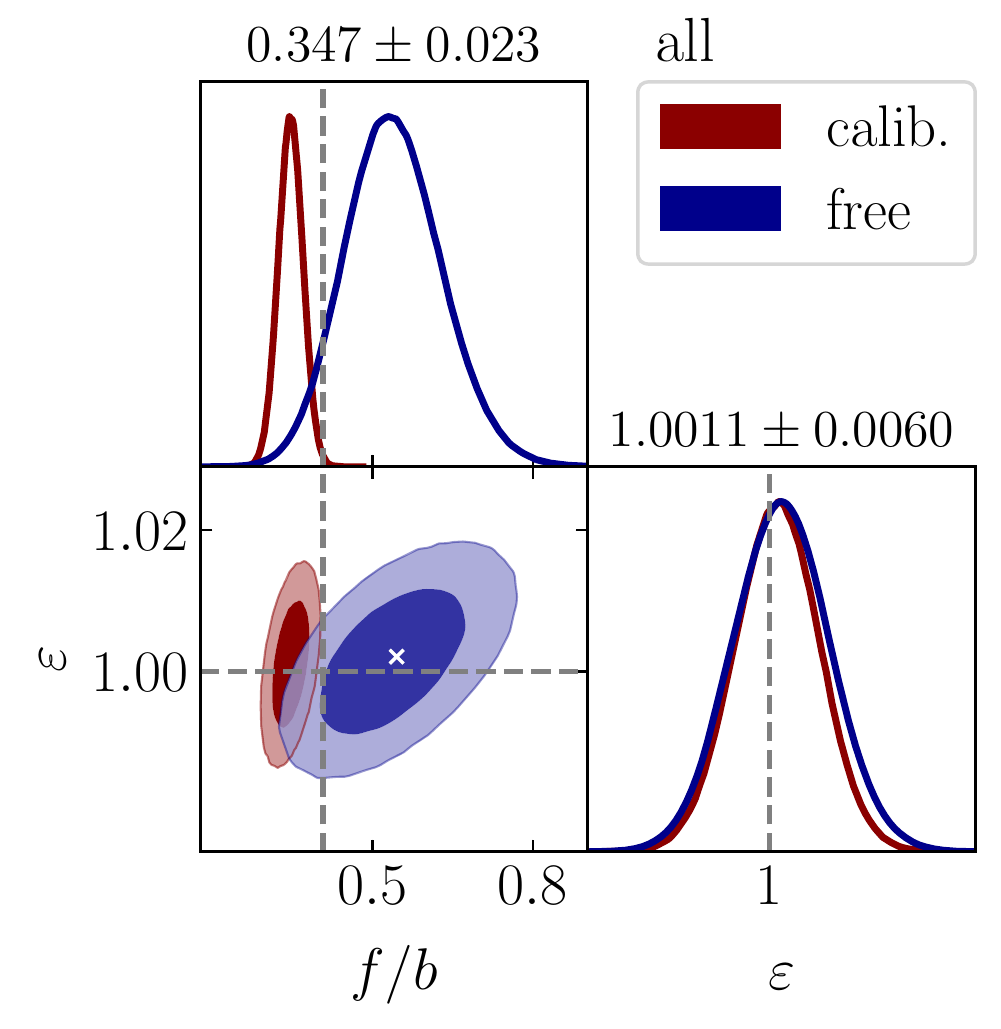}}
	\caption{As figures~\ref{fig:triangle_data} and~\ref{fig:triangle_data_Z}, but focused on the RSD and AP parameters $f/b$ and $\varepsilon$. The red contours show constraints when the theory model is calibrated with the PATCHY mocks to determine $\xi(r)$ and the values of the nuisance parameters $\mathcal{M}$ and $\mathcal{Q}$. The blue contours show the original constraints when $\xi(r)$, $\mathcal{M}$ and $\mathcal{Q}$ are left free to be jointly estimated from the BOSS data. The top of each column states the mean and standard deviation of the calibrated constraints, the corresponding void sample is indicated above the figure legend of each panel.}
	\label{fig:triangle_data_cal}
\end{figure}

The results are presented in figure~\ref{fig:triangle_data_cal}, showing their posterior distribution for each of our void samples. Evidently, the mock-calibrated analysis (calib.) significantly improves upon the constraints obtained without calibration (free). While the accuracy on the AP parameter $\varepsilon$ exhibits mild improvements of about $10\%$ to $30\%$, the error on $f/b$ shrinks by roughly a factor of~$4$. This is mainly due to the considerable anti-correlation between $f/b$ and $\mathcal{M}$ apparent in figures~\ref{fig:triangle_data} and~\ref{fig:triangle_data_Z}, which is removed when $\mathcal{M}$ is fixed to a fiducial value. We note, however, that the best-fit values for $\mathcal{M}$ in our uncalibrated analysis differ significantly between the observed data and the mocks. In particular, we found the values of $\mathcal{M}$ in the mocks to be higher than in the data by about $10\%$. This may be partly due to the higher bias parameter and the level of overdispersion in the PATCHY mocks as compared to the data, but more fundamentally the mismatch reveals that not all aspects of the data are understood precisely enough to be fully represented by the mocks.

Therefore, we caution the use of mocks for model calibration, as such an approach is prone to cause biased constraints on cosmology. This is evident from the significant shifts of the posteriors in figure~\ref{fig:triangle_data_cal} after performing the calibration. Another consequence is the underestimation of parameter uncertainties, which is caused by mistaking prior information from the mocks as the truth. The mocks merely represent many realizations of a single cosmological model with one fiducial parameter set and one fixed prescription of how dark matter halos are populated by galaxies (halo occupation distribution). A realistic model must therefore either take into account the dependence on these ingredients including their uncertainty, or constrain them from the data directly. Our approach follows the philosophy to exclusively rely on the observed data to obtain most robust constraints.

\section{Discussion\label{sec:discussion}}

\subsection{Parameter constraints \label{subsec:constraints}}
\begin{table}[b]
	\centering
	\caption{Constraints on RSD and AP parameters (mean values with $1\sigma$ errors) from \textsc{vide} voids in the final BOSS data (top rows). The middle rows show corresponding constraints after model-calibration on the PATCHY mocks, and the bottom rows provide Planck 2018~\cite{Planck2018} results as reference values, assuming a flat $\Lambda$CDM model. All constraints on $\Om$ in the last column assume flat $\Lambda$CDM as well.}\vspace{10pt}
	\label{tab:constraints}
	\centerline{
		\begin{tabular}{lccccc}
			\toprule
			Sample ($\bar{Z}$) & $f/b$           & $f\sigma_8$     & $\varepsilon$     & $\DA H/c$       & $\Om$\\
			\midrule
			low-z ($0.43$)     & $0.493\pm0.105$ & $0.590\pm0.125$ & $0.9996\pm0.0081$ & $0.485\pm0.004$ & $0.306\pm0.027$\\[3pt]
			high-z ($0.58$)    & $0.538\pm0.146$ & $0.594\pm0.162$ & $1.0100\pm0.0111$ & $0.702\pm0.008$ & $0.334\pm0.030$\\[3pt]
			all ($0.51$)       & $0.540\pm0.091$ & $0.621\pm0.104$ & $1.0017\pm0.0068$ & $0.588\pm0.004$ & $0.312\pm0.020$\\[3pt]
			\midrule
			low-z calib.       & $0.390\pm0.025$ & $0.554\pm0.036$ & $1.0134\pm0.0075$ & $0.492\pm0.004$ & $0.353\pm0.026$\\[3pt]
			high-z calib.      & $0.288\pm0.033$ & $0.379\pm0.043$ & $0.9953\pm0.0084$ & $0.691\pm0.006$ & $0.295\pm0.022$\\[3pt]
			all calib.         & $0.347\pm0.023$ & $0.474\pm0.031$ & $1.0011\pm0.0060$ & $0.588\pm0.003$ & $0.310\pm0.017$\\[3pt]
			\midrule
			low-z ref.         & $0.398\pm0.003$ & $0.476\pm0.006$ & $1.0025\pm0.0022$ & $0.487\pm0.001$ & $0.315\pm0.007$\\[3pt]
			high-z ref.        & $0.425\pm0.003$ & $0.470\pm0.005$ & $1.0031\pm0.0028$ & $0.697\pm0.002$ & $0.315\pm0.007$\\[3pt]
			all ref.           & $0.412\pm0.003$ & $0.474\pm0.006$ & $1.0028\pm0.0025$ & $0.589\pm0.001$ & $0.315\pm0.007$\\[3pt]
			\bottomrule
	\end{tabular}}
\end{table}

The final parameter constraints from our \textsc{vide} void samples found in the BOSS DR12 data are summarized in table~\ref{tab:constraints}. We distinguish between uncalibrated and mock-calibrated (calib.) samples, both for the subsets at low and high redshift (low-z, high-z), as well as the full redshift range (all). The table presents measured quantities of mean void redshift $\bar{Z}$, RSD parameter $f/b$ and AP parameter $\varepsilon$. Furthermore, it provides derived constraints on $f\sigma_8$, $\DA H$ and $\Om$. For $f\sigma_8$ we multiply our constraint on $f/b$ by $b\sigma_8$, with $b=1.85$ and $\sigma_8=0.8111$ from Planck 2018~\cite{Planck2018}. For the calibrated case we use $b=2.20$ from the mocks.

In principle the parameter combination $f\sigma_8$ could be constrained from voids directly, but only if the theory model can explicitly account for the dependence of the void-galaxy cross-correlation function $\xi(r)$ on $\sigma_8$. Sometimes the assumption $\xi(r)\propto\sigma_8$ is used without further justification, but evidently this must fail in the non-linear regime~\cite{Juszkiewicz2010} where $\xi(r)$ approaches values close to $-1$, due to the restriction $\xi(r)>-1$. The same argument applies to the density profiles of dark matter halos or galaxy clusters, which do not simply scale linearly with $\sigma_8$~\cite{Brown2020}. Moreover, while the value of $\sigma_8$ controls the amplitude of matter fluctuations and thus the formation of halos, its effect on voids identified in the distribution of galaxies that populate those halos is far from trivial. Another approach is to directly estimate $b\sigma_8$ via an integral over the projected galaxy auto-correlation function~\cite{Hawken2017}. However, the result contains non-linear contributions from small scales that must be accounted for, which again involves assumptions about a particular cosmological model. The covariance between measurements of $f/b$ and $b\sigma_8$ remains inaccessible to that approach as well.

\begin{figure}[t]
	\centering
	\resizebox{0.51\hsize}{!}{
		\includegraphics[trim=10 10 10 10]{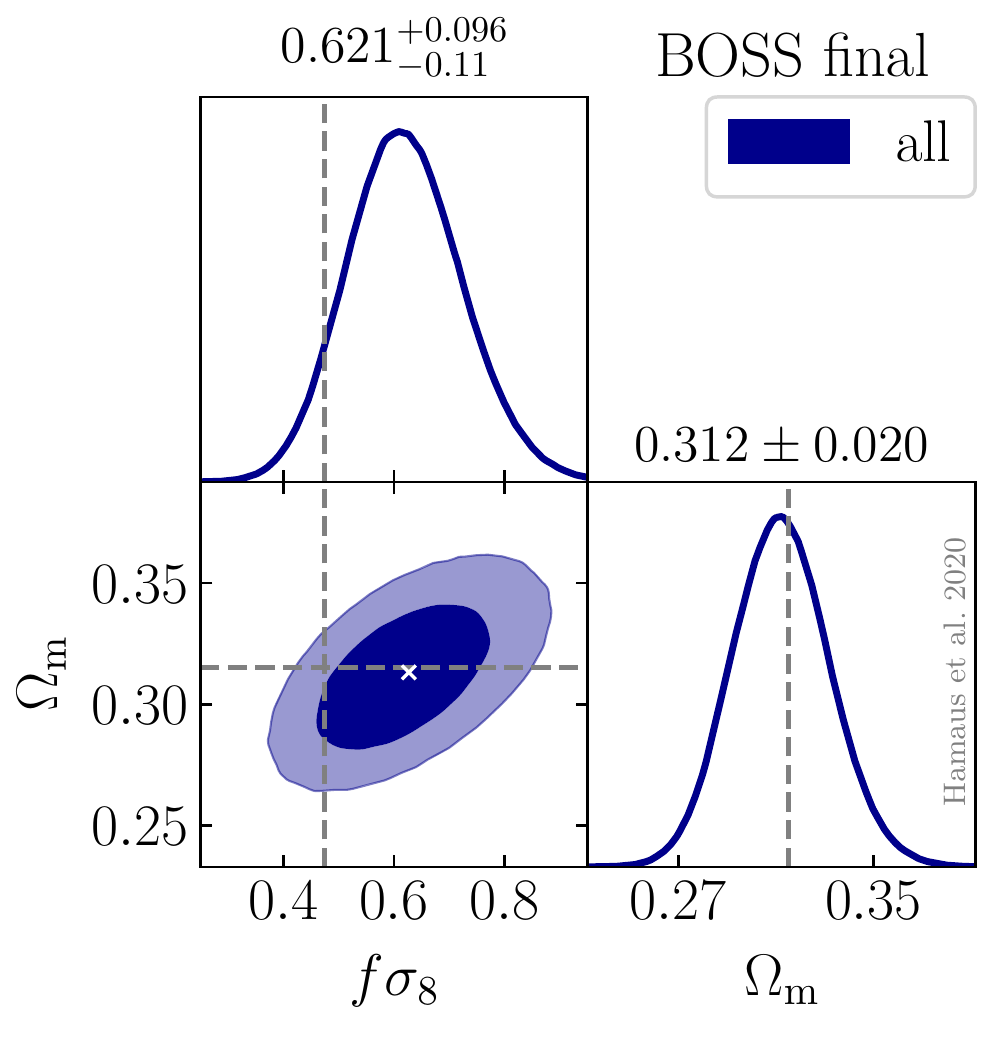}}
	\caption{Calibration-independent constraints on the cosmological parameters $f\sigma_8$ and $\Om$ from our full \textsc{VIDE} void sample in the final BOSS data. They are converted from the RSD and AP parameter posteriors shown in figure~\ref{fig:triangle_data}, as described in section~\ref{subsec:constraints}. A white cross indicates our best fit and dashed lines show mean parameter values from the Planck 2018 baseline results as a reference~\cite{Planck2018}.}
	\label{fig:triangle_data_fs8_Om}
\end{figure}

Finally, measurements involving the parameter $\sigma_8$ commonly ignore its implicit dependence on the Hubble parameter $h$ via the choice of $8\hMpc$ as reference scale, and therefore underestimate the uncertainty. Reference~\cite{Sanchez2020} argued to instead use $\sigma_{12}$ with $12$Mpc, which yields about the same value as $\sigma_8$ for a Planck-constrained value of $h$. For these reasons we decided to follow the simpler procedure described above to derive constraints on $f\sigma_8$, allowing us to compare existing results across the literature. The constraint on $\DA H$ can be obtained via equation~(\ref{epsilon}) by multiplying $\varepsilon$ and its error with $\DA'H'$ from our fiducial flat $\Lambda$CDM cosmology from section~\ref{subsec:galaxies}. In this case the only free cosmological parameter in the product $\DA H$ is $\Om$, so we can numerically invert this function to obtain the full posterior on $\Om$. Its mean and standard deviation are shown in the last column of table~\ref{tab:constraints}. Finally, we present our main result for the converted parameter constraints on $f\sigma_8$ and $\Om$ in figure~\ref{fig:triangle_data_fs8_Om}. It originates from the calibration-independent analysis of our full void sample in the final BOSS data at mean redshift $\bar{Z}=0.51$.

\subsection{Systematics tests\label{subsec:systematics}}

\subsubsection{Fiducial cosmology}
\begin{figure}[b]
	\centering
	\resizebox{\hsize}{!}{
		\includegraphics{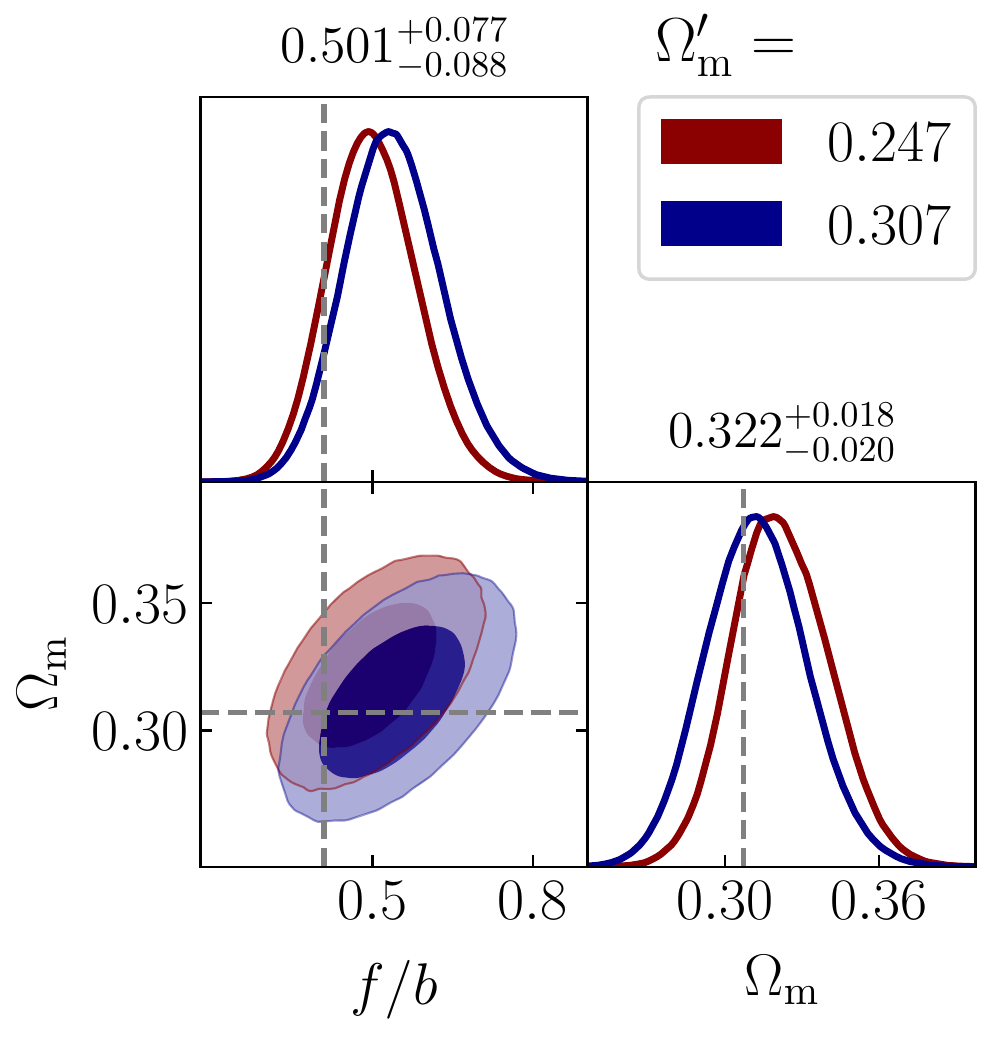}
		\includegraphics{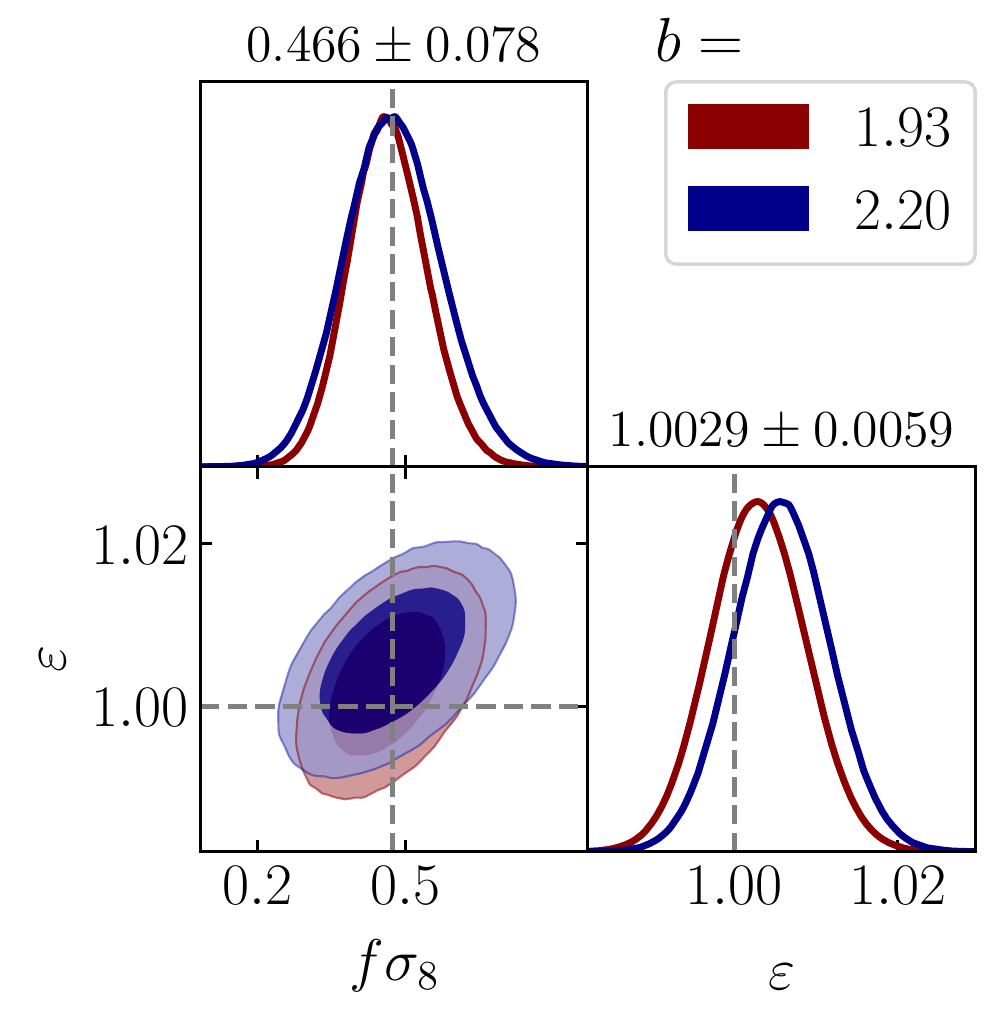}}
	\caption{LEFT: Impact of the fiducial parameter $\Om'$ on the final posterior distribution for $f/b$ and $\Om$ from all voids in the BOSS DR12 data. The top of each column states the mean and standard deviation obtained for the assumed value of $\Om'=0.247$ and dashed lines indicate fiducial values of the default cosmology with $\Om'=0.307$. RIGHT: Impact of the bias of the galaxy sample used in the cross-correlation with all voids from the PATCHY mocks on the derived posterior for $f\sigma_8$ and $\varepsilon$. The top of each column states the mean and standard deviation obtained for the new value $b=1.93$.}
	\label{fig:triangle_sys}
\end{figure}

In order to affirm the robustness of our results, we have performed a number of systematics tests on our analysis pipeline. One potential systematic can be a residual dependence on the fiducial cosmology we assumed in section~\ref{subsec:galaxies} when converting angular sky coordinates and redshifts into comoving space via equation~(\ref{x_comoving}). This conversion preserves the topology of large-scale structure, but in the presence of statistical noise due to sparse sampling of tracers it can have an impact on void identification~\cite{Mao2017}. We investigate how a change of the fiducial cosmology affects our final constraints on cosmological parameters by shifting the fiducial value for $\Om'=0.307$ to $0.247$. This shift amounts to three times the standard deviation we obtain from the posterior on $\Om=0.312\pm0.020$ in the uncalibrated analysis of all voids (see table~\ref{tab:constraints}). We then repeat our entire analysis including the void-finding procedure and sample the posterior on $f/b$ and $\Om$ assuming the new value for $\Om'$. The result is presented in the left panel of figure~\ref{fig:triangle_sys}, showing a very mild impact of the fiducial cosmology on the posterior parameter distribution. The resulting shifts of the posterior mean values are well within the $68\%$ credible regions of both cases and their relative accuracies practically remain unchanged, suggesting the impact of our fiducial cosmology to contribute a marginal systematic effect to our final constraints.

\subsubsection{Galaxy bias}
The bias of the galaxy sample we use to estimate the cross-correlation with voids can contribute another systematic effect on our final parameters. This especially so for the derived combination $f\sigma_8$, which we obtain via multiplying $f/b$ by the average bias $b$ of the galaxy sample, and the Planck-constrained $\sigma_8$ value (see section~\ref{subsec:constraints}). The BOSS data does not readily allow us to define sub-samples of galaxies with known bias values that differ from the sample average. However, the PATCHY mocks provide a bias parameter for every object in the catalog, so we can investigate its influence on our analysis pipeline. As a simple test, we selected $50\%$ of all PATCHY galaxies with a bias value below the median, which amounts to an average of $b=1.93$. Because the galaxy bias follows its own redshift evolution, we had to re-sample the random catalog in order for it to follow the same density-redshift trend as the selected galaxy sample. We then cross-correlate it with our original PATCHY void sample used in section~\ref{subsec:validation} and compare its posterior on $f\sigma_8$ and $\varepsilon$ to the original one from figure~\ref{fig:triangle_mock} in the right panel of figure~\ref{fig:triangle_sys}. The two constraints are very consistent with each other, suggesting that the final result on $f\sigma_8$ does not depend on the bias of the galaxy sample used for the cross-correlation.

\subsubsection{Estimator}
The main advantage of our clustering estimator from equation~(\ref{estimator}) is its simplicity, allowing a fast and precise evaluation of the void-galaxy cross-correlation function and its multipoles without angular binning. In order to assess its accuracy, we have compared it with the more common Landy-Szalay estimator~\cite{Landy1993}
\begin{equation}
\xi^s(\mathbf{s}) = \left(\frac{\langle\mathbf{X},\mathbf{x}\rangle(\mathbf{s})}{\langle\mathbf{X}\rangle\langle\mathbf{x}\rangle} -\frac{\langle\mathbf{X},\mathbf{x}_r\rangle(\mathbf{s})}{\langle\mathbf{X}\rangle\langle\mathbf{x}_r\rangle} -\frac{\langle\mathbf{X}_r,\mathbf{x}\rangle(\mathbf{s})}{\langle\mathbf{X}_r\rangle\langle\mathbf{x}\rangle} +\frac{\langle\mathbf{X}_r,\mathbf{x}_r\rangle(\mathbf{s})}{\langle\mathbf{X}_r\rangle\langle\mathbf{x}_r\rangle}\right)\left/
\left(\frac{\langle\mathbf{X}_r,\mathbf{x}_r\rangle(\mathbf{s})}{\langle\mathbf{X}_r\rangle\langle\mathbf{x}_r\rangle}\right.\right)\;,
\label{LS_estimator}
\end{equation}
which additionally involves the random void-center positions $\mathbf{X}_r$. From the PATCHY mocks we generate a sample of such void randoms by assigning the same angular and redshift distribution of its voids to a randomly generated set of points with $50$ times the number of objects (in analogy to the galaxy randoms, see section~\ref{subsec:galaxies}). We also assign an effective radius to each random void, with the same distribution as the one obtained in the mocks. This guarantees a consistent stacking procedure, as described in section~\ref{subsec:estimators}. We find that the additional terms $\langle\mathbf{X}_r,\mathbf{x}\rangle(\mathbf{s})/\langle\mathbf{X}_r\rangle\langle\mathbf{x}\rangle$ and $\langle\mathbf{X}_r,\mathbf{x}_r\rangle(\mathbf{s})/\langle\mathbf{X}_r\rangle\langle\mathbf{x}_r\rangle$ in the stacked void-galaxy correlation estimator from equation~(\ref{LS_estimator}) are independent of the direction and magnitude of~$\mathbf{s}$, in agreement with the findings of reference~\cite{Hamaus2017}. However, we notice the amplitude of both terms to exceed unity by roughly $20\%$, while their ratio remains very close to one with deviations in the order of $10^{-3}$. This results in a different overall normalization between equations~(\ref{estimator}) and~(\ref{LS_estimator}), making their amplitudes differ by a constant factor of about $1.2$. When we rescale one of the void-galaxy correlation functions by this number, we find the results from these two estimators to be virtually indistinguishable. Because we use the same estimator for $\xi^s(s_\parallel,s_\perp)$ and the projected correlation function $\xi^s_p(s_\perp)$, which is used to infer the real-space $\xi(r)$ via equation~(\ref{xi_d}), any such normalization constant gets absorbed on both sides of equation~(\ref{xi^s_lin}) and therefore has no effect on our model parameters.

\subsubsection{Covariance matrix}
\begin{figure}[t]
	\centering
	\resizebox{\hsize}{!}{
		\includegraphics[trim=5 10 5 5, clip]{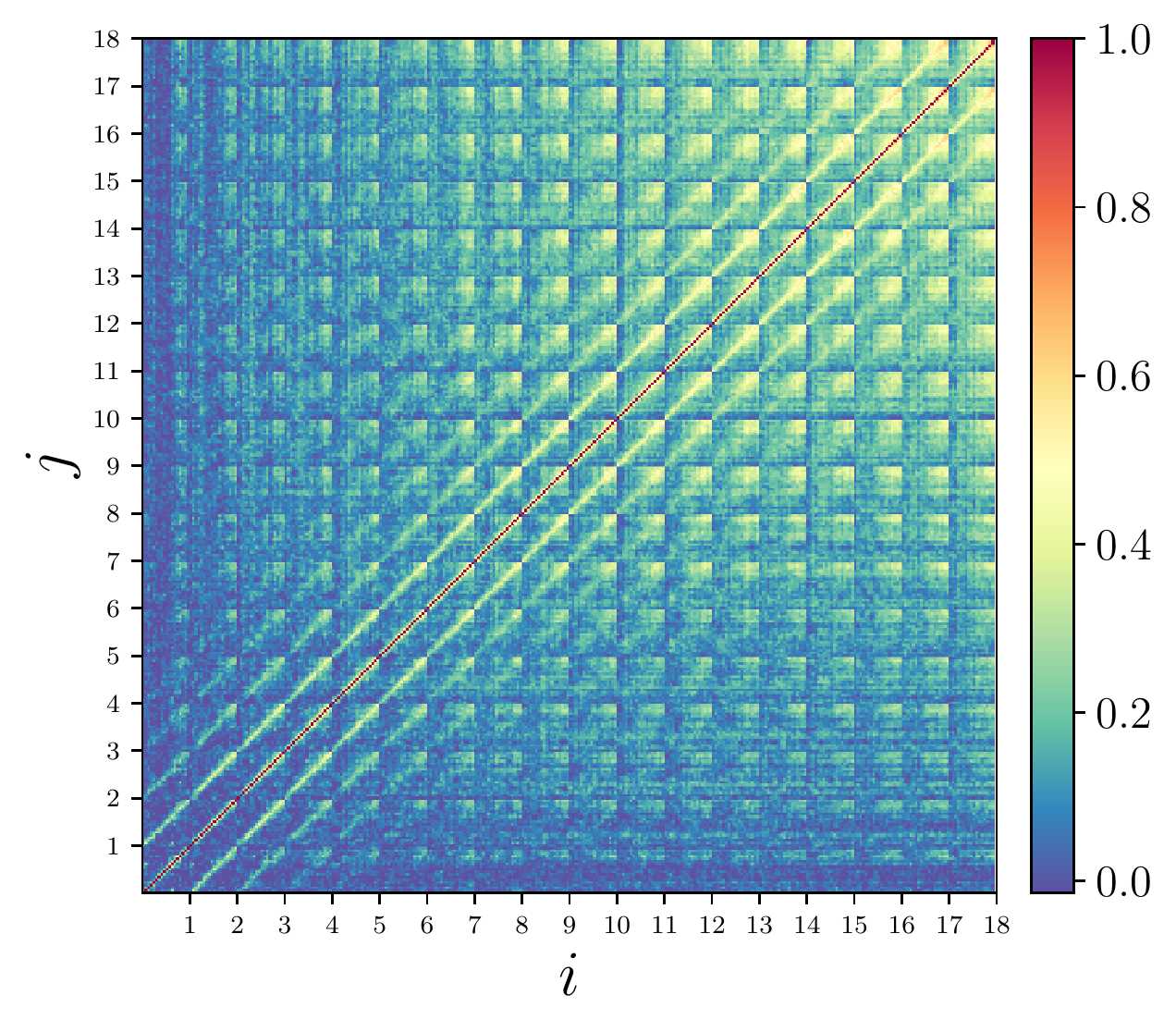}
		\includegraphics[trim=5 10 5 5, clip]{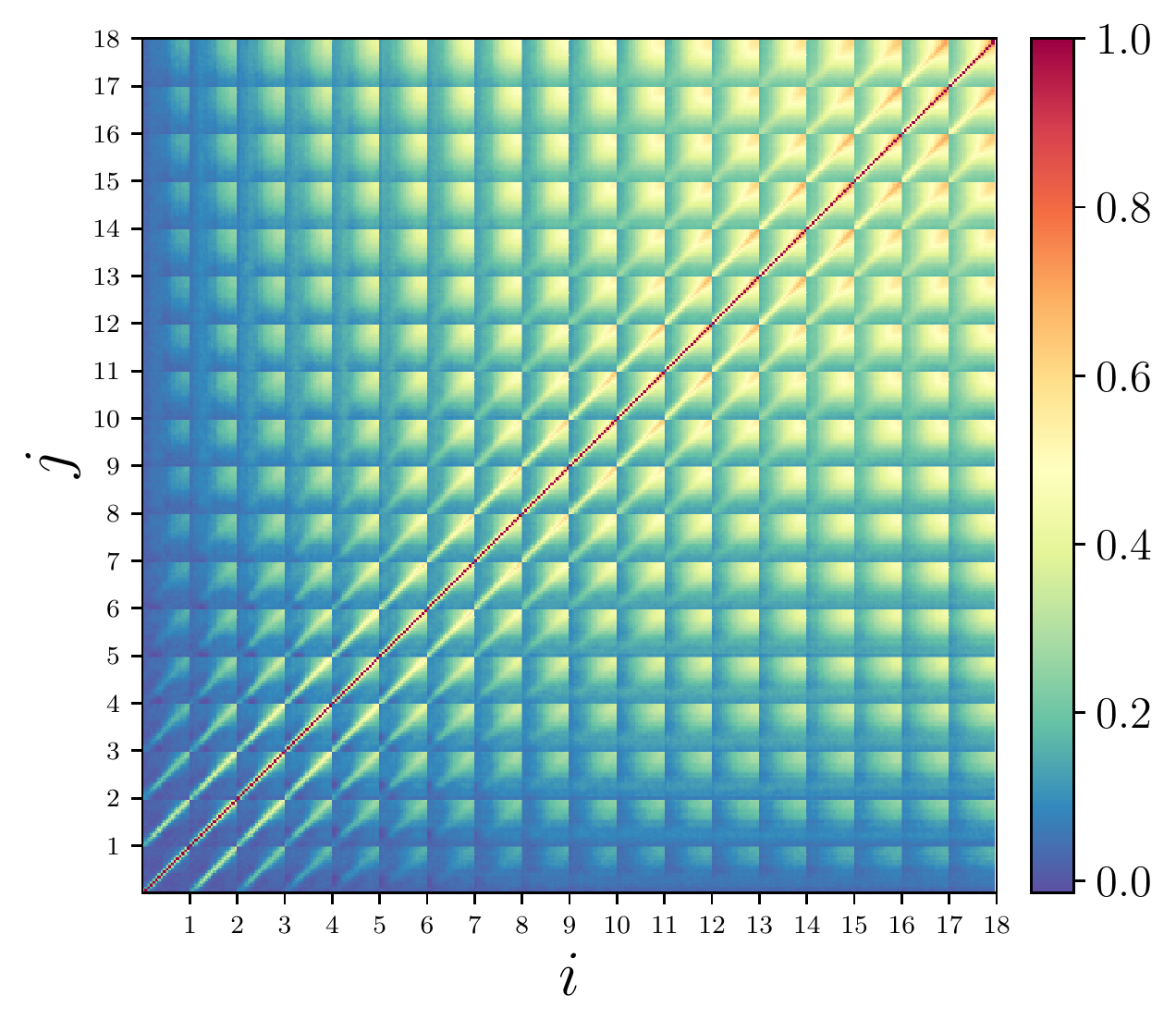}}
	\caption{Covariance matrix (normalized by its diagonal components) of the stacked void-galaxy cross-correlation function $\xi^s(s_\parallel,s_\perp)$ from the BOSS DR12 data (left) and the PATCHY mocks (right).}
	\label{fig:covariance}
\end{figure}
As a last consistency test we investigate the impact of the covariance matrix on our results. The left panel of figure~\ref{fig:covariance} shows the covariance matrix estimated using the jackknife technique on the BOSS data as described in section~\ref{subsec:estimators}, normalized by its diagonal components (i.e., the correlation matrix $\C_{ij}/\sqrt{\C_{ii}\C_{jj}}\,$). Note that this matrix contains $N_\mathrm{bin}^2=(18^2)^2$ elements for the covariance of the two-dimensional correlation function $\xi^s(s_\parallel,s_\perp)$. In order to overcome the statistical noise in the data covariance, we can measure the same quantity for all voids in our $N_\mathrm{m}=30$ independent mock catalogs. The result is shown in the right panel figure~\ref{fig:covariance}, featuring a very similar structure as for the real data. In our main analysis we have used the data covariance for the sake of maintaining an entirely calibration-independent approach. However, when exchanging it by the mock covariance in our likelihood from equation~(\ref{likelihood}), we obtain posteriors that are consistent with our previous results, which is why we do not show them again. This suggests that the estimation of the covariance matrix from the data itself provides a sufficiently precise method allowing us to obtain fully calibration-independent constraints on cosmology from the observed sample of voids.

\subsection{Comparison to previous work \label{subsec:comparison}}
\begin{figure}[t]
	\centering
	\resizebox{\hsize}{!}{
		\includegraphics[trim=30 50 60 80, clip]{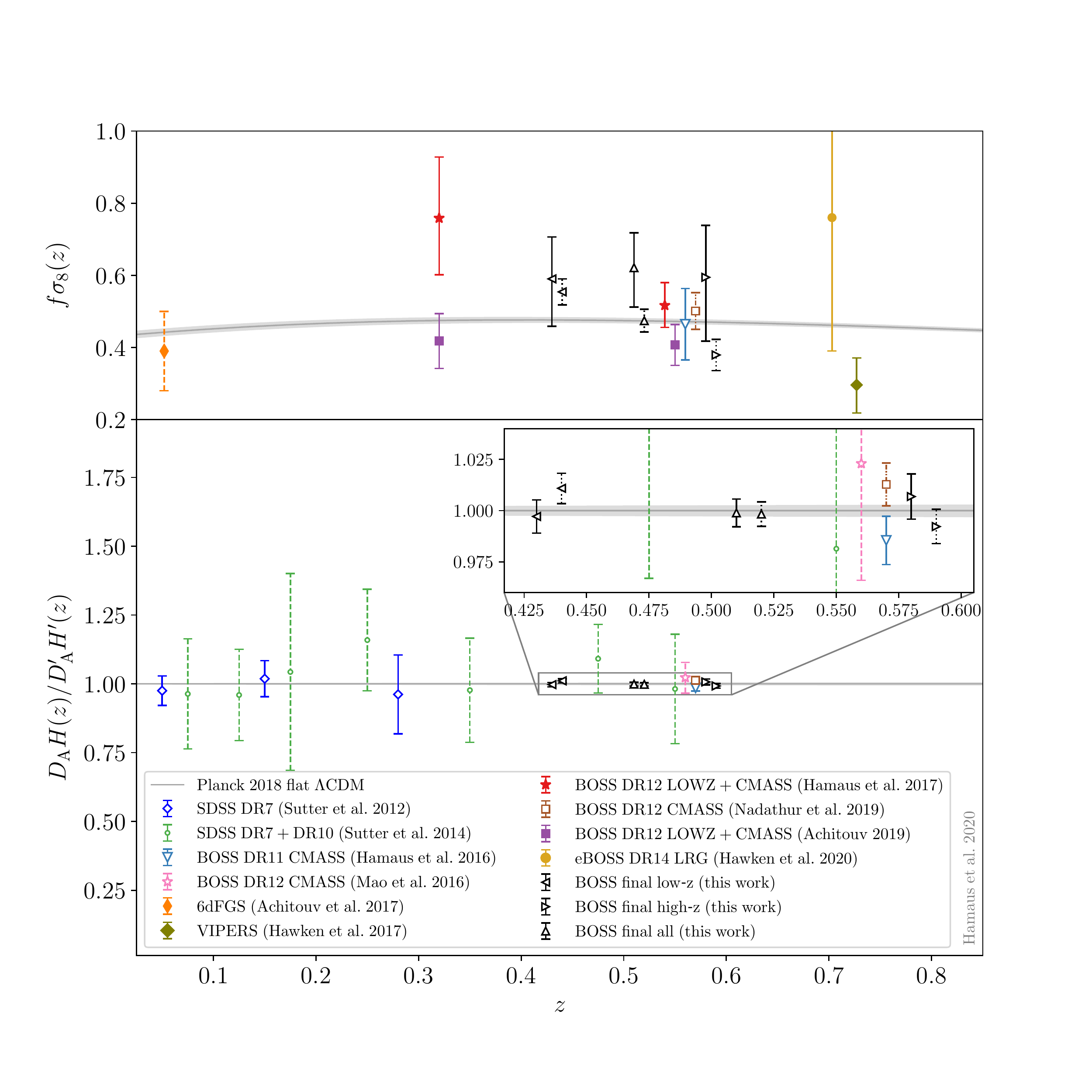}}
	\caption{Comparison of constraints on $f\sigma_8$ and $\DA H$ (mean values with $68\%$ confidence regions) obtained from cosmic voids in the literature, references are ordered chronologically in the figure legend. To improve readability, $\DA H$ is normalized by its reference value $\DA'H'$ in the Planck 2018 flat $\Lambda$CDM cosmology~\cite{Planck2018} (gray line with shaded error band). Filled markers indicate growth rate measurements without consideration of the AP effect, while open markers include the AP test. The line style of error bars indicates various degrees of model assumptions made: model-independent (solid), calibrated on simulations (dashed), calibrated on mocks (dotted), calibrated on simulations and mocks (dash-dotted). All the employed simulations and mocks assume a flat $\Lambda$CDM cosmology. Data points at similar redshifts have been slightly shifted horizontally to avoid overlap.}
	\label{fig:comparison}
\end{figure}
The observational AP test with cosmic voids has already experienced some history since it was first proposed by Lavaux and Wandelt in 2009~\cite{Lavaux2010} and its first measurement by Sutter et al. in 2012~\cite{Sutter2012b}. The early measurements of the AP effect did not yet account for RSD distortions by a physically motivated model, but they calibrated its impact using simulations~\cite{Sutter2014b,Mao2017}. The first joint RSD and AP analysis from observed cosmic voids has been published in 2016 based on BOSS DR11 data~\cite{Hamaus2016}. It demonstrated for the first time that a percent-level accuracy on $\varepsilon$ can be achieved with voids, making them the prime candidate for observational AP tests. Since then, a number of papers appeared that either focused on the RSD analysis of voids exclusively in order to constrain the growth rate~\cite{Achitouv2017a,Hawken2017,Hamaus2017,Achitouv2019,Hawken2020}, or performed further joint analyses including the AP effect~\cite{Nadathur2019b}.

We summarize the constraints on $f\sigma_8$ and $\DA H$ that have been obtained from voids throughout the literature in figure~\ref{fig:comparison}, including the results from this paper. Evidently, the different analysis techniques have progressed over time and achieved significant improvements of accuracy. Moreover, spectroscopic data from a number of surveys covering different redshift ranges has been exploited to this end, including 6dFGS~\cite{6dFGS}, BOSS~\cite{BOSS}, eBOSS~\cite{EBOSS}, SDSS~\cite{SDSS}, and VIPERS~\cite{VIPERS}. All of the published results are consistent with a flat $\Lambda$CDM cosmology, in agreement with the measurements by Planck~\cite{Planck2018}. However, some of the analyses have been calibrated using simulations and / or mocks to determine unknown model ingredients. If such external information has been used and has not been marginalized over, we indicate the calibrated results in figure~\ref{fig:comparison} by different line styles of error bars, as described in the caption. A comparison based on equal terms can only be made by taking this essential fact into account. In addition to this, there are a number of analysis choices that differ among the published results. In the following we provide a list of aspects that we have investigated in more detail and encountered to be relevant for our results.

\subsubsection{Void finding in real vs. redshift space}
Like most papers on the topic of void RSD in the literature, we define voids in observable redshift space. The recent analysis of reference~\cite{Nadathur2019b} advocates the use of reconstruction techniques to identify voids in real space instead. Their centers in real space are then correlated with the original galaxy positions in redshift space to estimate a hybrid real/redshift space void-galaxy cross-correlation function. We have investigated this approach using halo catalogs from $N$-body simulations and calculated the resulting two-point statistics. We confirm that this results in a more elongated shape of $\xi^s(s_\parallel,s_\perp)$ along the line of sight and a change of sign in its quadrupole at small void-centric separations. This can be readily understood from the illustration in figure~\ref{fig:voidstretch}: the separation vector between a void center in real space and one of the void's galaxies in redshift space is now given by the gray dashed line, which is more elongated along the line of sight because it contains a contribution from the velocity of the void center, $\tilde{\mathbf{s}}=\mathbf{r}+\mathbf{u}_\parallel+\mathbf{V}_\parallel$ (with velocities in units of $(1+z_h)/H$). Another consequence of this approach is that for void velocities with $|\mathbf{V}_\parallel|\gtrsim R$, a significant number of galaxies from neighboring voids in redshift space will be closer to the void center in real space than the void's own member galaxies. As a result, the void-galaxy cross-correlation function contains different contributions from galaxies of the same and neighboring voids, depending on the magnitude of $\mathbf{V}_\parallel$. As this effect is difficult to model from first principles, reference~\cite{Nadathur2019b} resorts to the use of mock catalogs to calibrate the form of this hybrid correlation function, and restricts its analysis to the largest $50\%$ of all voids.

The motivation for velocity field reconstruction was grounded on the claim that the velocities $\mathbf{V}$ of void centers cannot be accounted for in all existing models for the void-galaxy cross-correlation function in redshift space~\cite{Nadathur2019b}. This presumption is unfounded, as these models have actually been derived assuming local mass conservation relative to the motion of the void center~\cite{Paz2013,Hamaus2015,Cai2016}. While it is true that absolute void velocities $\mathbf{V}$ are difficult to predict, the same holds for the absolute galaxy velocities $\mathbf{v}$ in the vicinity of void centers. Both $\mathbf{V}$ and $\mathbf{v}$ contain bulk-flow contributions sourced by density fluctuations on scales beyond the void's extent. However, local mass conservation provides a very good prediction for their difference $\mathbf{u}$, as discussed in section~\ref{subsec:dynamic}. A consequence of this is a vanishing hexadecapole $\xi^s_4(s)$, as explained in reference~\cite{Cai2016} and confirmed by our analysis. We further note that the galaxy velocity field $\mathbf{v}$ is anisotropic around void centers in redshift space, as expected from figure~\ref{fig:voidstretch}. The model derived in section~\ref{subsec:correlation} merely assumes statistical isotropy of the field $\mathbf{u}$ in real space, which follows from the cosmological principle. Reference~\cite{Nadathur2019a} speculated about a potential selection effect in favor of voids with higher outflow velocities and hence lower observed central densities in redshift space. As explained in section~\ref{subsec:voids}, our void finder operates on local minima and their surrounding watershed basins, irrespective of any absolute density threshold. Such a selection effect therefore cannot affect voids identified with~\textsc{vide} or~\textsc{zobov}.

Another argument for the use of reconstruction was motivated by the impact of redshift-space distortions on the void-size function~\cite{Nadathur2019a}. We note that the effective radii for voids of any size are expected to change between real and redshift space due to dynamic distortions, as evident from figure~\ref{fig:voidstretch}. However, we only use the observed effective void radii as units to express all separations in either space, which leaves the mapping between $\mathbf{r}$ and $\mathbf{s}$ unchanged. A problematic impact of this mapping can be the destruction of voids from catastrophic redshift-space distortions, such as the FoG effect, or from shot noise due to the sparsity of tracers that may change the topology of watershed basins. Because the FoG effect is limited to scales of a few $\hMpc$ and smaller voids are defined by fewer tracers, this problem becomes more relevant for voids of relatively small size. We account for this potential systematic via marginalization over the nuisance parameters introduced in section~\ref{subsec:parameters}.

In conclusion, velocity field reconstruction is not required to account for the dynamic distortions of voids, as evident from this paper and numerous earlier works~\cite{Paz2013,Hamaus2016,Cai2016,Achitouv2017a,Hawken2017,Hamaus2017,Correa2019,Achitouv2019,Hawken2020}. The velocity field reconstruction technique merely offers an alternative approach to model RSDs around voids, in addition to the existing models. If reconstruction is used in conjunction with another RSD model, dynamic distortions are unnecessarily taken into account twice. The disadvantages of reconstruction include its dependence on a smoothing scale, assumptions on tracer bias and growth rate relations, as well as its sensitivity to survey edges and shot noise~(e.g., \cite{Sherwin2019,Philcox2020}). Last but not least, reconstruction makes the data a function of the theory model. Vice-versa, calibration of the theory model on survey mocks that are informed by the data generates an inverse dependence. If information from the mocks is used in the model, theory and data are intertwined to a degree that makes a rigorous likelihood analysis much more involved. Moreover, this practice forfeits the criteria necessary for an independent model validation. The authors of reference~\cite{Nadathur2019b} claim their analysis to be ``free of systematic errors'', but neglect its systematic dependence on the assumed mock cosmology.

\subsubsection{Void center definition}
Because voids are aspherical by nature, the definition of their centers is not unique. In observations, which typically provide the 3D locations (but not the 3D velocities) of tracers that outline each void, there are in practice two options: the point of minimum tracer density inside the void, or the geometric center defined by the void boundary. Minimum-density centers can be defined as maximally extended empty spheres in a tracer distribution~\cite{Zhao2016}, without requiring the sophistication of a watershed algorithm. The optimal choice of center definition depends on the specific type of application, so it is not possible to make general statements about this. However, for the sake of measuring geometric distortions via the AP effect it is desirable to enhance the amplitude of tracer fluctuations around their background density to increase the signal-to-noise ratio of anisotropic clustering measurements. As described in section~\ref{subsec:voids}, the geometric center (barycenter) retains information about the void boundary and thereby generates a pronounced compensation wall in its cross-correlation with galaxies at a separation of one effective void radius $R$. On the other hand, the minimum-density center produces a stronger negative amplitude of the void-galaxy cross-correlation function at small separations. The number of tracers in a shell of width $\mathrm{d}s$ grows as $s^2$ for a constant tracer density, and even faster for increasing density with $s$, as is the case inside voids. Therefore the coherent compensation walls around void barycenters serve as a lever arm to provide significantly higher signal-to-noise ratios for measurements of anisotropic clustering and hence the AP effect. We have checked this explicitly by repeating our analysis using minimum-density centers, which results in less pronounced compensation walls in $\xi^s(s_\parallel,s_\perp)$, a lower amplitude of its quadrupole, and an uncertainty on the AP parameter $\varepsilon$ of roughly double the size.

\subsubsection{Void stacking}
The method of void stacking is related to the previous aspect, as it affects the void-galaxy cross-correlation function in a similar way. Because voids are objects of finite extent, the correlation of their centers with tracers inside or outside their boundaries is qualitatively different~\cite{Hamaus2014a,Chan2014,Cai2016,Voivodic2020}. This is analogous to the halo model, which ascribes two different contributions to the clustering properties of matter particles, those inside the same halo, and those among different halos~\cite{Seljak2000,Peacock2000}. Therefore, in order to capture the characteristic clustering properties of tracers inside a sample of differently sized voids, one typically rescales the tracer coordinates by the effective radius of their respective host void, a method referred to as void stacking. This guarantees that the void boundaries coherently overlap at a separation of $s=R$, and thus creates a strong compensation wall feature in the stacked void-galaxy cross-correlation function. Without the rescaling procedure, compensation walls of different-size voids do not aggregate, which results in a smeared out correlation function with almost no feature remaining at $s=R$. This smearing in turn is disadvantageous for measurements of AP distortions, following the arguments discussed above. A similar effect can be caused by stacking voids of different evolutionary stages from a wide range of redshifts. For example, the properties of void galaxies are expected to be redshift-dependent~\cite{Kreckel2011a,Kreckel2011b}. This can be accounted for by splitting the void sample into redshift bins. However, in the BOSS DR12 data and the PATCHY mocks we find a very mild evolution with redshift, allowing us to average over the full void sample.

\subsubsection{Correlation function estimation}
It is common practice to estimate correlation functions via counts in shells, i.e. by counting the number of tracers and randoms inside a spherical shell of width $\mathrm{d}s$ at separation $s$. In the interiors of voids, however, the density of tracers is low by construction, which can result in shells with insufficiently low tracer counts to reliably estimate correlation functions that are intended to infer properties of the density field (such as its growth rate $f$). The previously discussed method on void stacking helps in this respect, as one may collect the tracers that fall into a given shell from all rescaled voids of the entire sample. The convergence of the estimator can then be assessed by increasing the void sample size, as we have done using mocks in section~\ref{subsec:validation}. Within our approach we find no dependence of the estimators on sample size, in support of the conclusion that our correlation function statistics have converged. However, if shells with very few or no tracers are encountered in every single void, the counts-in-shell estimator yields biased results, even in the limit of infinite sample size~\cite{Nadathur2015}.

This is particularly relevant for shells in the vicinity of the minimum-density center, which exhibits no nearby tracers by construction. As a consequence, the counts-in-shell estimator yields a value of $\xi=-1$ for all empty shells, regardless of the nature of the tracer distribution. In fact, this is even the case for empty shells in a random distribution of tracers, an example that reveals the limitations of this estimator most clearly. As its name suggests, the counts-in-shell estimator is only defined for non-zero object counts, but returns meaningless results otherwise. Towards larger scales, as soon as the first tracers are encountered in a shell at separation $s_\mathrm{d}$, the correlation estimate abruptly jumps to a value significantly higher than $\xi=-1$, and finally converges to a smooth curve at separations with sufficient tracer counts~\cite{Nadathur2019b}. If voids are not rescaled by their effective radius, this results in a kink in the average correlation function at $s=s_\mathrm{d}$ even for arbitrarily large sample sizes. A similar behavior can be observed when estimating the radial velocity profile of the tracers~\cite{Nadathur2019a}. The resulting bias in the counts-in-shell estimator on small scales breaks the validity of equations~(\ref{u(r)}) and~(\ref{xi(delta)}), which only apply in the continuous limit of high tracer counts, and can be misinterpreted as evidence for an intrinsic non-linearity or stochasticity of the tracer density field.

The shell at separation $s_\mathrm{d}$ indicates the discreteness limit of the tracer distribution. It is determined by the average density of tracers and therefore unrelated to cosmologically induced clustering statistics that can be measured on larger scales. Moreover, discreteness artifacts are notoriously difficult to model from first principles due to their unphysical nature, which leaves no option other than to calibrate them via mock catalogs. In reference~\cite{Nadathur2019a} it is argued that the void-galaxy cross-correlation function exhibits a ``feature'' both in its monopole and quadrupole at separation $s_\mathrm{d}$, which is calibrated on mocks to be used for AP distortion measurements in a later publication~\cite{Nadathur2019b}. It yet remains to be demonstrated whether such features at the discreteness limit of an estimator are of any use for the AP test. They similarly arise in scale-free Poisson distributions, which are insensitive to geometric distortions for the lack of spatial correlations and therefore satisfy the condition $\varepsilon=1$ in any coordinate system. Consequently, in such a scenario the AP test necessarily returns the fiducial cosmology and therefore becomes dominated by confirmation bias. Changing the fiducial cosmology provides a useful sanity check to exclude the presence of confirmation bias, as we have shown in section~\ref{subsec:systematics}.

\subsubsection{RSD model}
We have performed extensive tests to compare the existing RSD models for voids that are available in the literature. This essentially concerns the GSM for voids as proposed by Paz et al.~\cite{Paz2013}, the linear model of Cai et al.~\cite{Cai2016} used in this paper, and variants thereof. We find consistent results with the GSM, albeit with slightly weaker constraints on $f/b$ and $\varepsilon$ due to marginalization over the additional velocity dispersion parameter~$\sigma_v$. Moreover, as the GSM requires an integration over the pairwise velocity probability distribution function in every bin of $\xi^s(s_\parallel,s_\perp)$, it significantly slows down the model evaluation. We find the impact of velocity dispersion to marginally affect our fits to the data, so we settled on the simpler linear model from equation~(\ref{xi^s_lin2}).

Furthermore, we explored extensions of the linear model, such as the full non-linear expression~(\ref{xi^s_nonlin2}). We also tested the model extension proposed in equation (14) of reference~\cite{Nadathur2019a}, which contains terms of linear and second order in $\delta$. Note that in this model, every occurrence of the parameter $f$ is multiplied by a factor of $1+\xi(r)$, unfolding an additional degeneracy between the growth rate and the amplitude of $\xi(r)$, which depends on $b$ and $\sigma_8$. Moreover, that model requires the void mass density profile $\delta(r)$ from simulations in addition to $\xi(r)$ from the mocks, making it even more dependent on the calibration input. However, none of these extensions improves our fits to either data or mocks. We suspect that a rigorous model at the non-linear level must additionally involve an extension of the linear mass conservation relation that leads to equation~(\ref{u(r)}), as suggested by reference~\cite{Achitouv2017b}. Nevertheless, our analysis of the final BOSS data does not indicate any limitations of the simplest linear model from equation~(\ref{xi^s_lin}), in agreement with previous analyses~\cite{Hamaus2017,Achitouv2019}.

\section{Conclusion\label{sec:conclusion}}
We have presented a comprehensive cosmological analysis of the geometric and dynamic distortions of cosmic voids in the final BOSS dataset. The extracted information is condensed into constraints on two key quantities, the RSD parameter $f/b$, and the AP parameter $\varepsilon$. When calibrated on survey mocks, our analysis provides a relative accuracy of $6.6\%$ on $f/b$ and $0.60\%$ on $\varepsilon$ (at $68\%$ confidence level) from the full void sample at a mean redshift of $0.51$. This represents the tightest growth rate constraint obtained from voids in the literature. The AP result even represents the tightest constraint of its kind. However, as these results are calibrated by mock catalogs from a fixed fiducial cosmology, they need to be taken with a grain of salt. Without calibration we are still able to self-consistently model the data, obtaining a relative accuracy of $16.9\%$ on $f/b$ and $0.68\%$ on $\varepsilon$. While the weaker AP constraint still remains unrivaled, the degradation in the uncertainty on $f/b$ is mainly due to its strong anti-correlation with the amplitude of the real-space void-galaxy cross-correlation function $\xi(r)$, which we jointly infer from the data. We emphasize that these uncalibrated constraints are entirely independent from any prior model assumptions on cosmology, or structure formation involving baryonic components, and do not rely on mocks or simulations in any way. They exclusively emerge from the observed data and a linear-theory model that is derived from first principles. With the additional validation of this model on external survey mocks with much higher statistical power, these constraints are robust. The quality of the BOSS data even allows us to analyze sub-samples of voids in two redshift bins. This decreases the mean accuracy per bin by roughly a factor of $\sqrt{2}$, as expected for statistically independent samples.

We account for potential systematics in our analysis via a marginalization strategy. To this end we include two nuisance parameters in the model: $\mathcal{M}$ for modulating the amplitude of the deprojected real-space void-galaxy cross-correlation function $\xi(r)$, and $\mathcal{Q}$ for adjusting the quadrupole amplitude. The first parameter accounts for inaccuracies in the deprojection technique and a possible contamination of our void sample by random Poisson fluctuations that can be mistaken as voids with a shallow density profile. The second parameter accounts for anisotropic selection effects due to catastrophic RSDs (such as the FoG effect) or shot noise, that can affect the identification of voids. In the PATCHY mocks we find significant evidence for $\mathcal{M}>1$, and very marginal evidence for $\mathcal{Q}>1$, while in the BOSS data both parameters are consistent with unity (to within $68\%$ confidence) in our low-redshift bin. At higher redshift we observe a mild indication of $\mathcal{M}>1$ and $\mathcal{Q}<1$, but only at the significance level of $2\sigma$ at best. This observation leads us to draw the following conclusion: survey mocks do not necessarily account for all aspects of, and systematics in, the data. They typically represent different realizations drawn from one and the same set of cosmological and astrophysical parameters. Therefore, using the mocks for model calibration may lead to biased constraints on cosmology. This is particularly relevant in situations where model extensions from the fiducial mock cosmology are explored, such as curvature, the properties of dark energy, or massive neutrinos. In addition, this practice underestimates parameter uncertainties, by up to a factor of $4$ for constraints on growth from RSDs. 

As a final remark, we emphasize the important role that cosmic voids will play in the cosmological analysis of future datasets with a much larger scope. Both observatories from the ground~\cite{DESI,LSST} and from space~\cite{EUCLID,SPHEREX,WFIRST} will soon provide an unprecedented coverage of large-scale structure in the local Universe, increasing the available sample sizes of voids to at least an order of $10^5$ per survey. This implies that currently achievable error bars on RSD and AP parameters will be further reduced by a factor of a few, potentially allowing us to perform precision cosmology at the level of sub-percent accuracy that could bring about deep ramifications concerning the standard model of cosmology. In this work we have merely investigated flat $\Lambda$CDM, which exhibits no signs of inconsistency with the final BOSS data.

\begin{acknowledgments}
We thank David Spergel for reading over our manuscript and providing pertinent feedback and suggestions. NH would like to thank Kerstin Paech for exchange on coding best practices, Giorgia Pollina for useful suggestions to improve the quality of figures, and Enzo Branchini, Chia-Hsun Chuang, Carlos Correa, Luigi Guzzo, Martha Lippich, Nelson Padilla, Ariel S\'anchez, Ravi Sheth, Sergey Sibiryakov, Rien van de Weygaert, and Simon White for inspiring discussions about voids at the MIAPP 2019 workshop on ``Dynamics of Large Scale Structure Formation'' in Garching. NH and JW are supported by the Excellence Cluster ORIGINS, which is funded by the Deutsche Forschungsgemeinschaft (DFG, German Research Foundation) under Germany's Excellence Strategy -- EXC-2094 -- 390783311. AP is supported by NASA grant 15-WFIRST15-0008 to the Nancy Grace Roman Space Telescope Science Investigation Team ``Cosmology with the High Latitude Survey''. GL and BDW acknowledge financial support from the ANR BIG4 project, under reference ANR-16-CE23-0002. The Center for Computational Astrophysics is supported by the Simons Foundation.

Funding for SDSS-III~\cite{SDSS} has been provided by the Alfred P. Sloan Foundation, the Participating Institutions, the National Science Foundation, and the U.S. Department of Energy Office of Science. SDSS-III is managed by the Astrophysical Research Consortium for the Participating Institutions of the SDSS-III Collaboration including the University of Arizona, the Brazilian Participation Group, Brookhaven National Laboratory, Carnegie Mellon University, University of Florida, the French Participation Group, the German Participation Group, Harvard University, the Instituto de Astrofisica de Canarias, the Michigan State/Notre Dame/JINA Participation Group, Johns Hopkins University, Lawrence Berkeley National Laboratory, Max Planck Institute for Astrophysics, Max Planck Institute for Extraterrestrial Physics, New Mexico State University, New York University, Ohio State University, Pennsylvania State University, University of Portsmouth, Princeton University, the Spanish Participation Group, University of Tokyo, University of Utah, Vanderbilt University, University of Virginia, University of Washington, and Yale University.
\end{acknowledgments}

\bibliography{ms}
\bibliographystyle{JHEP}

\end{document}